\documentclass[12pt]{article}
\usepackage{amssymb}
\usepackage{mathtools}
\usepackage{amsmath}
\usepackage{amstext}
\usepackage{graphicx}
\usepackage{placeins}
\usepackage{epsfig}
\usepackage{verbatim} 
\usepackage{caption}
\usepackage{jheppub}
\usepackage{slashed}
\usepackage{color}
\usepackage{ulem}

\usepackage{enumitem}
\usepackage{subfigure}
\usepackage{bbm}	
\usepackage{parskip}
\usepackage{dsfont}

\usepackage{tikz}
\usepackage{tikz-feynman}
\usepackage{tkz-euclide}
\usetkzobj{all}

\usepackage[framemethod=TikZ]{mdframed}
\mdfdefinestyle{boxed}{%
linecolor=red,
roundcorner=5pt,
innerleftmargin=0,%
innerrightmargin=0,
backgroundcolor=gray!10,%
}

\def\bea{\begin{eqnarray}}
\def\eea{\end{eqnarray}}
\def\be{\begin{equation}}
\def\ee{\end{equation}}
\def\ba{\begin{array}}
\def\ea{\end{array}}

\newcommand{\calL}{{\cal L}}

\newcommand{\mpl}{{M_{\mathrm{Pl}}}}

\newcommand{\bfk}{{\bf  k}}
\newcommand{\bfp}{{\bf  p}}

\newcommand{\bfq}{{\bf  q}}
\newcommand{\bfx}{{\bf  x}}

\newcommand{\bfy}{{\bf  y}}

\def\d{{ d}}

\def\0{{\boldsymbol 0}}

\def\k{{\boldsymbol{k}}}

\begin{document}

\setlength\arraycolsep{2pt}

\renewcommand{\theequation}{\arabic{section}.\arabic{equation}}
\setcounter{page}{1}

\title{
On the time evolution of cosmological correlators
}

\author[a,b]{Sebasti\'{a}n C\'{e}spedes}
\author[c]{, Anne-Christine Davis}
\author[c,d]{and Scott Melville}
\affiliation[a]{ICTP, Strada Costiera 11, 34151 Trieste, Italy}
\affiliation[b]{Instituto de F\'{i}sica Te\'{o}rica UAM/CSIC, Calle Nicol\'{a}s Cabrera 13–15 Cantoblanco E-28049 Madrid, Spain}
\affiliation[c]{DAMTP, Center for Mathematical Sciences, University of Cambridge, CB3 0WA, UK}
\affiliation[d]{Emmanuel College, University of Cambridge, CB2 3AP, UK}

\emailAdd{a.c.davis@damtp.cam.ac.uk}
\emailAdd{sebastian.cespedes@uam.es}
\emailAdd{scott.melville@damtp.cam.ac.uk}

\abstract{
Developing our understanding of how correlations evolve during inflation is crucial if we are to extract information about the early Universe from our late-time observables. To that end, we revisit the time evolution of scalar field correlators on de Sitter spacetime in the Schr\"{o}dinger picture.
By direct manipulation of the Schr\"{o}dinger equation, we write down simple ``equations of motion" for the coefficients which determine the wavefunction. 
Rather than specify a particular interaction Hamiltonian, we assume only very basic properties (unitarity, de Sitter invariance and locality) to derive general consequences for the wavefunction's evolution. 
In particular, we identify a number of ``constants of motion'': properties of the initial state which are conserved by any unitary dynamics. We further constrain the time evolution by deriving constraints from the de Sitter isometries and show that these reduce to the familiar conformal Ward identities at late times. Finally, we show how the evolution of a state from the conformal boundary into the bulk can be described via a number of ``transfer functions'' which are analytic outside the horizon for any local interaction. These objects exhibit divergences for particular values of the scalar mass, and we show how such divergences can be removed by a renormalisation of the boundary wavefunction---this is equivalent to performing a ``Boundary Operator Expansion'' which expresses the bulk operators in terms of regulated boundary operators.
Altogether, this improved understanding of the wavefunction in the bulk of de Sitter complements recent advances from a purely boundary perspective, and reveals new structure in cosmological correlators.  
}


\date{\today}
\maketitle

\newpage
\section{Introduction}

Extracting imprints of the early inflationary Universe from our late-time observables is a central goal of modern cosmology---providing a new window into fundamental physics in the high-energy regime.  
In order to extract as much information as possible from these observables, a great deal of recent research has been devoted to better understanding how correlations are produced and evolve during inflation.   

In this work, we study the time evolution of cosmological correlators using the bulk Schr\"{o}dinger equation. 
Rather than specify a particular Hamiltonian, our aim is to use only very basic assumptions---such that the interactions are \emph{unitary}, \emph{de Sitter invariant} and \emph{local}---to derive general consequences for the wavefunction of the Universe. This model-independent approach to studying cosmological correlators is very much inspired by the successes enjoyed by the $S$-matrix programme on flat Minkowski space \cite{Eden, Chew}, in which similar foundational properties (unitarity, Lorentz invariance and locality) can be used to efficiently bootstrap scattering amplitudes without specifying a particular Lagrangian (see e.g. \cite{Benincasa:2013faa, Elvang:2013cua, Cheung:2017pzi} for recent reviews). 
The advantage of such an approach is that our results capture a large class of different models, and 
are not hindered by our ignorance of the detailed inflationary dynamics.

The early Universe is well-described by a period of quasi-de Sitter expansion \cite{Guth:1980zm,Starobinsky:1980te,Linde:1981mu,Albrecht:1982wi} in which the spontaneous breaking of time translations inevitably gives rise to a scalar excitation \cite{Creminelli:2006xe, Cheung:2007st}. Therefore, at the simplest level, inflationary correlators may be approximated by the correlators of a scalar field on a fixed de Sitter spacetime---this is the focus of our work. 

\paragraph{Evolution of the Wavefunction:}
One of the earliest approaches for computing correlation functions on de Sitter spacetime is the wavefunction formalism~\cite{Hartle:1983ai,Halliwell:1984eu}. In this formalism, the state of the system at time $\eta$ is described by a linear functional $\Psi_\eta [ \phi ]$, from which all equal-time correlation functions of the field $\phi$ can be efficiently computed (see for instance \cite{Maldacena:2002vr, Harlow:2011ke, Pimentel:2013gza}). 
Although modern computations of the wavefunction often favour path integral techniques \cite{Ghosh:2014kba,Anninos:2014lwa,Goon:2018fyu} (e.g. borrowing the bulk-to-boundary and bulk-to-bulk propagators of holography to compute the on-shell action), in this work we adopt the Schr\"{o}dinger picture, representing observables in terms of $\hat{\phi}$ and its canonical momentum $\hat{\Pi}$, and $\Psi_\eta [ \phi]$ is determined by solving the Schr\"{o}dinger equation. This picture naturally focuses on the interaction Hamiltonian rather than the Lagrangian, and so properties such as unitarity (hermiticity of the Hamiltonian) are made manifest\footnote{
The price to be paid for choosing this approach is that the symmetries (which were manifest in the Lagrangian) are now obscured. 
On Minkowski spacetime we are able to implement Lorentz invariance in such a powerful way (via off-shell mode functions) that nowadays the Lagrangian approach is used almost exclusively. However, for cosmology, since we are not yet able to implement de Sitter boosts in such a powerful way, there is merit in using a Hamiltonian description to benefit from the manifest unitarity.
}. 

\paragraph{Unitarity and Constants of Motion:}
Loosely speaking, the importance of a hermitian Hamiltonian lies in the conservation of probability. Since $U = e^{i \, \mathcal{H} \, t}$ is the operator which implements time evolution, if a state $| \Psi \rangle$ is initially properly normalised then at later times $\langle \Psi | U^\dagger U | \Psi \rangle$ remains normalised only if $U$ is unitary. Another way to phrase this is that $\int \mathcal{D} \phi \, | \Psi_\eta [\phi] |^2$ is a \textit{constant of motion}. Since $\Psi_\eta [ \phi]$ is a functional,  unitarity leads to the conservation of an infinite number of functions (one at each order in $\phi^n$)---we will call these constants of motion $\{ \beta_n \}$, and construct the first two explicitly. 
 
Previously, unitarity constraints have been derived in models of inflation by focusing on subhorizon scales, at which the background expansion can be ignored and scattering amplitudes can be constructed analogously to flat space \cite{Baumann:2011su, Baumann:2014cja, Baumann:2015nta, Grall:2020tqc}. One notable exception is the very recent work of \cite{Goodhew:2020hob}, which also uses properties of the wavefunction under unitary evolution to derive a ``Cosmological Optical Theorem''. Our constants of motion provide a major step beyond subhorizon scattering, giving a constraint which applies to the wavefunction on horizon scales (relevant for observation), and also a complementary way to understand the important result of \cite{Goodhew:2020hob}, since for the particular case of the Bunch-Davies initial condition the conservation of our $\beta_n$ directly implies the Cosmological Optical Theorem.

\paragraph{de Sitter Isometries:}
Although the wavefunction (and our above constants of motion) may be defined on any curved spacetime, we will focus primarily on an exact de Sitter background.  
The isometries of de Sitter spacetime are well-understood, and by now it is well-known that at late times the $(d+1)$-dimensional de Sitter symmetries approach those of the $d$-dimensional conformal group, and this places constraints on both the equal-time correlators and the wavefunction  as $\eta \to 0$ \cite{Antoniadis:2011ib, Creminelli:2011mw, Bzowski:2013sza, Kundu:2014gxa,Kundu:2015xta}. 
This lies at the heart of the recently proposed ``Cosmological Bootstrap'' program \cite{Arkani-Hamed:2015bza,Arkani-Hamed:2018kmz,Sleight:2019hfp,Baumann:2019oyu}, which aims to determine the boundary wavefunction by solving the conformal Ward identities.

However, less well-understood are the consequences of these symmetries (particularly de Sitter boosts) in the bulk, at finite $\eta$. One reason for this is that, while all de Sitter observers agree on the field eigenstate at $\eta = 0$ (since this timeslice is invariant under all de Sitter isometries), a field eigenstate at finite $\eta$ is generally observer-dependent, and therefore $\Psi_\eta [ \phi ] = \langle \phi | \Psi \rangle$ transforms in a more complicated way. By defining the Noether charges associated with dilations and de Sitter boosts, we will provide compact expressions for how the wavefunction changes under de Sitter transformations. This allows us to identify the states which are de Sitter invariant on any bulk timeslice (not just at the conformal boundary). 


\paragraph{Locality and Analyticity:}
Supposing one has successfully solved the conformal Ward identities and identified the boundary state at $\eta = 0$. How is this related to the bulk dynamics? Well, if we evolve the boundary state into the bulk, going a small $\eta$ from the conformal boundary, it should be possible to express this new state as a small perturbation from the boundary state. With this logic, we identify a set of ``transfer functions'', whose purpose is to map a boundary state (correlator) into a bulk state (correlator) at finite $\eta$. Providing the bulk interactions are local, these functions have the schematic form $1 + (- k \eta )^2 + (- k \eta)^4+...$ when $\eta$ is small and are smooth, analytic functions of the momenta of the fields in the correlator. 
When $| k \eta |$ approaches one (i.e. when a mode crosses the horizon), this series must be resummed and it is this resummation which produces non-analyticities in the transfer functions/bulk wavefunction (in spite of the Hamiltonian being an analytic function of $k$). 

One particular non-analyticity which the transfer functions always develop as a result of this resummation is a $1/k_T$ pole in the total energy, $k_T = \sum_j | \bfk_j|$, together with various so-called ``folded'' non-analyticities. 
In recent years there has been growing interest in the analytic structure of cosmological correlators, and in particular in the pole at $k_T = 0$ \cite{Maldacena:2011nz,Raju:2012zr,Arkani-Hamed:2015bza,Arkani-Hamed:2018kmz}, whose residue is related to the corresponding scattering amplitude on Minkowski space. 
We show that, when propagating a boundary state into the bulk, this pole only emerges on subhorizon scales $| k \eta| \gg 1$, and further show that the complementary limit $| k \eta| \ll 1 $ is free from non-analyticities providing both the interactions and the initial state are analytic (local) functions of the momenta.

\subsection*{Summary of Main Results}

We will work throughout with the following ansatz for the wavefunction,
\begin{align}
 \Psi_\eta [ \phi ] = \exp \left(  i \, g^{n-2} \, \Gamma_\eta [ g \phi ]  \right) \;\;\;\; \text{where} \;\;\;\; \Gamma_\eta[ \phi] = \sum_{n=2}^{\infty} \int_{\bfk_1 ... \bfk_n}  \, c_{\bfk_1 ... \bfk_n} (\eta) \, \phi_{\bfk_1} ... \phi_{\bfk_n} \; ,  
\end{align}
which reduces the functional $\Psi_\eta [ \phi]$ to a series of functions $c_{\bfk_1 ... \bfk_n} (\eta)$ (or $c_n$ for short), called \textit{wavefunction coefficients}. We focus on states that are approximately Gaussian, using $g$ as a weak coupling which controls our perturbative expansions. 
Our main results are:

\begin{itemize}

\item[(i)] Firstly, by direct manipulation of the Schr\"{o}dinger equation, we derive simple equations of motion for the time evolution of the wavefunction coefficients, $\partial_\eta c_n$. 
Explicit expressions for $\partial_\eta c_3$ and $\partial_\eta c_4$ are given in \eqref{eqn:HJc3} and \eqref{eqn:HJc4}.  
Solving this system of equations reproduces known results from the analogous path integral computation, only now at each order in $\phi^n$ a constant of integration must be specified---this encodes the initial condition for $\Psi_\eta[\phi]$, which can now be freely varied. 
Explicitly, the master equation from which any non-Gaussian $\partial_\eta c_n$ can be read off is,
\begin{align}
- \partial_\eta \left(
 \Gamma^I_{\eta} +  \frac{i}{2} \int_{\bfk}  \frac{ f_\bfk (\eta) }{ f^*_\bfk (\eta) }   \frac{\delta \Gamma^I_{\eta} }{ \delta \phi^I_{\bfk} } \frac{\delta \Gamma^I_{\eta} }{ \delta \phi^I_{- \bfk} } 
  \right)  = \mathcal{H}^I  - \frac{i}{2} \int_{\bfk} \frac{ f_{\bfk} (\eta) }{ f^*_{\bfk} (\eta) } \partial_\eta 
 \left( 
  \frac{\delta \Gamma^I_{\eta} }{\delta \phi^I_{\bfk} }  
  \frac{\delta \Gamma^I_{\eta} }{\delta \phi^I_{-\bfk} }    
  \right)    \; ,
  \label{eqn:intro_i}
\end{align}
where $\mathcal{H}$ is the interaction Hamiltonian and the superscript $I$ denotes that all fields should be normalised using the mode function, $ \phi_{\bfk} = \phi^I_{\bfk} f^*_{\bfk} (\eta) $.

\item[(ii)] 
We use unitarity of the interaction Hamiltonian to identify new \textit{constants of motion} at every order in $\phi^n$. For example, defining the following combinations,
\begin{align}
\frac{ \beta_{\bfk_1 \bfk_2 \bfk_3} }{  f_{\bfk_1}^* f_{\bfk_2}^* f_{\bfk_3}^* }  &=   c_{\bfk_1 \bfk_2 \bfk_3} - c^*_{\bar{\bfk_1} \bar{\bfk}_2 \bar{\bfk}_3}     \\
\frac{ \beta_{\bfk_1 \bfk_2 \bfk_3 \bfk_4} }{  f_{\bfk_1}^* f_{\bfk_2}^* f_{\bfk_3}^* f_{\bfk_4}^*  }  &= 
c_{\bfk_1 \bfk_2 \bfk_3 \bfk_4} - c^*_{\bar{\bfk_1} \bar{\bfk}_2 \bar{\bfk}_3 \bar{\bfk}_4}  
+ i \sum_{\rm perm.}^3 | f_{\bfk_s} |^2   \left(  c_{\bfk_1 \bfk_2 -\bfk_s} - c^*_{\bar{\bfk_1} \bar{\bfk}_2 -\bfk_s}  \right) \left( c_{\bfk_3 \bfk_4 \bfk_s} - c^*_{\bar{\bfk_3} \bar{\bfk}_4 \bfk_s}  \right)   \nonumber
\end{align}
where the argument $\bar{\bfk}$ is defined such that $f_{\bar{\bfk}} (\eta) = f^*_{\bfk} (\eta)$, the equations of motion for $c_3$ and $c_4$ become simply $\partial_\eta \beta_3 \propto \mathcal{H}_3 - \mathcal{H}^\dagger_3$ and $\partial_\eta \beta_4 \propto \mathcal{H}_4 - \mathcal{H}_4^{\dagger}$ (where $\mathcal{H}_3$ and $\mathcal{H}_4$ are the cubic and quartic terms in the interaction Hamiltonian). $\beta_3$ and $\beta_4$ are therefore constants of motion, preserved by any unitary dynamics (hermitian Hamiltonian).  
When restricting to the Bunch-Davies initial condition\footnote{
In the Bunch-Davies vacuum state, $\bar{\bfk}$ is given simply by $|\bar{\bfk}| = - |\bfk|$.
}, in which all $\beta_n = 0$, this reduces to the Cosmological Optical Theorem which has recently appeared in \cite{Goodhew:2020hob}.     
 
\item[(iii)]
Invariance under the de Sitter isometries can be imposed directly on the correlation functions by regarding $\langle \phi_{\bfk_1} ... \phi_{\bfk_n} \rangle (\eta)$ as the equal-time limit of $\langle \phi_{\bfk_1} ( \eta_1) ... \phi_{\bfk_n} ( \eta_n) \rangle$ in the Heisenberg picture. We show that this results in, 
\begin{align}
\left(  (n-1) d  - \eta \partial_\eta   + \sum_J  \bfk_J \cdot \partial_{\bfk_J}   \right)  \langle \phi_{\bfk_1} ... \phi_{\bfk_n}  \rangle'  &= 0    \\
\sum_J \left(   2 \left( d + \bfk_J \cdot \partial_{\bfk_J}  \right) \partial_{\bfk_J}  - \bfk_J \partial_{\bfk_J}^2     \right)  \langle \phi_{\bfk_1} ... \phi_{\bfk_n}  \rangle' &= \frac{2 \eta}{\Omega^{d-1}} \sum_J  \partial_{\bfk_J} \langle \phi_{\bfk_1} ... \Pi_{\bfk_J} ... \phi_{\bfk_n}  \rangle'   \, ,   \nonumber
\end{align}
where $\hat{\Pi}$ is the momentum conjugate to $\hat{\phi}$, and the prime indicates that the overall momentum-conserving $\delta$-function has been removed. Analogous constraints also apply to correlators involving arbitrary insertions of both $\hat{\phi}$ and $\hat{\Pi}$.

\item[(iv)]
The de Sitter isometries can also be implemented directly on the wavefunction via the conserved Noether charges $D$ and $\mathbf{K}$ (associated to dilations and boosts respectively). We show that this results in,
\begin{align}
\delta_D \Gamma_{\eta}  &=  - \eta \partial_\eta \Gamma_{\eta}  + \int_{\bfp}   \;  \phi_{-\bfp} \, \bfp \cdot \partial_{\bfp} \, \frac{\delta \Gamma_\eta }{ \delta \phi_{\bfp} }    \, ,  \\
 \delta_{\mathbf{K}} \Gamma_\eta  &=   \int_{\bfp} \left[ \frac{ \eta }{ \Omega^{d-1}} \frac{\delta \Gamma_\eta}{\delta \phi_{-\bfp}} \,  \frac{\partial}{ \partial \bfp } \, \frac{ \delta \Gamma_\eta }{\delta \phi_{\bfp}}    
 + \phi_{-\bfp} \left( 2 \bfp \cdot \partial_\bfp \frac{\partial}{\partial \bfp}  - \bfp \partial_{\bfp}^2     \right) \frac{\delta \Gamma_\eta}{\delta \phi_{\bfp}} 
   \right]    + 2 \eta \partial_{\bfq} \mathcal{H}^{\rm int}_{\bfq} |_{\bfq= 0} \; .  \nonumber 
\end{align}
where $\mathcal{H}_{\bfq}$ is the Fourier transform of the Hamiltonian density.
Like \eqref{eqn:intro_i}, these equations impose relations on every wavefunction coefficient, $c_n$.
Although the boost constraint in the bulk depends on the details of the interaction Hamiltonian $\mathcal{H}^{\rm int}$, this becomes a model-independent constraint at late times ($\eta \to 0$), where the interactions switch off, and one recovers the usual conformal Ward identities\footnote{
Except for special values of the scalar mass, at which the interactions do not turn off sufficiently quickly as $\eta \to 0$ and the conformal Ward identities must be corrected by anomalous terms.
}.

\item[(v)] Once a boundary value for the wavefunction is provided at $\eta = 0$, say $\Psi_{\eta =0} [ \phi]$ with $c_{\bfk_1 ... \bfk_n} (\eta = 0) = \alpha_{\bfk_1 ... \bfk_n}$, we can propagate this into the bulk using a set of transfer functions. Schematically, these act as the coefficients in the following expansion,
\begin{align}
 c_n (\eta ) = \alpha_n + \sum_{j=2}^{\infty} \frac{\delta c_n (\eta)}{ \delta \alpha_j}  \alpha_j + \frac{1}{2!} \sum_{i , j =2}^{\infty} \frac{\delta c_n (\eta)}{ \delta \alpha_i \delta \alpha_j} \alpha_i \alpha_j + ... \;\; .
\end{align}
\textit{Locality} of the bulk interactions corresponds to \textit{analyticity} of these transfer functions in every momenta $\bfk_j$ until the corresponding mode enters the horizon $(| k_j \eta | \sim 1)$, at which point non-analyticities may develop (e.g. the $1/k_T$ pole).  

\item[(vi)] For particular values of the scalar mass, the interactions do not turn off sufficiently quickly at late times and the wavefunction coefficients diverge. We show how such boundary divergences can be \textit{renormalised} by redefining the boundary condition at $\eta = 0$ (i.e. $\alpha_n \to \alpha_n^{\rm ren}$), and also provide an equivalent prescription in terms of a \textit{Boundary Operator Expansion}. 
This is a mapping between the bulk operators $\hat{\phi}$ and $\hat{\Pi}$ (with dilation weight $0$ and $d$ respectively) to the boundary operators $\hat{\varphi}$ and $\hat{\pi}$ (with dilation weight $\Delta$ and $d-\Delta$ respectively),
\begin{align}
\lim_{\eta \to 0} \left( \frac{ \hat{\phi} }{ (-\eta)^{\Delta}} \right) &=  \hat{\varphi}  + \sum_j Z_{\phi \mathcal{O}_j}  \, \hat{\mathcal{O}}_j [ \hat{\varphi}  , \hat{\pi} ] \; , \\
\lim_{\eta \to 0}  \left( (-\eta)^{\Delta} \hat{\Pi} \right) &=  \hat{\pi}  + \sum_j Z_{\Pi \mathcal{O}_j}  \, \hat{\mathcal{O}}_j [ \hat{\varphi}  , \hat{\pi} ] \;  ,  \nonumber 
\end{align}
where $\hat{\mathcal{O}}_j$ is a basis of composite boundary operators, and $\Delta$ is determined by the scalar field mass ($m^2 = \Delta (d- \Delta)$). Typically, dilation invariance sets every mixing coefficient $Z_{\phi \mathcal{O}_j}$ and $Z_{\Pi \mathcal{O}_j}$ to zero, but divergences can occur whenever there exists an operator with $\Delta_{\mathcal{O}_j} = \Delta$ or $\Delta_{\mathcal{O}_j} = d- \Delta$ and the corresponding mixing coefficient is non-zero. For instance, $\hat{\mathcal{O}} = \hat{\varphi}^n$ mixes into $\hat{\Pi}$ when $n \Delta = d - \Delta$ and into $\hat{\Phi}$ when $n\Delta = \Delta$ (these correspond to ultra-local and semi-local divergences respectively).

\end{itemize}

\begin{figure}[htbp!]
\centering
\includegraphics[width=0.95\textwidth]{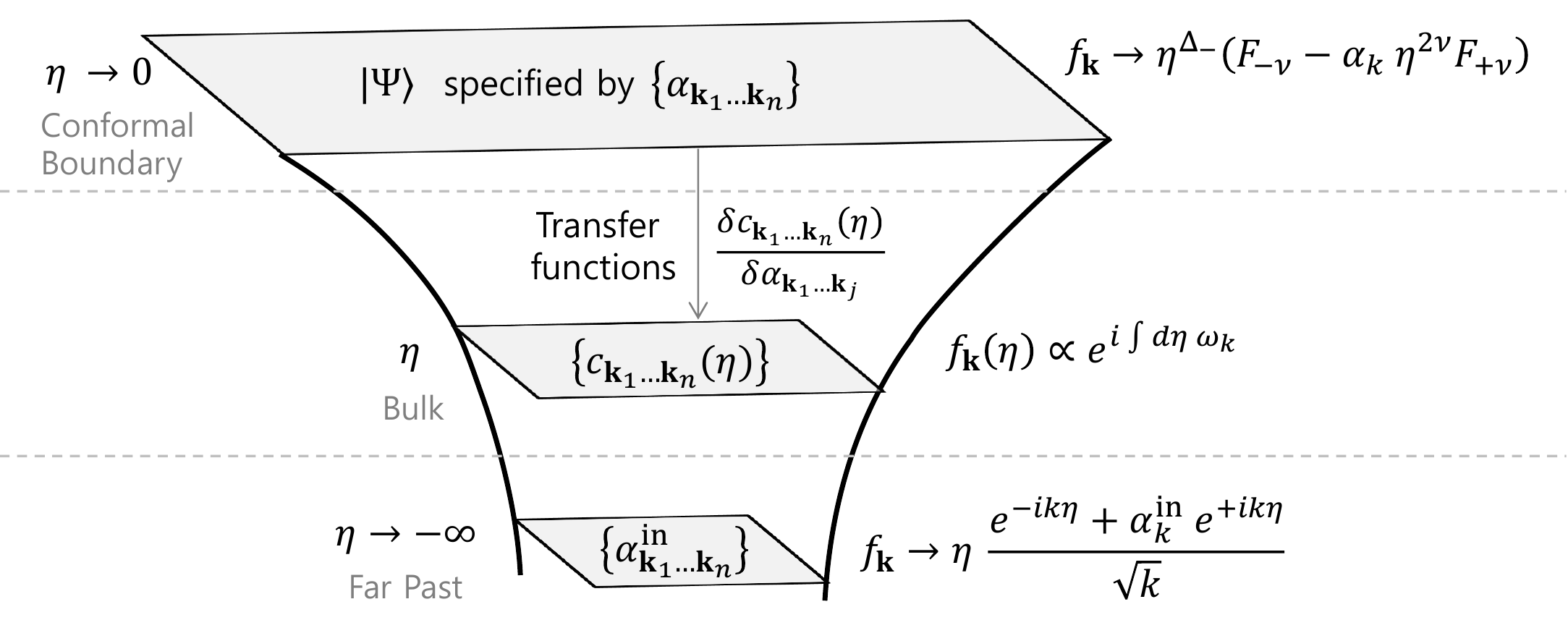}
\caption{
A cartoon of the expanding de Sitter spacetime. We refer to late times, $\eta \to 0$, as the conformal boundary, and denote the wavefunction coefficients there by $\alpha_n$. This boundary condition can be translated into a bulk wavefunction, with coefficients $c_n (\eta)$, by means of various ``transfer functions'' $\mathcal{I}_{\bfk_1 ... \bfk_n}^{\nu_1 ... \nu_n} (\eta)$ which we define in Section \ref{sec:superhorizon}---these objects are analytic (for any local Hamiltonian) until horizon crossing at $| k \eta | \sim 1$ where they develop non-analyticities (such as the $1/k_T$ pole). In the far past, $\eta \to -\infty$, we denote the wavefunction coefficients as $\alpha_n^{\rm in}$. Imposing the Bunch-Davies vacuum state in the far past corresponds simply to $\alpha_n^{\rm in} = 0$. 
}
\end{figure}

\subsection*{Synopsis and Conventions}
In section \ref{sec:timeEvolution}, we briefly review time evolution in the Schr\"{o}dinger picture, derive simple equations of motion for the wavefunction coefficients, and identify new constants of motion which are preserved by any unitary dynamics. In section \ref{sec:isometries}, we derive constraints on both the equal-time correlators and the wavefunction coefficients from the isometries of de Sitter spacetime. In section \ref{sec:superhorizon}, we show that the bulk wavefunction may be written in terms of a boundary condition at $\eta = 0$ and a number of transfer functions---locality of the bulk interactions is then manifest as analyticity of these functions outside the horizon---and show how to renormalise the IR divergences which can appear in the $\eta \to 0$ limit. 
In section \ref{sec:examples}, we discuss in detail three particular examples, namely a conformally coupled scalar, a massless scalar, and the EFT of inflation, before finally concluding in section~\ref{sec:discussion}. 

We work mostly in $d+1$ spacetime dimensions with metric signature $(-,+,...,+)$.
Bold variables are $d$-dimensional spatial vectors, and $\bfx \cdot \bfy = x^i \delta_{ij} y^j$. We will avoid using explicit indices on vectors, so that $\bfx_j$ can always be read as the position of the $j^{\rm th}$ field (and should not be confused with the $j$th component of $\bfx$). We also define $\bfk_s = \bfk_1 + \bfk_2$, $\bfk_t = \bfk_1 + \bfk_3$ and $\bfk_u = \bfk_1 + \bfk_4$ when describing exchange contributions to quartic correlators/wavefunction coefficients.
The Fourier transform is defined as $f_{\bfk} (\eta) = \int d^d \bfx \, e^{i \bfk \cdot \bfx} f_{\bfx} ( \eta )$ and commutes with functions of $\eta$. We denote the conformal weight by $\Delta = \frac{d}{2} - \nu$, where $\nu = \sqrt{ \frac{d^2}{4} - \frac{m^2}{H^2} }$ is real for light fields, and will write the shadow weight explicitly as $d  - \Delta$ (note that these are often referred to as $\Delta^-$ and $\Delta^+$).


\section{Time Evolution and Unitarity}
\label{sec:timeEvolution}

In this section, we will derive simple equations of motion for the coefficients appearing in the wavefunction, primarily for a scalar field on a conformally flat spacetime background. 
We will work in the Schr\"{o}dinger representation, in which states of the Hilbert space, $|\Psi \rangle$, are replaced by linear functionals of fields, $\Psi [\phi]$, which notionally act as $\langle \phi | \Psi \rangle$ and provide the overlap between the state and a classical (smooth) field configuration (see e.g. \cite{Jackiw, Luscher:1985iu, Long:1996wf} for reviews).
Observables built from the field operator $\hat{\phi}_{\bfk}$ and its canonical momentum $\hat{\Pi}_{\bfk}$ (which satisfy the commutation relation $
 [ \hat{\phi}_{\bfk} , \hat{\Pi}_{\bfk'} ] = i \delta^3 ( \bfk + \bfk') 
$)
are represented as $\phi_{\bfk}$ and $-i \delta/  \delta \phi_{-\bfk}$ acting on $\Psi_{\eta} [ \phi]$. 

Throughout this work we will focus on isotropic states of the form\footnote{
While this ansatz is general enough to capture any instantaneous vacuum state, it cannot describe $n$-particle states since these depend explicitly on the momenta of the particles (which spontaneously break rotational invariance).
},
\begin{align}
\Psi_\eta [ \phi ] = e^{i \Gamma_\eta [ \phi ]} \; , 
\label{eqn:PsiGamma}
\end{align}
where the functional $\Gamma_\eta [\phi]$ is approximately Gaussian, and can be expanded as,
\begin{align}
\Gamma_\eta [ \phi ] = 
 \frac{1}{2}  \int_{\bfk_1 \bfk_2}  c_{\bfk_1 \bfk_2} (\eta) \phi_{\bfk_1} \phi_{\bfk_2}  +
\sum_{n=3} \frac{1}{n!} \int_{\bfk_1 ... \bfk_n}  c_{\bfk_1 ... \bfk_n} ( \eta ) \phi_{\bfk_1} ... \phi_{\bfk_n} \; , 
 \label{eqn:coefficients}
\end{align}
where $\int_{\bfk_1 ... \bfk_n} = \int \prod_{i=1}^n d^3 \bfk_i \, \delta ( \sum_{i=1}^n \bfk_i )$ is an integral over $n$ conserved momenta, and the non-Gaussian coefficients $c_{\bfk_1 ... \bfk_n} (\eta)$ are assumed small (i.e. we assume throughout that each $c_n \sim g^{n-2}$ is suppressed by some weak coupling $g$). For brevity, we will often refer to $c_{\bfk_1 .. .\bfk_n} (\eta)$ as simply $c_n$ when its arguments are unimportant or otherwise clear from the context. 
This characterisation of the state is convenient because it allows any equal-time correlation function of $\phi_\bfk$ and $\Pi_\bfk$ to be read off straightforwardly, as we will show in section~\ref{sec:equalTime}. All of the dynamical information is effectively encoded in the $c_{\bfk_1 ... \bfk_n} (\eta)$. 

The time evolution of the coefficients $c_n (\eta)$ is governed by the Schr\"{o}dinger equation,
\begin{align}
i \partial_{\eta} \Psi_\eta [ \phi ] =  \mathcal{H} \left[ \phi,  -i \frac{\delta }{\delta \phi} \right] \Psi_{\eta} [ \phi ]  \, ,
\label{eqn:SchrodingerState}
\end{align}
where $ \mathcal{H} [ \phi , \Pi]$ is the Hamiltonian for the scalar field $\phi$. 
We will first describe how $\Psi_\eta [ \phi ]$ evolves in a free theory, then we will include the effects of small interactions in $\mathcal{H}$ in section \ref{sec:interacting}, arriving at a set of simple equations of motion which determine the time evolution of the $c_n (\eta)$. Finally, in section~\ref{sec:constants}, we use these equations to identify new constants of motion, which we call $\beta_n$, that are conserved in any scalar theory with unitary interactions.

\subsection{Free Evolution}
\label{sec:gaussian}

We begin by discussing a \textit{free} scalar field on a conformally flat spacetime, $ds^2 = \Omega^2 (\eta) \left( - d \eta^2 +  d \bfx^2 \right)$, with Hamiltonian given by,
\begin{align}
 \mathcal{H} [ \phi, \Pi ] = \int d^d \bfx \left(  \frac{1}{2 \Omega^{d-1} (\eta) }  \Pi_{\bfx}^2  + \frac{ \Omega^{d-1} (\eta) }{2} E^2_{\bfx} (\eta )  \phi_\bfx^2    \right) \; , 
 \label{eqn:Hfree}
\end{align}
where $E^2_k (\eta) =  k^2 +  m^2 \Omega^2 (\eta)$ in momentum space. It will also prove convenient to define the Gaussian width,
\begin{align}
c_{\bfk -\bfk} (\eta) =  i \Omega^{d-1} (\eta) \omega_k (\eta) \, ,
\label{eqn:wkdef}
\end{align} 
so that $\omega_k (\eta)$ has the same units as $E_k (\eta)$. 

The Schr\"{o}dinger equation \eqref{eqn:SchrodingerState} for the state $\Psi_\eta [\phi]$ given in \eqref{eqn:PsiGamma} becomes the Hamilton-Jacobi equation\footnote{
Note that the second functional derivative, $\delta^2/\delta \phi_{\bfk} \delta \phi_{\bfk} $ is formally divergent and requires renormalisation---this is the analogue of loop diagrams in the path integral approach. We will return to this point below when we discuss evolution in an interacting theory. 
} for $\Gamma_\eta [\phi]$,
\begin{equation}
- \partial_{\eta} \Gamma_{\eta} [ \phi ]  = \frac{1}{2} \int_{\bfk_1 \bfk_2}  \left\{ 
\frac{1}{\Omega^{d-1}}   \frac{\delta \Gamma_{\eta} [ \phi  ]}{\delta \phi_{\bfk_1}}   \frac{\delta \Gamma_{\eta} [ \phi  ]}{\delta \phi_{\bfk_2}}   +  \Omega^{d-1} E_{k_1}^2  \phi_{\bfk_1} \phi_{\bfk_2}   -  \frac{i}{\Omega^{d-1}}  \frac{\delta^2 \Gamma_{\eta} [ \phi ]}{\delta \phi_{\bfk_1} \delta \phi_{\bfk_2} }    \,
\right\}    \, .
\label{eqn:quantum_HamiltonJacobi}
\end{equation}
Expanding $\Gamma$ as in \eqref{eqn:coefficients}, this gives a first order differential equation for every $\partial_\eta c_{\bfk_1 ... \bfk_n} (\eta)$. 
In the absence of interactions, setting $c_n = 0$ for all $n > 2$ solves equation \eqref{eqn:quantum_HamiltonJacobi}---i.e. exactly Gaussian states remain Gaussian under this free evolution. 

\paragraph{Gaussian States:}
Consider an exactly Gaussian state, \eqref{eqn:coefficients} with $c_2 \neq 0$ and $c_{n>2}=0$. 
Such a state is annihilated by the operator,
\begin{align}
 \hat{A}_{\bfk} ( \eta ) = f_{\bfk}^* (\eta) \left(  - i c_{\bfk - \bfk} (\eta) \hat{\phi}_{-\bfk} + i \hat{\Pi}_{-\bfk}  \right) \; , 
 \label{eqn:adef}
\end{align}
where $f_{\bfk}^* (\eta)$ is an overall normalisation, chosen so that $2 f_{\bfk}^* (\eta) f_{-\bfk} (\eta) = 1/ \text{Im} \, c_{\bfk -\bfk}$ to ensure a commutation relation $ [ \hat{A}_{\bfk} (\eta) , \hat{A}_{\bfk'}^{\dagger} (\eta) ] =  \delta^3 ( \bfk - \bfk' )$ (note that a Hermitian operator in momentum space obeys $\phi_{\bfk}^\dagger = \phi_{-\bfk}$).
Equation \eqref{eqn:adef} can also be inverted to express the fields in terms of $\hat{A}$ and $\hat{A}^\dagger$,
\begin{align}
 \hat{\phi}_{\bfk} &= f_{-\bfk} (\eta) \hat{A}_{-\bfk} (\eta) + f_{\bfk}^* (\eta) \hat{A}^{\dagger}_{\bfk} (\eta)  \; ,    \label{eqn:pPia} \\
\frac{ i \hat{\Pi}_{\bfk} - i \text{Re} \, c_{\bfk - \bfk} (\eta) \, \hat{\phi}_{\bfk} }{ \text{Im} \, c_{\bfk -\bfk} (\eta) }   &=   f_{-\bfk} (\eta) \hat{A}_{-\bfk} (\eta)  -  f_{\bfk}^* (\eta) \hat{A}_{\bfk}^\dagger (\eta)    \; ,  \nonumber 
\end{align}
which will prove useful when we compute equal-time correlators in section~\ref{sec:equalTime}.

Any Gaussian state can be described in terms of these $\hat{A} (\eta)$ and $\hat{A}^\dagger (\eta)$ operators, whose time evolution is governed by $c_{\bfk -\bfk} (\eta)$ (or equivalently by the $\omega_k (\eta)$ given in \eqref{eqn:wkdef}). In a free theory, the Schr\"{o}dinger equation \eqref{eqn:quantum_HamiltonJacobi} gives,
\begin{equation}
 - \omega_k^2 (\eta)  + k^2 + m^2 \Omega^2  = \frac{  -i  \partial_\eta \left( \Omega^{d-1} \omega_k \right) }{  \Omega^{d-1} }   \,  . 
 \label{eqn:HJfree}
\end{equation}
For a Hermitian Hamiltonian (in this case a real $m^2$), the imaginary part of this equation uniquely fixes $\text{Im}\, \omega_{k}$ in terms of $\text{Re} \, \omega_{k}$,
\begin{align}
 -  \text{Im} \, \omega_{k} =  \frac{  \partial_\eta \left( \Omega^{d-1}  \text{Re} \, \omega_{k} \right) }{ 2 \Omega^{d-1} \text{Re} \, \omega_{k} }  =  \frac{\partial_\eta | f_{\bfk} | }{ | f_{\bfk} |}    \; . 
 \label{eqn:c2unitary}
\end{align} 
This is an example of a \textit{unitarity condition}: unitary dynamics (hermiticity of the Hamiltonian) requires relations between real and imaginary parts of the wavefunction coefficients. In this case, the damping of the mode functions ($\text{Im} \, \omega_k $) is controlled by their frequency ($\text{Re} \, \omega_k$). We will return to unitarity conditions in section~\ref{sec:constants}.

From the real part of the Schr\"{o}dinger equation \eqref{eqn:HJfree}, we see that $f_{\bfk} (\eta)$ solves the classical equations of motion, 
\begin{align}
- \frac{ \partial_\eta \left(  \Omega^{d-1} \partial_\eta f_{\bfk}    \right) }{ \Omega^{d-1} }  =   E_{\bfk}^2 \, f_{\bfk}  \; , 
\end{align}
providing it is normalised so that,
\begin{align}
2 \text{Im} \left[  f_{\bfk} (\eta) \partial_\eta f_{\bfk}^* (\eta)   \right] =  \Omega^{1-d} \, . 
\label{eqn:fNorm}
\end{align}
Consequently, the Gaussian width can be written as,
\begin{align}
 c_{\bfk -\bfk} (\eta) &= \frac{i}{ 2 | f_{\bfk} (\eta) |^2 } + \frac{\Omega^{d-1} \partial_\eta | f_{\bfk} (\eta) | }{ | f_{\bfk} (\eta) | }  \;  ,  \nonumber \\ 
\Rightarrow \;\;\;\;   \partial_\eta f^*_{\bfk} (\eta) &= i \omega_k (\eta) f^*_{\bfk} (\eta)  \; ,
\label{eqn:HJfreeMode}
\end{align}
and so \eqref{eqn:pPia} can be written as,
\begin{align}
\hat{\Pi}_{\bfk}  = \Omega^{d-1} \left( \partial_\eta f_{\bfk} (\eta) \hat{A}_{\bfk} (\eta) + \partial_\eta f_{\bfk}^* (\eta) \hat{A}^{\dagger} (\eta)  \right) \; , 
\end{align}
which coincides with the classical equation of motion, $\Pi (\eta) = \Omega^{d-1} \partial_{\eta} \phi_{\bfk} (\eta)$, in the Heisenberg picture.

We should stress at this stage that $\hat{A}$ and $\hat{A}^\dagger$ do \textit{not} necessarily diagonalise the Hamiltonian, and so a general Gaussian state is not necessarily a ground state (vacuum) of any $\mathcal{H} (\eta)$. 
Rather, at each time, it is only a particular family of Gaussian states that minimize the energy---those that satisfy the vacuum condition $\omega_k^2 (\eta_0) = k^2 + m^2 \Omega^2 (\eta_0)$ at time $\eta_0$ (equivalent to $\partial_\eta \, c_2  = 0$ instantaneously). 
At this time, $\hat{A}_{\bfk} (\eta_0)$ momentarily diagonalises the Hamiltonian, $\hat{\mathcal{H}} (\eta_0) = \int d^d \bfk \, \tfrac{1}{2} \sqrt{\Omega^{d-1} (\eta_0)} \, \omega_k (\eta_0) \, \hat{A}_{\bfk}^\dagger ( \eta_0 ) \hat{A}_{\bfk} (\eta_0)$, and the instantaneous ground state is therefore given by a Gaussian state in which $c_{\bfk - \bfk} ( \eta )$ has a boundary value $c_{\bfk -\bfk} ( \eta_0 ) = i  \Omega^{d-1} (\eta_0) \sqrt{  k^2 + m^2 \Omega^2 ( \eta_0 ) } $ at time $\eta_0$.   
Note that even under free time evolution, the width $c_{\bfk -\bfk} ( \eta)$ at later $\eta \neq \eta_0$ will in general not obey $\omega_k^2 (\eta) = k^2 + m^2 \Omega^2 ( \eta)$ and is therefore no longer a vacuum state. Since under free evolution the state remains Gaussian, there is always a linear operator \eqref{eqn:adef} which annihilates the state, but in general this operator does not diagonalise the instantaneous Hamiltonian.

\subsection{Interacting Evolution}
\label{sec:interacting}

Now consider a Hamiltonian with small interactions,
\begin{align}
 \mathcal{H} \left[    \phi , \Pi  \right] =  \mathcal{H}_{\rm free} [ \phi, \Pi ] + \mathcal{H}_{\rm int} [\phi, \Pi ]   \, . 
\end{align}
The interactions act as a source for the non-Gaussianities, and setting $c_{n} (\eta) = 0$ is no longer a possible solution---i.e. an initially Gaussian state will evolve into a non-Gaussian one. For example, the quartic coefficients evolves as,  
\begin{align}
- \partial_\eta c_{\bfk_1 \bfk_2 \bfk_3 \bfk_4} &=  \frac{\delta^4 \mathcal{H}_{\rm int}}{\delta \phi_{\bfk_1} ... \delta \phi_{\bfk_4}}     \label{eqn:HJeg} \\
&+ \frac{1}{\Omega^{d-1}} \Bigg\{   \sum_{j=1}^4 c_{\bfk_j ,- \bfk_j} c_{\bfk_1 \bfk_2 \bfk_3 \bfk_4}  +  \sum_{ \text{perm.} }^3 c_{\bfk_1 \bfk_2 -\bfk_s}  c_{\bfk_{3} \bfk_{4} \bfk_s }     
 - i  \int_{\bfp} c_{\bfk_1 ... \bfk_4 \bfp, -\bfp}  \Bigg\}   \, ,  \nonumber
\end{align}
where $\bfk_s = \bfk_1 + \bfk_2$, $\bfk_t = \bfk_1 + \bfk_3$ and $\bfk_u = \bfk_1 + \bfk_4$ are the three permutations which appear in the $c_3 c_3$ sum. 
In addition to the time-evolution from $\mathcal{H}_{\rm int}$, the existing non-Gaussian features of the wavefunction also contribute to the time-evolution. 
The coefficients $c_n (\eta)$ mix with each other (even in a completely free theory like \eqref{eqn:Hfree}) and therefore evolve in a non-trivial way. 
Our goal in this section will be to study \eqref{eqn:quantum_HamiltonJacobi} in more detail and express the time evolution of the $c_n (\eta)$ as simply as possible.

\paragraph{Connection to Feynman-Witten diagrams:}
Before developing relations like \eqref{eqn:HJeg} further, it is worth stressing that solving this system of differential equations for $c_n$ is equivalent to conventional path integral techniques, which express the $c_{n}$ as integrals over products of the bulk-to-bulk propagator\footnote{
Explicitly, this propagator can be expressed in terms of mode functions as,
\begin{align}
 G^{\eta_*}_\bfk ( \eta_1, \eta_2) = \frac{i}{2} \left( f_{\bfk} ( \eta_1) - \frac{ f_{\bfk} ( \eta_* ) }{ f_{\bfk}^* (\eta_*) } f^*_{\bfk} ( \eta_1) \right) f^*_{\bfk} ( \eta_2 ) \Theta ( \eta_1 - \eta_2  ) + \left( \eta_1 \leftrightarrow \eta_2 \right) \; ,
\end{align}
where $f_k (\eta)$ has been normalised so that $ \Omega^{d-1} f_{\bfk}^* \overset{\leftrightarrow}{\partial}_\eta f_{-\bfk}  =- i$.
}, $G_{\bfk}^{\eta_*} ( \eta_1 , \eta_2)$ and bulk-to-boundary propagator\footnote{
Explicitly, this propagator can be expressed in terms of mode functions as,
\begin{align}
 K^{\eta_*}_\bfk ( \eta ) = \frac{ f^*_{\bfk} (\eta) }{ f^*_{\bfk} (\eta_*) }  \; . 
\end{align}
}, $ K_{\bfk}^{\eta_*} ( \eta )$. 
For instance, the exchange contribution to the wavefunction coefficient $c_4$ can be expressed at leading order as the Feynman-Witten diagram,
\FloatBarrier
\begin{figure}[htbp!]
\begin{tikzpicture}[baseline=-0.6cm]
			\begin{feynman}
				\vertex (a1);
				\vertex [below=1.5cm of a1] (b1);
				\vertex [below=1.5cm of b1] (c1);
				\vertex [left=1.5cm of a1] (a2);
				\vertex [below=1.5cm of a2] (b2);
				\vertex [below=1.5cm of b2] (c2);
				\vertex [left=2.0cm of a2] (a3);
				\vertex [below=1.5cm of a3] (b3);
				\vertex [below=1.5cm of b3] (c3);
				\vertex [left=1.5cm of a3] (a4);
				\vertex [below=1.5cm of a4] (b4);
				\vertex [below=1.5cm of b4] (c4);
				\vertex [left=1.5cm of a4] (a5);
				\vertex [below=1.5cm of a5] (b5);
				\vertex [below=1.5cm of b5] (c5);
				\vertex [left=1.5cm of a5] (a6);
				\vertex [below=1.5cm of a6] (b6);
				\vertex [below=1.5cm of b6] (c6);
				\vertex [left=1.5cm of a6] (a7);
				\vertex [below=1.5cm of a7] (b7);
				\vertex [below=1.5cm of b7] (c7);
				\vertex [left=1.5cm of a7] (a8);
				\vertex [below=1.5cm of a8] (b8);
				\vertex [below=1.5cm of b8] (c8);
				\node at (-11.5,  -0.0) {$\eta_*$};
				\node at (-11.5,  -1.5) {$\eta_1$};
				\node at (-11.5,  -3.0) {$\eta_2$};
				\node at (-9.8,  -0.75) {$K_{\bfk_1}^{\eta_*} (\eta_1)$};
				\node at (-6.2,  -0.75) {$K_{\bfk_2}^{\eta_*} (\eta_1)$};
				\node at (-3.55,  -0.75) {$K_{\bfk_3}^{\eta_*} (\eta_2)$};
				\node at (-0.9,  -0.75) {$K_{\bfk_4}^{\eta_*} (\eta_2)$};
				\node at (-7.75,  -2.3) {$G_{\bfk_1+\bfk_2}^{\eta_*} (\eta_1, \eta_2)$};
				\node at (-5.5,  +0.5) {$c_{\bfk_1 \bfk_2 \bfk_3 \bfk_4} (\eta_*) $};

				\diagram*{
				(a8) -- [scalar] (a1),
				(b8) -- [scalar] (b1),
				(c8) -- [scalar] (c1),
				(c3) -- (a2),
				(c3) -- (a4),
				(b6) -- (a5),
				(b6) -- (a7),
				(b6) -- [photon] (c3),
				};				
			\end{feynman}		
		\end{tikzpicture}
\end{figure}
\FloatBarrier
\begin{equation}
  = \int_{-\infty}^{\eta_*} d \eta_1 \int_{-\infty}^{\eta_*} d \eta_2 \;\; K_{\bfk_1}^{\eta_*} ( \eta_1  ) K_{\bfk_2}^{\eta_*} ( \eta_1  ) G_{\bfk_1 + \bfk_2}^{\eta_*} ( \eta_1 , \eta_2  ) K_{\bfk_3}^{\eta_*} ( \eta_2  ) K_{\bfk_4}^{\eta_*} ( \eta_2  ) \; , 
  \label{eqn:4ptLeg}
\end{equation}
where we have imposed boundary conditions at $\eta \to -\infty$. Rather than compute this integral directly (which is difficult even for simple spacetime backgrounds), notice that by considering a small variation in the position $\eta_*$ of our timeslice, the propagators obey,
\begin{align}
 \partial_{\eta_*} K_{\bfk}^{\eta_*} ( \eta )  &= - c_{\bfk -\bfk} (\eta_*) K_{\bfk}^{\eta_*} ( \eta ) \nonumber  \\
 \partial_{\eta_*} G_{\bfk}^{\eta_*} ( \eta_1, \eta_2 ) &= \frac{1}{2} K_{\bfk}^{\eta_*} ( \eta_1 ) K_{\bfk}^{\eta_*} ( \eta_2 )  \label{eqn:dG}
\end{align}
and the four-point wavefunction coefficient decomposes into two-point and three-point coefficients, as in \eqref{eqn:HJeg}. In fact, armed with \eqref{eqn:dG}, it is straightforward to show that $\partial_{\eta_*}$ of the conventional expressions for $c_n ( \eta_*)$ in terms of $K_\bfk^{\eta_*}$ and $G_{\bfk}^{\eta_*}$ always coincides\footnote{
It is, of course, also possible to go the other way, writing the general solution to the first order HJ equation in terms of an integrating factor, which produces integral expressions like \eqref{eqn:4ptLeg}.
} with the Hamilton-Jacobi equation \eqref{eqn:HJint}---this is not surprising, since both approaches are solving the same underlying Schr\"{o}dinger equation for $\Psi_\eta [ \phi ]$. 
  
\paragraph{Interaction Picture:}
We will now take steps to simplify \eqref{eqn:HJeg}. First, note that solving \eqref{eqn:HJeg} requires inverting $\partial_\eta + \sum_{j=1}^n c_{\bfk_j -\bfk_j} (\eta)$. Diagrammatically, this contribution from $c_{\bfk -\bfk}$ is associated with the propagation of the external legs, rather than any nonlinear interaction. To simplify matters, we can use our knowledge of $\mathcal{H}_{\rm free}$ to transform into the \textit{interaction picture}, performing a unitary transformation at each time to define a new operator $\hat{\phi}^I_{\bfk} ( \eta ) = \hat{\phi}_{\bfk} / f_{\bfk}^* (\eta)$. This corresponds to rescaling the wavefunction coefficients, 
\begin{equation}
 c_{\bfk_1 ... \bfk_n} ( \eta ) = \frac{  c^I_{\bfk_1 ... \bfk_n} ( \eta ) }{ f_{\bfk_1}^* ( \eta) ... f^*_{\bfk_n} (\eta)  } \; , 
\end{equation}
so that for example \eqref{eqn:HJeg}  becomes,
\begin{align}
- \partial_\eta c^I_{\bfk_1 \bfk_2 \bfk_3 \bfk_4} -  \frac{\delta^4 \mathcal{H}^I_{\rm int}}{\delta \phi^I_{\bfk_1} ... \delta \phi^I_{\bfk_4}}  &=  \frac{1}{\Omega^{d-1}} \Bigg\{   \sum_{ ij }^3 \frac{ c^I_{\bfk_i \bfk_j \bfk_{ij}}  c^I_{\bfk_{i'} \bfk_{j'} \bfk_{i'j'}}  }{ f^*_{\bfk_{ij}} f^*_{\bfk_{i'j'}}}       
 - i  \int_{\bfp} \frac{ c^I_{\bfk_1 ... \bfk_4 \bfp, -\bfp} }{ f^*_{\bfp} f^*_{-\bfp}  }   \Bigg\}  \, , 
\end{align}
and the term $\sum c_2 c_4$ associated with the free propagation has been removed. The interaction picture Hamiltonian is $ \mathcal{H}^I_{\rm int} [  \phi^I ] = \mathcal{H}_{\rm int} [ f_{\bfk}^* \phi^I_{\bfk}  ] $---for instance the interaction $ \lambda \phi_{\bfk_1} ... \phi_{\bfk_n}$ in $\mathcal{H}_{\rm int}$ would correspond to the time-dependent interaction $\lambda f^*_{\bfk_1} ... f^*_{\bfk_n} \phi^I_{\bfk_1} ... \phi^I_{\bfk_n}$ in $\mathcal{H}_{\rm int}^I$. 

In general, if we  split the effective action into its free and interacting parts,
 \begin{equation}
  \Gamma_\eta [ \phi ] = \Gamma_{\eta  , \, \text{free}} [ \phi ] + \Gamma_{ \eta , \,  \text{int}} [ \phi  ] \, ,
 \end{equation}
where $\Gamma_{\eta , \, \text{free} } [ \phi ] =  \int_{\bfk}  \frac{1}{2} \ c_{\bfk_1 \bfk_2} (\eta) \,  \phi_{\bfk_1} \phi_{\bfk_2} $, the Schr\"{o}dinger equation \eqref{eqn:SchrodingerState} becomes,
\vspace{+0.5\baselineskip}
\begin{mdframed}[style=boxed]
\vspace{-0.5\baselineskip}
\begin{align}
-\partial_\eta \Gamma^I_{\eta , \text{int} } [ \phi^I  ]    =  \mathcal{H}_{\rm int}^I +  \frac{1}{2} \int_{\bfk_1 \bfk_2} \frac{ \Omega^{1-d} }{f^*_{\bfk_1} f^*_{\bfk_2} } \left\{ 
 \left( \frac{\delta}{\delta \phi_{\bfk}^I} \Gamma^I_{\eta \, \text{int}} [ \phi^I  ] \right)^2  -   i   \frac{\delta^2 \Gamma^I_{\eta \, \text{int}} [ \phi^I ] }{\delta \phi_{\bfk}^I \delta \phi_{\bfk}^I}    \, 
\right\} \, ,
\label{eqn:HJint}
\end{align}
\end{mdframed}
where in the interaction picture, $\Gamma^I_{\eta , \text{int}} [ \phi^I_{\bfk} ] = \Gamma_{\eta , \, \text{int} } \left[  f^*_{\bfk} (\eta) \, \phi^I_{\bfk}  \right]$. 
This corresponds to the evolution of $\Psi^I = e^{i \Gamma_{\rm int}^I}$ generated by $\mathcal{H}^I_{\rm int}$, which is now decoupled from the free evolution (contained implicitly within $f_{\bfk}$, which are determined from \eqref{eqn:HJfree} and \eqref{eqn:HJfreeMode}). 

\begin{figure}
\centering
\includegraphics[width=0.9\textwidth]{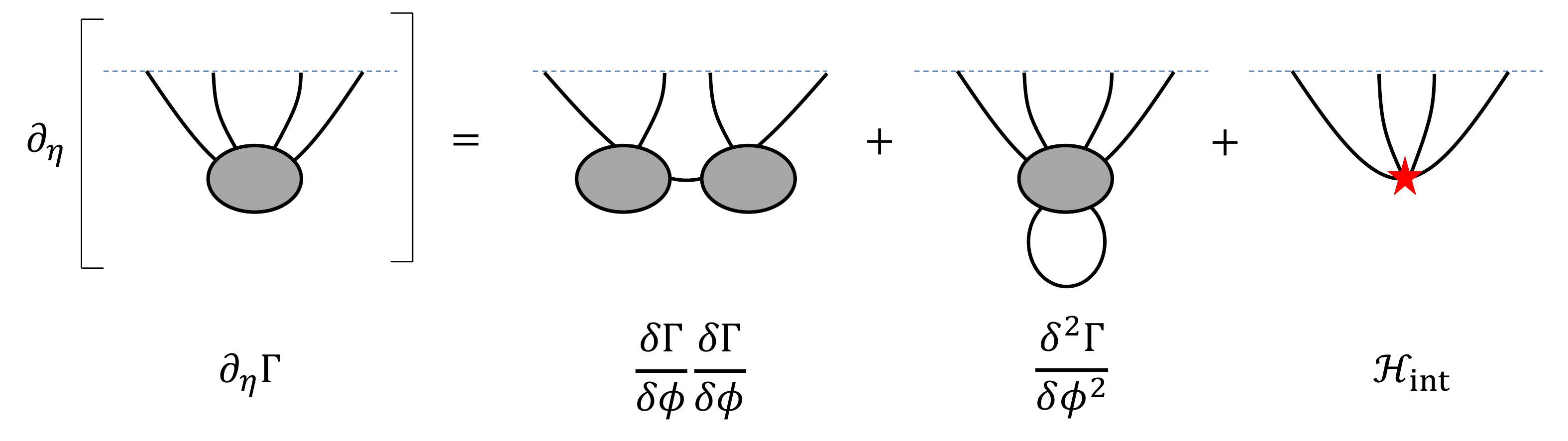}
\caption{The Hamilton-Jacobi equations for the wavefunction phase, $\Gamma$. The time derivative is given by all possible ways of splitting the interaction into two pieces (exchange contribution), plus all possible ways of contracting a single higher-point coefficient into a loop (loop contribution), plus any interactions in the Hamiltonian (contact contribution). 
}
\label{fig:HJeqn}
\end{figure}

Each of the terms in \eqref{eqn:HJint} can be expressed diagrammatically, as shown in figure~\ref{fig:HJeqn}. There are three distinct source terms on the right-hand side: a contact contribution from $\mathcal{H}^I_{\rm int}$, an exchange contribution from $(\delta \Gamma / \delta \phi )^2$, and a loop contribution from $\delta^2 \Gamma / \delta \phi^2$. We will now discuss each of these in turn.

\paragraph{Loop Contributions:}
Note that if each field is given a small coupling, so that the wavefunction phase is written as\footnote{
This is motivated by the power counting rules familiar from flat space, which guarantee that this scaling with $g$ is radiatively stable.
} $\frac{1}{g^2} \Gamma [  g \phi_{\bfk} ]$, then the final term in \eqref{eqn:HJint} is $\mathcal{O} (g^2)$ suppressed relative to the other terms. This $\delta^2 \Gamma_{\eta} [\phi] / \delta \phi_{\bfk}^2$ term is formally divergent but can be treated in a perturbative expansion in $g$ (this corresponds to the usual loop expansion).
For a general Hamiltonian (with momentum-dependent interactions), the Schr\"{o}dinger equation at weak coupling can therefore be written as,
\begin{align}
 - \partial_\eta \Gamma_\eta [ \phi ] = \mathcal{H} \left[  \phi , \frac{ \delta \Gamma_\eta [\phi ]}{ \delta \phi }  \right]  \; , 
 \label{eqn:HJweak}
\end{align}
which coincides with the classical Hamilton-Jacobi equation, and therefore $\Gamma_\eta [ \phi ]$ coincides with the classical on-shell action at tree level. 
We postpone any further discussion of loop contributions to the Hamilton-Jacobi equation for the future, and in this work we will assume throughout that interaction strengths are sufficiently weak that all terms in $\delta^2 \Gamma / \delta \phi^2$ can be neglected. 

\paragraph{Contact Contributions:}
We refer to the contribution of a $\phi^n$ interaction in $\mathcal{H}_{\rm int}$ to the $n$-point coefficient $c_n$ as a \textit{contact} contribution. The most general set of interactions that can appear in $\mathcal{H}_{\rm int}$ at this order can be written as,
\begin{align}
\{\;  V_0 \phi^n \; , \; V_1 \phi^{n-1} \Pi \; , \; ... \; , \;  V_j \phi^{n-j} \Pi^j \; , \; ...  \; , \; V_{n} \Pi^n \; \} \,  ,
\label{eqn:nbasis}
\end{align}
where each $V_j ( \eta, \bfk_1 , ... , \bfk_n)$ is a function of both time and momenta. In local, rotationally invariant, theories these functions are analytic in the momenta---this is most easily seen in the corresponding Lagrangian, in which local interactions either have contracted pairs of spatial derivatives (which give $\bfk_i \cdot \bfk_j$), or a single time derivative $\dot \phi$ (which produces the $\Pi$ dependent terms), or more time derivatives (which can be reduced to $\phi$ and $\dot \phi$, with analytic coefficients, using the equations of motion). 

In the Hamilton-Jacobi equation, each of the terms in \eqref{eqn:nbasis} contribute to $c_n$ as $V_j ( c_2 )^j$ (i.e. imagine taking $n$ functional derivatives with respect to $\phi$ and then set both $\phi$ and $\delta \Gamma / \delta \phi$ equal to zero). Therefore for contact contributions we can treat $\Pi_{\bfk}$ as $c_{\bfk -\bfk} (\eta) \phi_{\bfk}$ when acting on the wavefunction at time $\eta$ (at leading order in the small coupling). The simplest example of this is the cubic coefficient, which is sourced solely by contact contributions,
\vspace{+0.5\baselineskip}
\begin{mdframed}[style=boxed]
\vspace{-0.5\baselineskip}
\begin{align}
 - \partial_\eta c^I_{\bfk_1 \bfk_2 \bfk_3} = \prod_{j=1}^3  f^*_{\bfk_j} (\eta) \frac{\delta^3 \mathcal{H} [ \phi ,  c_2 (\eta) \phi ]}{\delta \phi_{\bfk_1} \delta \phi_{\bfk_2} \delta \phi_{\bfk_3}}  \Big|_{\phi = 0} \; . 
 \label{eqn:HJc3}
\end{align}
\end{mdframed}
The most general time evolution
for the bispectrum is therefore encoded in four independent functions of $\bfk$,
\begin{align}
 - \partial_\eta \hat{c}_{\bfk_1 \bfk_2 \bfk_3} = \sum_{s_j}  V_{s_1 s_2 s_3} ( \eta , \bfk_j) \partial_\eta^{s_1} f^*_{\bfk_1} \partial_\eta^{s_1} f^*_{\bfk_2} \partial_\eta^{s_1} f^*_{\bfk_3}  \; , 
 \label{eqn:H3}
\end{align}
where the $s_j$ are either $0$ or $1$.

\paragraph{Exchange Contributions:}
Now we turn our attention to the term in \eqref{eqn:HJint} which is quadratic in $\delta \Gamma / \delta \phi$, which we refer to as an \textit{exchange} contribution. The factor of $\Omega^{1-d} / f_{\bfk_1}^* f_{\bfk_2}^*$ now plays the role of the ``propagator'' for the wavefunction coefficients $c^I_{\bfk_1 .. \bfk_n} $. However, this is not always straightforward to integrate---in particular, if $c^I_{\bfk_1 \bfk_2 \bfk_3}$ is already quite complicated, then the exchange contribution to \eqref{eqn:HJeg} will be very difficult to integrate.  
Fortunately, the exchange terms can be simplified by noting that 
$\Omega^{1-d}/f^*_{\bfk} f^*_{\bfk} = i \partial_\eta \left[  f_{\bfk} /f^*_{\bfk}  \right]$, so for example by shifting the wavefunction coefficient,
\begin{align}
 C_{\bfk_1 \bfk_2 \bfk_3 \bfk_4} (\eta) =  c_{\bfk_1 \bfk_2 \bfk_3 \bfk_4} (\eta) + i \sum_{\text{perm.}}^3  c_{\bfk_1 \bfk_2 - \bfk_s} (\eta) c_{\bfk_3 \bfk_4 \bfk_s} (\eta)  | f_{\bfk_s} (\eta) |^2
 \label{eqn:bigC}
\end{align}
we arrive at a simpler HJ equation,
\begin{align}
- \partial_\eta C_{\bfk_1 \bfk_2 \bfk_3 \bfk_4}^I = f_{\bfk_1}^* f_{\bfk_2}^* f_{\bfk_3}^* f_{\bfk_4}^*  
 \frac{ \delta^4 \mathcal{H}_{\rm int} }{\delta \phi_{\bfk_1} ... \delta \phi_{\bfk_4} }  - i \sum_{\text{perm.}}^6  c^I_{\bfk_1 \bfk_2 -\bfk_s}   \frac{f_{\bfk_s}}{f_{\bfk_s}^*} \partial_\eta c^I_{\bfk_3 \bfk_4 \bfk_s}   + \text{loops}
 \label{eqn:C4}
\end{align}
in which the exchange contribution to the time evolution is now \textit{linear} (rather than quadratic) in $c_3$, since \eqref{eqn:HJc3} can be used to replace $\partial_\eta c^I_{\bfk_1 \bfk_2 \bfk_3}$ with $\mathcal{H}$. 

This simplification can be achieved in general by subtracting a boundary term from the time integral, 
\begin{align}
\tilde{\Gamma}^I_{\eta \, \rm int} = \Gamma^I_{\eta \, \rm int} +  \frac{i}{2} \int_{\bfk} \frac{\delta \Gamma^I_{\eta \, \rm int} }{ \delta \phi_{\bfk} } \frac{\delta \Gamma^I_{\eta  \, \rm int} }{ \delta \phi_{- \bfk} }  \frac{ f_\bfk (\eta) }{ f^*_\bfk (\eta) }
\label{eqn:Gammatilde}
\end{align}
which produces, 
\begin{align}
- \partial_\eta \tilde{\Gamma}^I_{\eta \, \text{int} } [ \phi^I  ]   = \mathcal{H}_{\rm int}^I [ \phi^I ]  - \frac{i}{2} \int_{\bfk} \frac{ f_{\bfk} }{ f^*_{\bfk} } \partial_\eta 
 \left( 
  \frac{\delta \Gamma^I_{\eta \, \text{int}} }{\delta \phi_{\bfk}^I}  
  \frac{\delta \Gamma^I_{\eta \, \text{int}} }{\delta \phi_{-\bfk}^I}    
  \right)   + \text{loops} \; .
  \label{eqn:HJtilde}
\end{align}
In practice, \eqref{eqn:HJtilde} is often easier to integrate than \eqref{eqn:HJint}, and its solution can then be used to find $\Gamma_{\eta , \text{int}}$ using \eqref{eqn:Gammatilde} (which no longer contains any time integrals).  

In fact, \eqref{eqn:C4} can be simplified even further. Consider for instance a simple $g \phi^3$ interaction in $\mathcal{H}_{\rm int}$. Then each exchange contribution can be written as,
\begin{align}
  c^I_{\bfk_1 \bfk_2 -\bfk_s}   \frac{f_{\bfk_s}}{f_{\bfk_s}^*} \partial_\eta c^I_{\bfk_3 \bfk_4 \bfk_s}
  &= -  c_{\bfk_1 \bfk_2 -\bfk_s}^I \,g   f_{\bfk_s} f^*_{\bfk_3} f^*_{\bfk_4}   \nonumber \\
 &=  c_{\bfk_1 \bfk_2 -\bfk_s}^I \, \partial_\eta c_{\bfk_3 \bfk_4 \bar{\bfk}_s }^I 
 \label{eqn:c4c3dc3}
\end{align}
where we have defined $\bar{\bfk}_s$ such that $f_{\bar{\bfk}} ( \eta) = f_{\bfk}^* (\eta)$.
If we had considered instead the most general cubic interaction in $\mathcal{H}_{\rm int}$, given in \eqref{eqn:H3}, we would have found several terms from $\delta^3 \mathcal{H}_{\rm int} / \delta \phi^3$, which always combine into the combination \eqref{eqn:c4c3dc3}. For instance, the next simplest interaction is $V_1 \phi^2 \Pi$, and so replacing $\Pi$ with $\delta \Gamma / \delta \phi$ leads to a contribution from $\delta^4 \mathcal{H}_{\rm int}/\delta \phi^4$, which combines with the $c^I_{\bfk_1 \bfk_2 \bfk_s} \partial_\eta c^I_{\bfk_3 \bfk_4 \bfk_s}$ term to give, 
 \begin{align}
  - \partial_\eta C_{\bfk_1 \bfk_2 \bfk_3 \bfk_4}^I &=  
 \sum_{\text{perm.}}^6  c^I_{\bfk_1  \bfk_2 -\bfk_s} V_1 ( \bfk_3, \bfk_4, \bfk_s) \frac{ f_{\bfk_3}^* f_{\bfk_4}^*   }{ f_{\bfk_s}^* }
\nonumber  \\
 &+  i \sum_{\text{perm.}}^6  c^I_{\bfk_1 \bfk_2 -\bfk_s}  V_1 ( \bfk_3, \bfk_4, \bfk_s) \Omega^{d-1} \left(  \frac{ f_{\bfk_3}^* f_{\bfk_4}^* }{f_{\bfk_s}^*}  f_{\bfk_s} \partial_\eta f_{\bfk_s}^*   +  \partial_\eta f_{\bfk_3}^* f_{\bfk_4}^* f_{\bfk_s} + f_{\bfk_3}^* \partial_\eta f_{\bfk_4}^* f_{\bfk_s}   \right)  \nonumber  \\
 &= - i \sum_{\rm perm.}^6 c^I_{\bfk_1 \bfk_2 -\bfk_s} \partial_\eta c^I_{\bfk_3 \bfk_4 \bar{\bfk}_s}
 \end{align}
where we have made use of the normalisation $f_\bfk \partial_\eta f_\bfk^* - f_{\bfk}^* \partial_\eta f_{\bfk} = i \Omega^{1-d}$. Therefore the quartic wavefunction coefficient can be found from the following relation, 
\vspace{+0.5\baselineskip}
\begin{mdframed}[style=boxed]
\vspace{-0.5\baselineskip}
\begin{align}
 - \partial_\eta C^I_{\bfk_1 \bfk_2 \bfk_3 \bfk_4} =  - i \sum_{\text{perm.}}^6  c^I_{\bfk_1 \bfk_2 -\bfk_s} \partial_\eta c^I_{\bfk_3 \bfk_4 \bar{\bfk}_s}  + \prod_{j=1}^4  f^*_{\bfk_j} (\eta)  \frac{\delta}{\delta \phi_{\bfk_j} } \, \mathcal{H} [ \phi,   c_2 \phi ] \Big|_{\phi = 0} \; . 
 \label{eqn:HJc4}
\end{align}
\end{mdframed}
The equations of motion, \eqref{eqn:HJc3} and \eqref{eqn:HJc4}, describe how the non-Gaussianities in the wavefunction evolve with time in response to a Hamiltonian $\mathcal{H}$. 
We will now show that, by manipulating these equations, one can construct constants of motion which are preserved by any unitary dynamics.

\subsection{Constants of Motion from Unitarity}
\label{sec:constants}

During the final stages of this work, \cite{Goodhew:2020hob} appeared, in which the properties of de Sitter bulk-to-bulk and bulk-to-boundary propagators (under a careful analytic continuation to negative values of $| \mathbf{k}_j|$) are used to establish a ``Cosmological Optical Theorem''. 
We believe that our relations \eqref{eqn:HJc3} and \eqref{eqn:HJc4} shed light on this important result. 

It will prove useful to define the discontinuity,
\begin{align}
\text{Disc} \left[  c_{\bfk_1 ... \bfk_n}   \right] = c_{\bfk_1 ... \bfk_4} - c^*_{\bar{\bfk}_1 ... \bar{\bfk}_4}
\label{eqn:Disc}
\end{align}
where $\bar{\bfk}$ is again the continuation of the momentum which achieves $ f_{\bar{\bfk}}=f_{\bfk}^*$. This has the advantage that $\text{Disc} \, f_{\bfk} = 0$ by construction, and similarly $\text{Disc} \, c_2 = 0$. 
For instance, in a general Gaussian state on pure de Sitter, with $\Omega = 1/(-H \eta)$, the Bunch-Davies mode function is,
\begin{align}
 f_\bfk (\eta) =\frac{ \sqrt{\pi}  e^{- i \frac{\pi}{2} ( \nu - 1 ) } }{2 \sqrt{H} }  \Omega^{-d/2} H^{(2)}_{\nu} ( - k \eta )  \; , 
 \label{eqn:modefdS}
\end{align}
which has been normalised so that $ f_{\bfk} \overset{\leftrightarrow}{\partial_\eta} f_{\bfk}^* = i \Omega^{1-d}$ and where we have chosen the overall phase\footnote{
Alternatively, one could use the de Sitter invariant mode function,
\begin{align}
 \hat{f}_{\bfk} (\eta) = \frac{i \sqrt{\pi} k^\nu }{2 \sqrt{H} }  \Omega^{-d/2} H^{(2)}_\nu ( - k \eta )
\end{align}
which is annihilated by de Sitter boosts and has weight $\Delta$ under dilations.
} so that $\bar{\bfk}$ corresponds to the replacement $| \bar{\bfk} | = e^{-i \pi} | \bfk |^*$, since $H_\nu^{(2)} ( z e^{- i \pi} ) = - e^{i \pi \nu} H_\nu^{(2) *} ( z^* )$ (for $\text{Im} (z) < 0$).
For more general initial states or background spacetimes, $\bar{\bfk}$ may be more complicated, but one can always keep in mind the simple de Sitter example \eqref{eqn:modefdS}.  

The discontinuity \eqref{eqn:Disc} is useful for the following reason. For a pure potential interaction, e.g. $V_{000} \phi^3$, then unitarity requires $\text{Im} \, V_{000} = 0$ and so the imaginary part of $c_3$ is fixed in terms of its real part \textit{independently} of the interaction (much like \eqref{eqn:c2unitary} for $c_2$),  
\begin{align}
- \partial_\eta \text{Im} \, c_{\bfk_1 \bfk_2 \bfk_3} = \text{Im} \left(  i  \sum_j  \omega_{k_j} \, c_{\bfk_1 \bfk_2 \bfk_3} \right) \;\;\;\; \text{when} \;\;\;\; \mathcal{H}_{\rm int} = V_{000} \phi^3 \; .
\label{eqn:c3unitaryeg}
\end{align}
 However, taking the imaginary part in this way will not remove interactions like $V_{001} \phi^2 \Pi$ from \eqref{eqn:HJc3}, since they depend on $c_2$ and thus $\text{Im} \, \left( c_2 V_{001} \right)$ can be non-zero even in a unitary theory. It is this subtlety that the $\text{Disc}$ in  \eqref{eqn:Disc} is overcoming, since\footnote{
Recall that a local $V_{001}$ is analytic in the $\bfk_j$ and therefore does not depend on any $k_j = \sqrt{\bfk_j \cdot \bfk_j}$, so $\text{Disc} ( c_2 V_{001}  ) = c_2 \text{Im} \, V_{001}$ vanishes by unitarity.  
 } $\text{Disc} \left(  c_2 V_{001}  \right) = 0$. The extension of \eqref{eqn:c3unitaryeg}, which removes every cubic interaction \eqref{eqn:H3} from the equation of motion \eqref{eqn:HJc3}, is,
 \begin{align}
- \partial_\eta \text{Disc} \, c_{\bfk_1 \bfk_2 \bfk_3}  = i \sum \omega_{k_j} \, \text{Disc} \, c_{\bfk_1 \bfk_2 \bfk_3} \;\; \text{for any} \;\; \mathcal{H}_{\rm int} \;  . 
 \end{align}
Since factors of $f_{\bfk} (\eta)$ commute with the $\text{Disc}$, this can also be written simply as,
\begin{align}
 \partial_\eta \, \text{Disc} \, c^I_{\bfk_1 \bfk_2 \bfk_3} = 0 \; , 
 \label{eqn:c3Disc}
\end{align}
and therefore $\text{Disc} \, c^I_{\bfk_1 \bfk_2 \bfk_3}$ is a \textit{constant of motion}. In fact, this argument applies to any contact contribution to any $c_n$ coefficient: if exchange and loop contributions are neglected, then $\text{Disc} \, c_n^I$ is always conserved by a unitary evolution. 

For $n>3$, the $c_n$ coefficient generically has exchange interactions which also contribute to the $\text{Disc}$. For example, \eqref{eqn:HJc4} can be written as,
\begin{align}
 - \partial_\eta \text{Disc} \left[ C^I_{\bfk_1 \bfk_2 \bfk_3 \bfk_4}  -  i \sum_{\text{perm.}}^3 c^I_{\bfk_1 \bfk_2 -\bfk_s} c^{I\, *}_{ \bar{\bfk}_3 \bar{\bfk}_4 \bfk_s}   \right] =  \prod_{j=1}^4  f^*_{\bfk_j}  \frac{\delta}{\delta \phi_{\bfk_j} } \, \left( 
  \mathcal{H} [ \phi, c_2 \phi ]  -  \mathcal{H}^\dagger [ \phi, c_2 \phi ]
  \right) \Big|_{\phi = 0}
  \label{eqn:c4Disc}
\end{align}
where $\mathcal{H}^\dagger$ is the Hermitian conjugate of the Hamiltonian, and we have again used that $c_{\bar{\bfk} -\bar{\bfk}}^* = c_{\bfk -\bfk} $. A unitary time evolution (Hermitian Hamiltonian), therefore requires that the right-hand side of this equation vanishes, and therefore that this particular discontinuity is also a constant of motion. 

In summary, the Hamilton-Jacobi relations \eqref{eqn:HJc3} and \eqref{eqn:HJc4}, together with unitarity of the interaction Hamiltonian, establish that at each order in $\phi$ there is one additional constant of motion, which we shall call $\beta_n$. At cubic and quartic order, these are given explicitly by\footnote{
Note that on Minkowski or de Sitter with \eqref{eqn:modefdS}, if $c_{\bfk_1 \bfk_2 \bfk_3} = c_3 ( k_1, k_2, k_3)$ is written in terms of the magnitudes of $\bfk_j$ only (which can always be done without loss of generality using $\bfk_1 + \bfk_2 + \bfk_3=0$), then $c_{\bar{\bfk}_1 \bar{\bfk}_2  \bfk_3} = c_3 (-k_1, -k_2, k_3)$ is the same function with the signs of $| \bfk_1|$ and $| \bfk_2|$ reversed. 
}, 
\vspace{+0.5\baselineskip}
\begin{mdframed}[style=boxed]
\vspace{-0.5\baselineskip}
\begin{align}
\frac{ \beta_{\bfk_1 \bfk_2 \bfk_3} }{  f_{\bfk_1}^* f_{\bfk_2}^* f_{\bfk_3}^* }  &=   c_{\bfk_1 \bfk_2 \bfk_3} - c^*_{\bar{\bfk_1} \bar{\bfk}_2 \bar{\bfk}_3}    \label{eqn:constants} \\
\frac{ \beta_{\bfk_1 \bfk_2 \bfk_3 \bfk_4} }{  f_{\bfk_1}^* f_{\bfk_2}^* f_{\bfk_3}^* f_{\bfk_4}^*  }  &= 
c_{\bfk_1 \bfk_2 \bfk_3 \bfk_4} - c^*_{\bar{\bfk_1} \bar{\bfk}_2 \bar{\bfk}_3 \bar{\bfk}_4}  
+ i \sum_{\rm perm.}^3 | f_{\bfk_s} |^2   \left(  c_{\bfk_1 \bfk_2 -\bfk_s} - c^*_{\bar{\bfk_1} \bar{\bfk}_2 -\bfk_s}  \right) \left( c_{\bfk_3 \bfk_4 \bfk_s} - c^*_{\bar{\bfk_3} \bar{\bfk}_4 \bfk_s}  \right)   \nonumber
\end{align}
\end{mdframed}
since our equations of motion \eqref{eqn:c3Disc} and \eqref{eqn:c4Disc} set $\partial_\eta \beta_{\bfk_1 \bfk_2 \bfk_3}=0$ and $\partial_\eta \beta_{\bfk_1 \bfk_2 \bfk_3 \bfk_4}=0$ for any hermitian Hamiltonian.
This is somewhat analogous to classical mechanics, in which the Hamilton-Jacobi approach identifies a pair of constants (corresponding to the initial position and initial velocity) for each degree of freedom. 
Once the boundary value for the wavefunction $\Psi [\phi]$ has been specified, \eqref{eqn:constants} allows the constant $\beta_n$ to be calculated immediately (without the need for any time integration)---they are a property of the initial state. In \cite{Goodhew:2020hob}, a Bunch-Davies vacuum state was assumed in the past---setting $\alpha_n^{\rm in} = 0$ in this way for the initial state then sets $\beta_n = 0$ for all time. Now we see that in fact \textit{any} initial state in which $\beta_n = 0$ will have $\beta_n = 0$ for all time, and more generally that an arbitrary initial condition (specified at an arbitrary time in the bulk or on the boundary) will likewise have a conserved (in general non-zero) set of $\beta_n$.  


\section{de Sitter Isometries}
\label{sec:isometries}

While the preceding formalism may be applied to a scalar field on any conformally flat spacetime, we will now focus on the particular case of a (at least quasi-)de Sitter background. This means that, providing the state is initially de Sitter invariant, time evolution will produce a state at later times which is also de Sitter invariant. This provides powerful constraints on the wavefunction and its evolution, which we now describe.  

We will work in the expanding Poincare patch,   $ds^2=\frac{1}{H^2\eta^2}(-d\eta^2+d \bfx^2)$, which is most relevant for cosmology. Here, the Hubble constant $H$ is the inverse of the de Sitter radius and the conformal time $\eta$ runs from $-\infty$ in the past to $0$ in the asymptotic future.
First we will briefly review the generators of the de Sitter isometries and their associated Noether charges in \eqref{sec:generators}, then their simple action on equal-time correlation functions in \eqref{sec:equalTime}, and finally how they can be implemented directly on wavefunction coefficients in \eqref{sec:isometry}. While the action of these generators is well-known near the conformal boundary at $\eta \to 0$ (where they reduce to the $d$-dimensional conformal group), the way that they constrain the correlators and wavefunction in the bulk is less widely appreciated. Our aim is to describe these constraints in a similar fashion to our Hamilton-Jacobi equations from Section~\ref{sec:timeEvolution}, providing a further set of differential equations which can be used to determine properties of the $c_n$ coefficients.

\subsection{Symmetry Generators}
\label{sec:generators}

In addition to spatial translations/rotations and dilations,
\begin{align}
 \eta \to \alpha \eta \;\;\;\; , \;\;\;\; \bfx \to \alpha \, \bfx \; , 
 \label{eqn:Dil}
\end{align}
de Sitter spacetime has an additional $d$ ``boost'' isometries,
\begin{align}
 \eta \to  \gamma \, \eta \;\;\;\; , \;\;\;\; \bfx \to \gamma \left(  \bfx  +  \mathbf{b} \cdot (  \eta^2 - |\bfx|^2 )   \right)     \label{eqn:SCT}  \\
\text{where} \;\;\;\;  \gamma = \left( 1 - 2 \mathbf{b} \cdot \bfx - |\mathbf{b}|^2 ( \eta^2 - | \bfx |^2  )   \right)^{-1}  \; ,  \nonumber
\end{align}
parametrised here by the constant ($d$-dimensional )vector $\mathbf{b}$. 
The infinitesimal versions of \eqref{eqn:Dil} and \eqref{eqn:SCT} are generated by\footnote{
Note that writing $x^\mu = (\eta, \bfx)$, with the understanding that Greek indices are raised/lowered with $\eta_{\mu\nu}$, these generators
\begin{align}
D = - x^\mu \partial_\mu   \;\;\;\; \text{and} \;\;\;\;
K_i = 2 x_i  x^\nu \partial_\nu  - x^2 \partial_i
\end{align}
can be thought of as the usual time translations and boosts of Minkowski space with the time direction replaced by $t \to  x^\mu x_\mu$. 
},
\begin{align}
D &=  -\eta\partial_{\eta}- \bfx \cdot \partial_{\bfx}   \label{deSitter:SI}\\
\mathbf{K} &= - 2 \bfx D  + (  \eta^2 -  \bfx^2 )  \partial_{\bfx}  \, . 
\label{deSitter:SCT}
\end{align}
Transforming the generators \eqref{deSitter:SI} and \eqref{deSitter:SCT} to momentum space is straightforward (see, for instance, \cite{Maldacena:2011nz}), and amounts to replacing $\vec{x}\to-i\partial_{\vec{k}}$ and $\partial_{\vec{x}}\to -i\vec{k}$ (taking suitable care with products like $\bfx \cdot \partial_{\bfx}$)\footnote{
Note that, when acting on a function of $|\bfk|$ only, the boost generator simplifies to,
\begin{align}
\mathbf{K} &= \frac{\bfk}{k} \left[  2 ( d - 1  -\eta \partial_\eta  ) \partial_k  - k \left(  \eta^2  +  \partial_k^2 \right)     \right]  \; . 
\end{align}
},
\begin{align}
D &= d  - \eta\partial_{\eta} + \bfk \cdot \partial_{\bfk}  \label{deSitter:SImom}  \\
\mathbf{K} &= 2 D \partial_{\mathbf{k}}  - \bfk \, \left(  \eta^2 +  \partial_{\bfk}^2   \right)  \; . \label{deSitter:SCTmom}
\end{align}

\paragraph{Noether Currents:}
Invariance of the scalar field action $S[\phi] = \int d^4 x \sqrt{-g} \mathcal{L}$ under these symmetries gives rise to conserved Noether currents\footnote{
These may be recognised as the components of the Noether stress-energy and angular momentum tensors (associated to translations and boosts) from Minkowski space, with time replaced by $x_\mu x^\mu$. 
},
\begin{align}
 J_D^0 &=   \frac{\delta \mathcal{L}}{\delta \partial_\eta \phi} \hat{D}[ \phi ] + \eta \mathcal{L} \;\;\;\; , \;\;\;\;   
 & J_D^i &=  - \frac{\delta \mathcal{L}}{\delta \partial_i \phi} \hat{D} [ \phi ] + \bfx^i \mathcal{L}  \\ 
  J_{K_i}^0 &= \frac{\delta \mathcal{L}}{\delta \partial_\eta \phi} \hat{K}_i [\phi]   - 2 \bfx_i \eta \mathcal{L}  \;\;\;\; , \;\;\;\;  &J_{K_i}^j &=   \frac{\delta \mathcal{L}}{\delta \partial_j \phi}  \hat{K}_i [\phi]    - \left(  \delta_i^j (  \eta^2 - \bfx^2)  + 2 \bfx_i \bfx^j \right) \mathcal{L}
\end{align}
Once promoted to operators (and normally ordered appropriately), these can be used to implement \eqref{deSitter:SI} and \eqref{deSitter:SCT} directly on the wavefunction. For instance, dilations in the quantum theory are implemented by,
\begin{align}
 \hat{Q}_D = \;\;  : \int_{\bfx} \left( - \hat{\Pi}_{\bfx} \, \bfx \cdot \partial_{\bfx} \hat{\phi}_{\bfx}    - \eta \, \hat{\mathcal{H}}_\bfx  \right) :
 \label{eqn:QDx}
\end{align}
and similarly for de Sitter boosts,
\begin{align}
 \hat{Q}_{K_i} =  \;\;  : \int_{\bfx} \left(  \hat{\Pi}_{\bfx} \left( 2 \bfx_i \, \bfx \cdot \partial_{\bfx}  +  (\eta^2 - \bfx^2) \partial_{\bfx_i}  \right) \hat{\phi}_\bfx      + 2 \bfx_i \eta \hat{\mathcal{H}}_\bfx      \right) : 
 \label{eqn:QKx}
\end{align}
where we have defined the Hamiltonian density, $ \int d^d \bfx \, \mathcal{H}_{\bfx} = \mathcal{H} $, and used $: \hat{\mathcal{O}} :$ to highlight the normal ordering.

\subsection{Equal-Time Correlators}
\label{sec:equalTime}

Equal-time correlators in approximately Gaussian states can be computed either by using \eqref{eqn:adef} to express $\hat{\phi}$ and $\hat{\Pi}$ in terms of $\hat{A}$ and $\hat{A}^\dagger$ as in usual canonical quantisation, or equivalently by inserting a functional integral over a complete set of field eigenstates\footnote{
Note that this equation is implicitly normalised with a factor of $\langle \Psi_\eta | \Psi_\eta \rangle = \int \mathcal{D}\,  \mathcal{\phi} | \Psi_\eta |^2$. 
},
\begin{align}
 \langle \Psi_\eta | \; \mathcal{O} [ \hat{\phi} , \hat{\Pi} ] \; | \Psi_\eta \rangle = \int \mathcal{D} \phi \; \; \Psi_\eta^* [ \phi ] \;  \mathcal{O} \left[ \phi ,  \frac{1}{i}\frac{\delta}{\delta \phi} \right] \; \Psi_\eta [\phi]  \; . 
\end{align} 
Note that this is not an integral over paths---the $\phi$ here is a function of spatial momentum  only---but rather an average over all possible field realisations on a fixed hypersurface, weighted by how likely each is given that the system is in the state $\Psi_\eta [ \eta ]$. 

\paragraph{Field Correlators:}
For instance, to leading order in the non-Gaussianity (i.e. assuming weak coupling), the first few equal-time correlators of the scalar field $\hat{\phi}$ are,  
\begin{align}
 \langle  \phi_{\bfk} \phi_{-\bfk}   \rangle'   &=  | f_k |^2   \label{eqn:phiphic2}   \\
\frac{ \langle  \phi_{\bfk_1} \phi_{\bfk_2} \phi_{\bfk_3}  \rangle' }{ | f_{k_1} |^2 | f_{k_2} |^2 | f_{k_3} |^2 }  &=  i \left( c_{\bfk_1 \bfk_2 \bfk_3} - c^*_{\bfk_1 \bfk_2 \bfk_3} \right)   \, .   \label{eqn:phiphiphic3}    \\
\frac{ \langle  \phi_{\bfk_1} \phi_{\bfk_2} \phi_{\bfk_3} \phi_{\bfk_4} \rangle' }{ | f_{k_1} |^2 | f_{k_2} |^2 | f_{k_3} |^2 | f_{k_4} |^2 }  &=  i \left( c_{\bfk_1 \bfk_2 \bfk_3 \bfk_4} - c^*_{\bfk_1 \bfk_2 \bfk_3 \bfk_4}  \right)  -  | f_{ k_{s} } |^2  2 i \text{Im} \, c_{\bfk_1 \bfk_2 -\bfk_s} 2 i \text{Im} \,  c_{\bfk_3 \bfk_4 \bfk_s}  \nonumber \\
&-  | f_{ k_{t} } |^2  2i \text{Im} \, c_{\bfk_1 \bfk_3 -\bfk_t} 2i \text{Im} \,  c_{\bfk_2 \bfk_4 \bfk_t} 
  -  | f_{ k_{u} } |^2  2i \text{Im} \, c_{\bfk_1 \bfk_4 -\bfk_u} 2i \text{Im} \,  c_{\bfk_2 \bfk_3 \bfk_u}     \, .   \label{eqn:phiphiphic4}    
\end{align}
where $\langle \mathcal{O} \rangle'$ is $\langle \Psi_\eta |  \hat{\mathcal{O} }| \Psi_\eta \rangle$ with the overall momentum-conserving $\delta$-function removed. 
Note that the phase of the wavefunction (the $\text{Re} \, c_{\bfk_1 ... \bfk_n}$) does not affect these observables---any observable $\mathcal{O} [ \phi]$ which depends only on $\phi$ is sensitive only to the magnitude $| \Psi_{\eta} |^2$. 

\paragraph{Momentum Correlators:}
While the phase of the wavefunction does not contribute to correlators of $\hat{\phi}$, it does contribute to their time derivatives (much like the rate of change of the phase determines the velocity for non-relativistic particles). In the Schr\"{o}dinger picture, although $\hat{\phi}$ is not explicitly time dependent, one can form correlators of the canonical momentum $\hat{\Pi}_{\bfk} $---for instance the quadratic correlators are,
\begin{align}
 \langle \Pi_{\bfk} \phi_{- \bfk} \rangle'  &=  \frac{i}{2} +  | f_k |^2 \, \text{Re} \, c_{\bfk - \bfk}   \; ,  \\
 \langle \Pi_{\bfk} \Pi_{- \bfk} \rangle'  &=  \frac{1}{4 | f_k |^2 } +  | f_k |^2 \left(  \text{Re} \, c_{\bfk - \bfk}  \right)^2   \; . 
 \label{eqn:quadratic2}
\end{align}
and depend on $\text{Re} \, c_2$. Similarly, cubic correlators containing $\Pi_{\bfk}$ now depend on $\text{Re} \, c_3$, 
\begin{align}
\frac{ \langle  \Pi_{\bfk_1} \phi_{\bfk_2} \phi_{\bfk_3}  \rangle' }{ | f_{k_1} |^2 | f_{k_2} |^2 | f_{k_3} |^2 }  &=  i \left( c_{\bfk_1 \bfk_2 \bfk_3} c^*_{\bfk_1- \bfk_1} - c^*_{\bfk_1 \bfk_2 \bfk_3} c_{\bfk_1 - \bfk_1} \right)   \, .   
\label{eqn:PiPiPi}
\end{align}
and so on. 

However, recall that $\text{Re}\, c_2$ is due to the damping of the mode functions $\partial_\eta | f_{\bfk} (\eta) |$. This results in an additional contribution to the canonical momentum, as can be seen from \eqref{eqn:pPia}. For example, if $| \Psi_\eta \rangle$ was the time evolution of a vacuum state, then the momentum variance in this state would be larger than naively expected from the Heisenberg uncertainty relation, $\Delta \phi \Delta \Pi = \hbar /2$. Due to $\text{Re} \, c_2$, the vacuum is effectively squeezed by the time evolution, and this unnecessarily complicates the equal-time correlation functions of $\hat{\Pi}$.

\paragraph{Removing the Damping:}
Fortunately, this damping can be removed from $\Psi_{\eta} [\phi]$ by performing the (time-dependent) unitary transformation,
\begin{align}
 \hat{B} (\eta) = \text{exp} \left( - i \int_{\bfk_1 \bfk_2} \frac{\text{Re} \, c_{\bfk -\bfk} (\eta) }{2} \hat{\phi}_{\bfk_1} \hat{\phi}_{\bfk_2}    \right)
 \label{eqn:F}
\end{align} 
which explicitly shifts the momentum,
\begin{align}
\tilde{\Pi}_\bfk (\eta) :=  \hat{B}^{\dagger} (\eta) \hat{\Pi}_\bfk \hat{B} (\eta) = \hat{\Pi}_{\bfk} - \text{Re} \, c_{\bfk -\bfk} (\eta) \hat{\phi}_{\bfk} \, , 
\label{eqn:Pitilde}
\end{align}
for which the quadratic correlators are now $\langle \hat{\phi}_{\bfk} \tilde{\Pi}_{\bfk} \rangle = i/2$ and $\Delta \phi \Delta \tilde{\Pi} = \hbar/2$, saturating the Heisenberg uncertainty bound.
%
%
Once the damping has been removed by \eqref{eqn:Pitilde}, the cubic correlators are now simply,
\begin{align}
\frac{ \langle \phi_{\bfk_1} \phi_{\bfk_2} \phi_{\bfk_3}  \rangle' } { |f_{k_1} |^2 | f_{k_2} |^2 | f_{k_3} |^2 }  &=  i \left( c_{\bfk_1 \bfk_2 \bfk_3} - c^*_{\bfk_1 \bfk_2 \bfk_3}  \right)   \, , \\
\left( \frac{2}{i} \right) \frac{ \langle  \tilde{\Pi}_{\bfk_1} \phi_{\bfk_2} \phi_{\bfk_3}  \rangle' } {  | f_{k_2} |^2 | f_{k_3} |^2 }  &=  - i  \left( c_{\bfk_1 \bfk_2 \bfk_3} + c^*_{\bfk_1 \bfk_2 \bfk_3}  \right)   \, ,   \\
\left( \frac{2}{i} \right)^2 \frac{ \langle  \tilde{\Pi}_{\bfk_1} \tilde{\Pi}_{\bfk_2} \phi_{\bfk_3}  \rangle' }{  | f_{k_3} |^2  }  &=   i \left( c_{\bfk_1 \bfk_2 \bfk_3} - c^*_{\bfk_1 \bfk_2 \bfk_3 }  \right)    \\
\left( \frac{2}{i} \right)^3  \langle  \tilde{\Pi}_{\bfk_1} \tilde{\Pi}_{\bfk_2} \tilde{\Pi}_{\bfk_3}  \rangle'   &= - i \left( c_{\bfk_1 \bfk_2 \bfk_3} + c^*_{\bfk_1 \bfk_2 \bfk_3 }  \right)  \, . 
\end{align}
and are determined solely by $\text{Re}\, c_3$ or $\text{Im} \, c_3$, depending on whether there is an even or odd number of momenta in the correlator\footnote{
In an approximately Gaussian state, because there is always a linear combination of $\phi$ and $\Pi$ for which $\hat{A} | \Psi \rangle = 0$, there are only two independent cubic correlators (namely $AAA^\dagger$ and $A A^\dagger A^\dagger$)---this is consistent with the wavefunction carrying only two independent functions at this order, $\text{Re} c_3$ and $\text{Im} c_3$. At quartic order, there are two new independent functions in the wavefunction, $\text{Re} c_4$ and $\text{Im} c_4$, consistent with inserting either an additional $A$ or an additional $A^\dagger$ into the correlator. 
}. 
The quartic correlators follow the same pattern, and are all related to \eqref{eqn:phiphiphic4} (up to an overall normalisation) by substituting $\text{Im} \, c_n$ for $\text{Re}\, c_n$ when an odd number of the momenta are carried by $\tilde{\Pi}$'s. For the sake of concreteness, they are given explicitly by, 
\begin{align}
\left( \frac{2}{i} \right) \frac{ \langle  \tilde{\Pi}_{\bfk_1} \phi_{\bfk_2} \phi_{\bfk_3} \phi_{\bfk_4} \rangle' }{  | f_{k_2} |^2 | f_{k_3} |^2 | f_{k_4} |^2 }  &=  - i \left( c_{\bfk_1 \bfk_2 \bfk_3 \bfk_4} + c^*_{\bfk_1 \bfk_2 \bfk_3 \bfk_4}  \right) 
 -  | f_{ k_{s} } |^2  2 \text{Re} \, c_{\bfk_1 \bfk_2 -\bfk_s} 2 i \text{Im} \,  c_{\bfk_3 \bfk_4 \bfk_s}  \nonumber \\
&-  | f_{ k_{t} } |^2  2 \text{Re} \, c_{\bfk_1 \bfk_3 -\bfk_t} 2i \text{Im} \,  c_{\bfk_2 \bfk_4 \bfk_t} 
  -  | f_{ k_{u} } |^2  2 \text{Re} \, c_{\bfk_1 \bfk_4 -\bfk_u} 2i \text{Im} \,  c_{\bfk_2 \bfk_3 \bfk_u} 
   \label{eqn:Piphiphiphi} \\
\left( \frac{2}{i} \right)^2 \frac{ \langle  \tilde{\Pi}_{\bfk_1} \tilde{\Pi}_{\bfk_2} \phi_{\bfk_3} \phi_{\bfk_4} \rangle' }{  | f_{k_3} |^2 | f_{k_4} |^2 }  &=   i \left( c_{\bfk_1 \bfk_2 \bfk_3 \bfk_4} - c^*_{\bfk_1 \bfk_2 \bfk_3 \bfk_4}  \right) 
 -  | f_{ k_{s} } |^2  2 \text{Im} \, c_{\bfk_1 \bfk_2 -\bfk_s} 2 i \text{Im} \,  c_{\bfk_3 \bfk_4 \bfk_s}  \nonumber \\
&-  | f_{ k_{t} } |^2  2 \text{Re} \, c_{\bfk_1 \bfk_3 -\bfk_t} 2 \text{Re} \,  c_{\bfk_2 \bfk_4 \bfk_t} 
  -  | f_{ k_{u} } |^2  2 \text{Re} \, c_{\bfk_1 \bfk_4 -\bfk_u} 2 \text{Re} \,  c_{\bfk_2 \bfk_3 \bfk_u} 
    \\   
   \left( \frac{2}{i} \right)^3 \frac{ \langle  \tilde{\Pi}_{\bfk_1} \tilde{\Pi}_{\bfk_2} \tilde{\Pi}_{\bfk_3}  \phi_{\bfk_4} \rangle' }{  | f_{k_4} |^2 }  &= -  i \left( c_{\bfk_1 \bfk_2 \bfk_3 \bfk_4} + c^*_{\bfk_1 \bfk_2 \bfk_3 \bfk_4}  \right) 
 -  | f_{ k_{s} } |^2  2i \text{Im} \, c_{\bfk_1 \bfk_2 -\bfk_s} 2  \text{Re} \,  c_{\bfk_3 \bfk_4 \bfk_s}  \nonumber \\
&-  | f_{ k_{t} } |^2  2i  \text{Im} \, c_{\bfk_1 \bfk_3 -\bfk_t} 2 \text{Re} \,  c_{\bfk_2 \bfk_4 \bfk_t} 
  -  | f_{ k_{u} } |^2  2 \text{Re} \, c_{\bfk_1 \bfk_4 -\bfk_u} 2i \text{Im} \,  c_{\bfk_2 \bfk_3 \bfk_u} 
    \\
       \left( \frac{2}{i} \right)^4 \langle  \tilde{\Pi}_{\bfk_1} \tilde{\Pi}_{\bfk_2} \tilde{\Pi}_{\bfk_3}  \tilde{\Pi}_{\bfk_4} \rangle'   &=   i \left( c_{\bfk_1 \bfk_2 \bfk_3 \bfk_4} - c^*_{\bfk_1 \bfk_2 \bfk_3 \bfk_4}  \right) 
 -  | f_{ k_{s} } |^2  2i \text{Im} \, c_{\bfk_1 \bfk_2 -\bfk_s} 2i  \text{Im} \,  c_{\bfk_3 \bfk_4 \bfk_s}  \nonumber \\
&-  | f_{ k_{t} } |^2  2i  \text{Im} \, c_{\bfk_1 \bfk_3 -\bfk_t} 2 i \text{Im} \,  c_{\bfk_2 \bfk_4 \bfk_t} 
  -  | f_{ k_{u} } |^2  2 i \text{Im} \, c_{\bfk_1 \bfk_4 -\bfk_u} 2i \text{Im} \,  c_{\bfk_2 \bfk_3 \bfk_u} 
\end{align}

To summarise, once the Hamilton-Jacobi equation of section~\ref{sec:gaussian} has been solved for (both the real and imaginary parts of) the wavefunction coefficients $c_n$, they can be translated straightforwardly into the equal-time correlation functions of $\phi$ and $\tilde{\Pi}$.  


\paragraph{de Sitter Invariance:}
Equal-time correlators are generally not manifestly invariant under spacetime isometries, because the restriction of a correlation like $\langle \phi_{\bfk_1} (\eta_1 ) ... \phi_{\bfk_n} (\eta_n) \rangle$ to ``equal times'' is not a frame-independent procedure---different observers will construct different equal-time correlators.
However, the underlying dynamics is still invariant under the isometries, and this should leave some imprint on the equal-time correlators. 

To find this constraint, first consider the unequal-time in-in correlator in the Heisenberg picture,
 $\langle \phi_{\bfk_1} (\eta_1 ) ... \phi_{\bfk_n} ( \eta_n) \rangle$. Such an object is invariant under the isometries providing that,
\begin{align}
 \sum_J D_J  [ \langle \phi_{\bfk_1} (\eta_1 ) ... \phi_{\bfk_n} ( \eta_n) \rangle ] &=  0  \label{eqn:Dinvar} \\
 \sum_J \mathbf{K}_J [  \langle \phi_{\bfk_1} (\eta_1 ) ... \phi_{\bfk_n} ( \eta_n) \rangle ] &= 0   \, .  \label{eqn:Kinvar}
\end{align}  
where $D_J$ and $\mathbf{K}_J$ are given by \eqref{deSitter:SImom} and \eqref{deSitter:SCTmom} with $(\eta, \bfk)$ replaced by $(\eta_J, \bfk_J)$. 
Once the overall momentum-conserving $\delta$-function has been removed, this requires that,
\begin{align}
0 &= \left[  (n-1) d    +  \sum_{j=1}^n \left( - \eta_j \partial_{\eta_j}  + \bfk_j \cdot \partial_{\bfk_j}  \right)     \right] \langle \phi_{\bfk_1} (\eta_1 ) ... \phi_{\bfk_n} ( \eta_n) \rangle'  \nonumber  \\
0 &=  \sum_{j=1}^n \left[  2 (  d  - \eta_j \partial_{\eta_j} + \bfk_j \cdot \partial_{\bfk_j}  ) \partial_{\bfk_j}  - \bfk_j \left(  \partial_{\bfk_j}^2    + \eta_j^2  \right)    \right] \langle \phi_{\bfk_1} (\eta_1 ) ... \phi_{\bfk_n} ( \eta_n) \rangle' \, ,
\label{eqn:unequalK}
\end{align}
which differ from \eqref{eqn:Dinvar} and \eqref{eqn:Kinvar} only by the scaling weight ($-d$) of the $\delta$-function\footnote{
Note that for $\mathbf{K}$, we have dropped the $\eta^2 \bfk$ term (which vanishes through momentum conservation), and used the fact that the action of the $\partial_k$ derivatives on the $\delta$ function also vanishes by dilation and rotation invariance (as described in Appendix D of \cite{Maldacena:2011nz}).
}. Analogous relations hold with any number of $\hat{\phi}$'s replaced with $\hat{\Pi}$. 

Now, since the equal-time limit of $\langle \phi_{\bfk_1} (\eta_1 ) ... \phi_{\bfk_n} ( \eta_n) \rangle$ is nothing but the $\langle \phi_{\bfk_1} ... \phi_{\bfk_n} \rangle ( \eta )$ given above in the Schr\"{o}dinger picture, the equal-time limit of \eqref{eqn:unequalK} provides the constraint corresponding to bulk de Sitter invariance. 
Since de Sitter boosts involve $\sum_J \eta_J \partial_{\eta_J} \partial_{\bfk_J}$, the equal-time limit must be taken with care (as this cannot be extracted from $\langle \phi_{\bfk_1} ... \phi_{\bfk_n} \rangle$ alone)---each $\partial_\eta \phi$ in \eqref{eqn:unequalK} must first be replaced with the canonical momentum $\Pi$ before taking the equal-time limit. This results in the ``equal-time'' version of the de Sitter isometries, 
\vspace{+0.5\baselineskip}
\begin{mdframed}[style=boxed]
\vspace{-0.5\baselineskip}
\begin{align}
\left(  (n-1) d  - \eta \partial_\eta   + \sum_J  \bfk_J \cdot \partial_{\bfk_J}   \right)  \langle \phi_{\bfk_1} ... \phi_{\bfk_n}  \rangle'  &= 0   \label{eqn:equalTimeIsometries} \\
\sum_J \left(   2 \left( d + \bfk_J \cdot \partial_{\bfk_J}  \right) \partial_{\bfk_J}  - \bfk_J \partial_{\bfk_J}^2     \right)  \langle \phi_{\bfk_1} ... \phi_{\bfk_n}  \rangle' &= \frac{2 \eta}{\Omega^{d-1}} \sum_J  \partial_{\bfk_J} \langle \phi_{\bfk_1} ... \Pi_{\bfk_J} ... \phi_{\bfk_n}  \rangle'   \, ,   \nonumber
\end{align}
\end{mdframed} 
where we have assumed weak coupling so that $\Pi = \Omega^{d-1} \partial_\eta \phi + \mathcal{O}(g)$ in the Heisenberg picture.  

We note for later convenience that when $\langle \phi_{\bfk_1} (\eta_1 ) ... \phi_{\bfk_n} ( \eta_n) \rangle'$ depends only on the magnitude of the vectors, $k_J = | \bfk_J |$ (which is the case for the two-point and three-point correlators, and also all $\phi^n$ contact contributions), then \eqref{eqn:unequalK} can be written more simply as,
 \begin{align}
  0 &=   \left( K_I - K_J  \right)  \langle \phi_{\bfk_1} (\eta_1 ) ... \phi_{\bfk_n} ( \eta_n) \rangle' \label{eqn:Ksimple} \\ 
%
%
\text{where} \;\;\;\; K_J &=  \frac{1}{k_J} \left( d + 1  - 2 D_J  \right)    \partial_{k_J} + \partial_{k_J}^2 + \eta_J^2  \, .  \nonumber 
\end{align}
for any pair $(I, J)$ of fields. The equal time limit of \eqref{eqn:Ksimple} (again taking care to replace $\partial_\eta \phi$ with $\Pi$) results in a constraint on $\langle \phi_{\bfk_1} ... \phi_{\bfk_n} \rangle (\eta)$ which is often easier to implement than \eqref{eqn:equalTimeIsometries}. 

Given a set of equal-time correlators, \eqref{eqn:equalTimeIsometries} may be used to check whether they were produced by a de Sitter invariant state. Now, while these correlators can be written in terms of the $c_n$, we will see later in explicit examples that \eqref{eqn:equalTimeIsometries} applied to just one correlator (say $\langle \phi^n \rangle$) is actually not sufficient to guarantee that the whole state is de Sitter invariant, and it can be cumbersome to apply \eqref{eqn:equalTimeIsometries} to every correlator of both $\phi$ and $\Pi$. Ideally, we would instead have a simple constraint written directly in terms of the $c_n$ coefficients.  

While the wavefunction coefficients can themselves be written as an equal-time correlator,
\begin{align}
 c_{\bfk_1 ... \bfk_n} (\eta) \propto  \langle  \phi_{\bfk} = 0  |  \Pi_{\bfk_1} ... \Pi_{\bfk_n} |  \Psi \rangle  (\eta)
 \label{eqn:cnEqualTime}
\end{align} 
involving a field eigenstate (defined such that $\hat{\phi}_{\bfk}   | \phi_{\bfk} = g_{\bfk}  \rangle = g_\bfk | \phi_{\bfk} = g_{\bfk} \rangle$ for all $\bfk$), since the different $\hat{\phi}_{\bfk_J}$ only commute at equal times it does not seem possible to construct this state using the equal-time limit of some unequal-time correlator, as we did above for $\langle \phi_{\bfk_1} ... \phi_{\bfk_n} \rangle (\eta)$. 
In fact, we will now use the Noether charges constructed in section \ref{sec:generators} in order to implement dilations and boosts directly on $| \Psi_\eta \rangle$.

\subsection{Wavefunction Coefficients}
\label{sec:isometry}

\paragraph{Dilation Constraints:}
A dilation acts on the wavefunction as $\hat{Q}_{D} | \Psi \rangle$, where the charge $\hat{Q}_{D}$ is defined in \eqref{eqn:QDx}. Transforming to momentum space, this symmetry generator shifts the wavefunction phase by,
\vspace{+0.5\baselineskip}
\begin{mdframed}[style=boxed]
\vspace{-0.5\baselineskip}
\begin{align}
\delta_D \Gamma_{\eta}  =  - \eta \partial_\eta \Gamma_{\eta}  + \int_{\bfp}   \;  \phi_{-\bfp} \, \bfp \cdot \partial_{\bfp} \, \frac{\delta \Gamma_\eta }{ \delta \phi_{\bfp} }    \, .
\label{eqn:DGamma}
\end{align}
\end{mdframed}
Written in terms of the $c_n (\eta)$ coefficients (taking care to also account for the $\partial_{\bfp}$ acting on the overall momentum conserving $\delta$-function),  
\begin{align}
\delta_D \, c_{\bfk_1 ... \bfk_n} (\eta)  =  \left( - d  - \eta \partial_\eta + \sum_J  \; \bfk_J \cdot \partial_{\bfk_J} \right) \, c_{\bfk_1 ... \bfk_n}   (\eta) \, .
\label{eqn:Dilc1}
\end{align}
Note that, crucially, this differs from the way that dilations are implemented on correlators \eqref{eqn:equalTimeIsometries} by the replacement $d - \eta \partial_\eta \to - \eta \partial_\eta$. This will be important when we study the conformal boundary at $\eta \to 0$, where $\eta \partial_\eta$ acting on a $\phi$ correlator will give the scaling dimension $\Delta$, while $\eta \partial_\eta$ acting on a $c_n$ coefficient will give the shadow weight $d - \Delta$.  

Note that although $\sum_J D_J$ contains derivatives with respect to unequal times, since $\sum \partial_{\eta_J} f (\eta_1, \eta_2, ..., \eta_n) = \partial_\eta f (\eta,\eta,..., \eta)$ it is possible to write the dilation constraint simply in terms of $\partial_\eta c_n (\eta)$. 
This can even be combined with \eqref{eqn:HJeg}, the equation of motion for $\partial_\eta \Gamma_\eta$, and written directly in terms of wavefunction coefficients,
\begin{align}
\delta_D c_{\bfk_1 ... \bfk_n} (\eta) = \left[   - d + \sum_J \left(  \frac{\eta c_{\bfk_J -\bfk_J} (\eta) }{\Omega^{d-1}}    +  \bfk_J \cdot \partial_{\bfk_J}         \right)   \right] c_{\bfk_1 ... \bfk_n} +  \eta \frac{\delta^n \mathcal{H}_{\rm int} }{\delta \phi_{\bfk_1} ... \delta \phi_{\bfk_n} }
\label{eqn:Dilc}
\end{align}
plus exchange contributions.

\paragraph{Boost Constraints:}
A boost acts on the wavefunction as $ \hat{Q}_{\mathbf{K}} | \Psi \rangle $, where $\hat{Q}_{\mathbf{K}}$ is defined in \eqref{eqn:QKx}. Transforming to momentum space, this symmetry generator shifts the wavefunction phase by, 
\vspace{+0.5\baselineskip}
\begin{mdframed}[style=boxed]
\vspace{-0.5\baselineskip}
\begin{align}
 \delta_{\mathbf{K}} \Gamma_\eta =   \int_{\bfp} \left[ \frac{ \eta }{ \Omega^{d-1}} \frac{\delta \Gamma_\eta}{\delta \phi_{-\bfp}} \,  \frac{\partial}{ \partial \bfp } \, \frac{ \delta \Gamma_\eta }{\delta \phi_{\bfp}}    
 + \phi_{-\bfp} \left( 2 \bfp \cdot \partial_\bfp \frac{\partial}{\partial \bfp}  - \bfp \partial_{\bfp}^2     \right) \frac{\delta \Gamma_\eta}{\delta \phi_{\bfp}} 
   \right]    +  2 \eta \partial_{\bfq} \mathcal{H}^{\rm int}_{\bfq} |_{\bfq= 0}
   \label{eqn:KGamma}
\end{align}
\end{mdframed}
plus terms quadratic in $\phi$ (which do not affect the wavefunction coefficients $c_n$ with $n>2$). We have used the Fourier transform of the Hamiltonian density, $\mathcal{H}_{\bfq} = \int d^d \bfx \, e^{i \bfq \cdot \bfx} \, \mathcal{H}_{\bfx}$, to express $\bfx \, \mathcal{H}_{\bfx}$ as $\partial_{\bfq} \mathcal{H}_{\bfq} |_{\bfq= 0}$. 
Unlike the dilation constraint, which resembles the dilation operator with $\sum \partial_{\eta_J}$ replaced by $\partial_\eta$, it is not possible to write the boost constraint simply in terms of $\partial_\eta \Gamma$, since $\sum_J \partial_{\bfk_J} \partial_{\eta_J}$ acting on an unequal-time object cannot be replaced by a single $\partial_\eta$ acting on an equal-time object. 
In terms of wavefunction coefficients\footnote{
Note that if the coefficient is a function of the $|\bfk_J|$ only, then this simplifies to,
\begin{align}
\delta_{\mathbf{K}} c_{k_1 ... k_n} (\eta) = \sum_{J=1}^n \frac{\bfk_J}{k_J} \left[  2 \left( \frac{\eta c_{\bfk_J - \bfk_J} }{\Omega^{d-1}}  - 1  \right) \partial_{k_J} + k_J \partial_{k_J}^2     \right] c_{k_1 ... k_n} (\eta)
\label{eqn:KcSph}
\end{align}
plus exchange and interaction terms. 
}, 
\begin{align}
 \delta_{\mathbf{K}} \, c_{\bfk_1 ... \bfk_n} (\eta) = \sum_{J=1}^n \left[ 2  \left( \frac{\eta \, c_{\bfk_J - \bfk_J} (\eta)}{\Omega^{d-1}}  + \bfk_J \cdot \partial_{\bfk_J}   \right)  \frac{\partial}{\partial \bfk_J} - \bfk_J \partial_{\bfk_J}^2     \right] c_{\bfk_1 ... \bfk_n} (\eta)
 \label{eqn:Kc}
\end{align}
plus exchange and interaction terms. 

\paragraph{$\alpha$ Vacua:}
Armed with these de Sitter isometries, it is now straightforward to determine which states are de Sitter invariant. For instance, consider the Gaussian states described in section~\ref{sec:gaussian}. 
On a de Sitter background, we can solve \eqref{eqn:HJfree} for $c_{2}$,
\begin{equation}
 c_{\bfk, -\bfk} (\eta)  = - \Omega^d \, H \left( 
  \Delta - k \eta \frac{ ( H^{(2)} +\alpha^{\rm in}_{\bfk} H^{(1)})_{\nu-1}(-k \eta)}{(H^{(2)} + \alpha^{\rm in}_{\bfk} H^{(1)} )_{\nu}(-k \eta)} 
\right)  
  \label{eqn:deSiterFree}
\end{equation}
with $\alpha_{\bfk}^{\rm in}$ a constant function of $\bfk$, and $\nu = \sqrt{ \frac{d^2}{4} - \frac{m^2}{H^2}}$ is the order of the Hankel functions. The associated mode function \eqref{eqn:HJfreeMode} is,
\begin{align}
f_{\bfk} (\eta) = N(k) \Omega^{-d/2} \left(  H^{(1)}_\nu ( - k \eta) + \left( \alpha^{\rm in}_\bfk \right)^* H^{(2)}_{\nu} ( - k \eta ) \right) \, , 
\label{eqn:modefalpha}
\end{align} 
where the overall normalisation $N (k)$ does not affect any observable\footnote{
By this we mean that the physical wavefunction coefficients, $c_n$, only ever depend on ratios of mode functions, $f_{\bfk} (\eta_2) / f_{\bfk} ( \eta_1)$, and their derivatives. Of course, when writing expressions like (\ref{eqn:phiphic2}--\ref{eqn:phiphiphic4}) in section~\ref{sec:equalTime}, we have assumed that $f_{\bfk}$ is normalized as in \eqref{eqn:fNorm} in order to simplify expressions. 
Unless stated otherwise, we will always choose $N(k)$ such that the commutation relations are canonically normalised \eqref{eqn:fNorm}, and such that $|\bar{\bfk}| = - |\bfk|$ when $\alpha_{\bfk}^{\rm in} = 0$, as in \eqref{eqn:modefdS}. 
}. 
 
Specifying an initial value of $c_{2}$ fixes $\alpha^{\rm in}_{\bfk}$, which could in principle be an arbitrarily complicated function of $\bfk$. However, if the initial state is to respect the de Sitter isometries, this uniquely fixes $\alpha_\bfk^{\rm in}$ up to a constant. For example, in the asymptotic past, $\eta \to -\infty$, invariance under dilations requires that $\partial_{k} \alpha^{\rm in}_{\bfk}=0$ (independently of $\mathcal{H}_{\rm int}$, which becomes unimportant in this limit). Since rotational invariance restricts $\alpha_{\bfk}^{\rm in}$ to be a function of $k$ only, this shows that the only Gaussian states which are de Sitter invariant can be parametrised by a single complex constant, $\alpha^{\rm in}$. These states correspond to the well-known $\alpha$-vacua, originally derived in \cite{Allen:1985ux} from studying properties of two-point Green's functions.

\paragraph{Anomalies:}
For particular values of the scalar field mass, the interactions can persist until late times and lead to divergences in the wavefunction coefficients. This requires renormalisation, and in particular the renormalisation of the boundary term in Noether's theorem leads to additional (anomalous) contributions to the above Ward identities as $\eta \to 0$. Unlike on Minkowski spacetime, where the only such divergences arise from loops (from the momentum $p \to \infty$ limit, in which the theory breaks down), on de Sitter spacetime these divergences can arise at tree level (from the late time $\eta \to 0$ limit, in which the volume element diverges).
We will discuss this subtlety further in the next section, where we study the behaviour of the bulk coefficients $c_n (\eta)$ in the limit $\eta \to 0$. 

\section{Locality and Analyticity on Superhorizon Scales}
\label{sec:superhorizon}

We will begin this section by describing how the de Sitter isometry conditions in the bulk (given in section \ref{sec:isometries}) become conformal Ward identities near the boundary---this is the essence of the recent Cosmological Bootstrap program \cite{Arkani-Hamed:2015bza,Arkani-Hamed:2018kmz,Sleight:2019hfp,Baumann:2019oyu} (see also \cite{Antoniadis:2011ib, Creminelli:2011mw, Bzowski:2013sza, Kundu:2014gxa,Kundu:2015xta}). 

Then in section \ref{sec:smallEta}, we discuss how the bulk wavefunction coefficients may be expressed as a power series in $\eta$ near the boundary at $\eta = 0$, assuming some generic value of the scalar field mass. We refer to the coefficients in this series as ``transfer functions'', since they are the objects which propagate the boundary data into the bulk spacetime. For local interactions, these transfer functions are analytic in the momenta of the fields outside the horizon (i.e. for every $|k \eta| < 1$), and only once $| k \eta |$ exceeds one do they develop non-analyticities (such as a $1/k_T$ pole in the total energy). This can be summarised by saying that, once a boundary value $c_n ( \eta \to 0) = \alpha_n$ is specified for the state $\Psi_{\eta \to 0} [\phi]$, the resulting bulk wavefunction coefficients $c_{\bfk_1 ... \bfk_n} (\eta)$ have non-analyticities in momenta due to the following three sources\footnote{
At tree level, this seems to exhaust all possible sources of non-analyticity. Further non-analyticities can develop once loop corrections are included.
},
\begin{itemize}

 \item[(a)] From non-analyticities in the boundary value of the wavefunction---this can be interpreted as non-localities in the initial state (for instance, imposing the Bunch-Davies initial condition in the far past produces non-analytic $\alpha_n$ at late times),
 
 \item[(b)] From non-analyticities in the interaction Hamiltonian---this would be due to the presence of non-local interactions, 
 
 \item[(c)] From the resummation which takes place in the transfer function once $| k \eta | \sim 1$ for a particular mode---this can be interpreted as the field beginning to oscillate.

\end{itemize} 

We will also find that in particular cases (in which $\nu$ is rational) this series solution becomes ill-defined, and this signals the presence of divergences which must be renormalised. This is discussed systematically in section \ref{sec:33}, first by renormalising the boundary condition for $\Psi_{\eta \to 0} [\phi]$ (analogous to the usual renormalisation of the boundary value for $\phi (\eta\to 0)$ in holography), and then from the viewpoint of boundary operator mixing (analogous to the usual renormalisation of composite operators in flat space QFT).  

Throughout this section, we will allow for the $n$ fields which multiply $c_n (\eta)$ to have different masses (conformal weights), since this both achieves a greater level of generality (the results apply to any number of distinct scalar fields) and also makes the origin of various terms clear notationally.

\subsection{The Conformal Boundary}
\label{sec:boundary}
Near the $\eta\to 0$ boundary of the expanding Poincare patch, the de Sitter isometries \eqref{deSitter:SI} and \eqref{deSitter:SCT} become conformal symmetries,
\begin{align}
D_j  \;\; &\to \;\; d - \Delta_j +\bf{k}_j \cdot \partial_{\bfk_j}     &(\text{Dil.})
 \label{eq:Dil}
 \\
\mathbf{K}_j \;\; &\to \;\; 2 (d - \Delta_j + \mathbf{k}_j \cdot\partial_{\mathbf{k}_j} ) \partial_{\mathbf{k}_j} - \mathbf{k}_j \partial_{\mathbf{k}_j}^2   &(\text{SCT})
 \label{eq:SCT}
\end{align}
namely dilations and special conformal transformations\footnote{
When acting on a function which depends on the $|\bfk|$ only, the action of SCT simplifies to,
\begin{align}
 \mathbf{K} \;\; \to \;\;  \frac{ \bfk_j}{k_j} \left[ 2 (  d - 1 - \Delta_j )\partial_{k_j} + k_j \partial_{k_j}^2 \right] \; . 
\end{align}
} (SCT).
These symmetries provide a set of Ward identites on the wavefunction coefficients which uniquely determine the two- and three-point function (up to constant coefficients), and which can be used to bootstrap four (and higher) point functions. We will now show how our bulk constraints on the $c_n ( \eta)$ from de Sitter dilations \eqref{eqn:Dilc} and de Sitter boosts \eqref{eqn:Kc} are related to these conformal Ward identities.

\paragraph{Free Evolution:}
First, let us ignore the effect of the bulk interactions and consider the free evolution given by \eqref{eqn:Hfree}. Given an initial condition for the wavefunction at any time, the resulting wavefunction coefficients are $c_{\bfk_1 ... \bfk_n} (\eta) \propto \alpha^{\rm in}_{\bfk_1 ... \bfk_n} / f_{\bfk_1} (\eta) ... f_{\bfk_n} (\eta)$. The operators $\delta_D c_n$ and $\delta_{\mathbf{K}} c_n$ defined in \eqref{eqn:Dilc} and \eqref{eqn:Kc} have the interesting property,
\begin{align}
k_1^\nu f_{\bfk_1} (\eta) ... k_n^\nu f_{\bfk_n} (\eta)  \delta_{D}  \left[  \frac{   \alpha^{\rm in}_{\bfk_1 ... \bfk_n}  }{ k_1^\nu f_{\bfk_1} (\eta)  ... k_1^\nu f_{\bfk_n} (\eta)   }   \right] = \left[ - d  +  \sum_J \left(   \Delta_J   + \bfk_J \cdot \partial_{\bfk_J}  \right)   \right]  \alpha^{\rm in}_{\bfk_1 ... \bfk_n}
 \label{eqn:Dila}
\end{align}
\begin{align}
k_1^\nu f_{\bfk_1} (\eta) ... k_n^\nu f_{\bfk_n} (\eta)  \delta_{\mathbf{K}}  \left[  \frac{ \alpha^{\rm in}_{ \bfk_1 ... \bfk_n } }{ k_1^\nu f_{\bfk_1} (\eta)  ... k_n^\nu f_{\bfk_n} (\eta)   }  \right] = \sum_J \left[  2  \left( \Delta_J   + \bfk_J \cdot \partial_{\bfk_J}   \right)  \frac{\partial}{\partial \bfk_J} - \bfk_J \partial_{\bfk_J}^2    \right]  \alpha^{\rm in}_{\bfk_1 ... \bfk_n}
 \label{eqn:Ka}
\end{align}
since the mode functions are themselves eigenstates of $D$ and $\mathbf{K}$. While $\delta_D$ and $\delta_{\mathbf{K}}$ are explicitly $\eta$-dependent, they reduce to the conformal constraints from dilations and SCT when acting on $\alpha_n^{\rm in}$, which has the conformal weight of a correlator of primary operators\footnote{
Recall that an overall momentum-conserving $\delta$-function has been removed from $\alpha_{\bfk_1 ... \bfk_n}$, so \eqref{eqn:Dila} really does correspond to $\sum_j D_j$ with $d-\Delta_j$ in place of $\Delta_j$.
} each of weight $d-\Delta_j$.
This shows that de Sitter invariance of $c_n (\eta)$ in the bulk is equivalent to conformal invariance of the initial condition $\alpha_n^{\rm in}$, regardless of the timeslice on which $\alpha^{\rm in}_n$ is specified (it need not be the boundary condition at $\eta = 0$). 

Once interactions are included, this need no longer be the case, since $c_n (\eta)$ is now generally a more complicated function of time and the operators $\delta_D$ and $\delta_{\mathbf{K}}$ acquire additional terms in $\mathcal{H}_{\rm int}$. However, in the limit $\eta \to 0$, providing the interactions turn off sufficiently fast, one recovers the conclusion that $c_n ( \eta \to 0)$ is de Sitter invariant only if the boundary value $\alpha_n$ (now supplied strictly at $\eta = 0$) is conformal.

\paragraph{Boundary Coefficients:}
Near the conformal boundary, we have the following scaling with time,
\begin{align}
c_{\bfk_1 ... \bfk_n} \sim \frac{1}{\eta^{ \sum_J \Delta_J}}  \;\;\;\;\text{and} \;\;\;\; \eta c_{\bfk_J -\bfk_J} = \frac{\Delta_J}{ ( - H \eta)^{d-1}} + ... \;\;\;\; \text{and} \;\;\;\; \mathcal{H}_{\rm int} \sim \sqrt{-g} = \frac{1}{ (- H \eta)^{d+1}} \; , 
\label{eqn:scaling}
\end{align}
where we write the momentum (and conformal weight) of each of the $n$ fields multiplying $c_n$ as $\bfk_1$ (and $\Delta_1$), $\bfk_2$ (and $\Delta_2$)..., $\bfk_n$ (and $\Delta_n$). Providing $\sum_J \Delta_J > d$, the interactions become subdominant as $\eta \to 0$.
Consequently, the bulk dilation constraint \eqref{eqn:Dilc} and bulk boost constraint \eqref{eqn:Kc} become at late times,
\begin{align}
\lim_{\eta \to 0} \delta_D c_{\bfk_1 ... \bfk_n} (\eta)  &=  \left[ - d  +  \sum_J \left(   \Delta_J   + \bfk_J \cdot \partial_{\bfk_J}  \right)   \right]  c_{\bfk_1 ... \bfk_n} (\eta = 0) \; , 
\label{eqn:DilcConformal}   \\
 \lim_{\eta \to 0} \delta_K c_{\bfk_1 ... \bfk_n} (\eta) &= \sum_J \left[ 2 \left(  \Delta_J + \bfk_J \cdot \partial_{\bfk_J}  \right) \partial_{\bfk_J} - \bfk_J \partial_{\bfk_J}^2   \right] c_{\bfk_1 ... \bfk_n} (\eta = 0) \; .
\label{eqn:KcConformal}
\end{align}
and coincide with the action of dilations and SCT in a $d$-dimensional CFT correlator (of operators with scaling dimension $d-\Delta_J$). 

Given \eqref{eqn:scaling}, the limit $\lim_{\eta \to 0} c_{\bfk_1 ... \bfk_n} ( \eta)$ is not regular. Instead, we will define the boundary value of the wavefunction coefficient via\footnote{
The scaling \eqref{eqn:andef} will also emerge naturally in section~\ref{sec:33} when we express the bulk operator $\hat{\phi}$ in terms of the boundary operator $\hat{\varphi}$. Since $\lim_{\eta \to 0} \left( \hat{\phi} / (-\eta)^{\Delta} \right) = \hat{\varphi}$, the $\alpha_n$ are naturally the wavefunction coefficients of $\Psi_{\eta=0} [ \varphi]$. 
},
\begin{align}
\alpha_n = \lim_{\eta \to 0} \left[  ( - \eta )^{\sum_J \Delta_J} c_n (\eta) \right]
\label{eqn:andef}
\end{align}
The coefficients $\alpha_{\bfk_1 ... \bfk_n}$ must then obey the conformal Ward identities \eqref{eqn:DilcConformal} and \eqref{eqn:KcConformal}, which can be used to determine which boundary conditions for the wavefunction respect the spacetime isometries. 

Although \eqref{eqn:DilcConformal} and \eqref{eqn:KcConformal} are themselves well-known, to the best of our knowledge the bulk analogues \eqref{eqn:Dilc} and \eqref{eqn:Kc} are novel---the fact that \eqref{eqn:Dilc} and \eqref{eqn:Kc} reduce to the expected conformal Ward identities as $\eta \to 0$ is an important sanity check.

\paragraph{Anomalous Ward Identities:}
Before moving on to discuss the late-time limit of the $c_n$ coefficients in more detail, we must highlight an important caveat to the conformal constraints \eqref{eqn:DilcConformal} and \eqref{eqn:KcConformal}. 
In section~\ref{sec:smallEta}, we will encounter divergences in the late-time limit of the wavefunction coefficients---these divergences can be renormalised via a (formally singular) redefinition of the boundary condition for the $\Psi_{\eta = 0}$ (the $\alpha_n$ in \eqref{eqn:andef}), as we will describe in section~\ref{sec:33}. This can be viewed as a Boundary Operator Expansion (BOE), in which the bulk operators $\hat{\phi}$ and $\hat{\Pi}$ are rewritten in terms of boundary operators $\hat{\varphi}$ and $\hat{\pi}$---the coefficients of this expansion depend on both a small regulator (e.g. $\delta = d-3$ in dimensional regularisation, or a hard cutoff at time $\eta_{\delta}$) and an RG scale $\mu$, introduced for dimensional consistency.   
The renormalisation thus introduces anomalous terms into the conformal Ward identities, through this new scale $\mu$. 

This is best illustrated with a simple example. Consider the cubic wavefunction coefficient of conformally coupled scalars (which we discuss in more detail in section \ref{sec:examples}). Setting $\alpha_2^{\rm in} = 0$ and $\alpha_3^{\rm in} = 0$ in the past (i.e. Bunch-Davies initial condition) produces a boundary coefficient,
\begin{align}
 \alpha_{\bfk_1 \bfk_2 \bfk_3} \propto  \frac{1}{ 3 - d } - \text{log} \left( i k_T \right)
\end{align}
where $k_T = k_1 + k_2 + k_3$. This function is perfectly invariant under $d$-dimensional dilations \eqref{eq:Dil} (recall that $\Delta = 1$ for a conformally coupled scalar),
\begin{align}
 D_{d \text{ dim.}}  \left[ \alpha_{\bfk_1 \bfk_2 \bfk_3} \right] = \left( -d + 3 + \sum_J \bfk_J \cdot \partial_{\bfk_J} \right) \alpha_{\bfk_1 \bfk_2 \bfk_3} = 0 ,
 \label{eqn:egDd}
\end{align} 
however once the divergence is subtracted and $d \to 3$, the resulting (finite) $\alpha_3$ is not invariant under $3$-dimensional dilations,
\begin{align}
 D_{3 \text{ dim.}} \left[   \lim_{d\to 3} \left( \alpha_{\bfk_1 \bfk_2 \bfk_3} - \frac{1}{3-d} \right)  \right] =  \sum_J \bfk_J \cdot \partial_{\bfk_J}  \left[ - \log \left( k_T \right) \right] = -1  \neq 0 \; .
 \label{eqn:egD3}
\end{align}
In general, the $3$-dimensional dilations acting on a regulated $\alpha_n$ will pick out the residue of the $1 / ( d - n \Delta ) $ pole. The discrepancy between \eqref{eqn:egDd} and \eqref{eqn:egD3} can be understood simply by noting that, in $d$ dimensions, the coefficient $\alpha_{\bfk_1 \bfk_2 \bfk_3}$ has a mass dimension $[\alpha_3 ] = d-3$, 
and so simply subtracting $1/(3-d)$ is not dimensionally consistent. Rather, we should define,
\begin{align}
 \alpha^{\rm ren}_3 (\mu ) = \alpha_3 - \frac{ \mu^{d-3} }{3-d}  = -\log \left( \frac{i k_T}{\mu} \right) + \mathcal{O} ( d - 3 ) \, , 
 \label{eqn:egaren}
\end{align}
where $\mu$ is an arbitrary scale introduced on dimensional grounds. This renormalised coefficient now satisfies $D_{3 \text{ dim}} \left[ \alpha_3 \right]  = -\mu \partial_\mu \alpha_3$, and we interpret the right-hand side as an anomalous contribution to the dilation Ward identity. 

In general, the $\alpha_n$ satisfy the anomalous form of the Ward identities\footnote{
See e.g. \cite{Bzowski:2015pba} for a careful derivation of the anomalous contributions to the conformal Ward identity for 3-point correlators in position space.
},
\begin{align}
&\left[   -d + \sum_{J} \left( \Delta_J + \bfk_J \cdot \partial_{\bfk_J}  \right) \right] \alpha_{\bfk_1 ... \bfk_n}^{\rm ren}   
=   \mathcal{A}_n  \,    \\
&\sum_{J} \left[   2  \left( \Delta_J + \bfk_J \cdot \partial_{\bfk_J}  \right) \partial_{\bfk_J} - \bfk_J \partial_{\bfk_J}^2  \right] \alpha_{\bfk_1 ...\bfk_n}^{\rm ren}  =  - \sum_{J} \, 2 \partial_{\bfk_J} \mathcal{A}_n
\label{eqn:anomaly}
\end{align}
where $\mathcal{A}_n$ is a known function of the $\alpha_j$ (e.g. $\mathcal{A}_3 = \mu \partial_\mu \alpha_3$ in the above example).  

\subsection{Analyticity from Locality}
\label{sec:smallEta}

Now we turn our attention to the behaviour of $c_n (\eta)$ at small $\eta$, close to the conformal boundary. Given a boundary condition for the state $\Psi_{\eta = 0}$ (i.e. a set of $\alpha_n$ coefficients), obtained for instance from a conformal bootstrap approach or even from a direct measurement of the end of inflation, what does that tell us about the bulk evolution? To answer this question, we will now solve the equations of motion for the $c_n (\eta)$ perturbatively at small $\eta$, beginning with the quadratic coefficient $c_2$ in the free theory. Throughout this section we will work in units in which $H=1$, so that e.g. $\Omega = 1/(-\eta)$. 

\paragraph{Mode Functions:}
The classical equations of motion at small $\eta$ enforce either $f_k ( \eta) \propto (- \eta)^{\Delta}$ or $f_k (\eta) \propto (-\eta)^{d - \Delta}$. Choosing $(-\eta)^{\Delta}$ (which dominates for light fields since $\Delta < d - \Delta$), the classical equation of motion then has a series solution, $f_k (\eta) \propto ( - \eta )^{\Delta} F_{-\nu} ( - k \eta )$, where $F_{-\nu}$ is an analytic function\footnote{
Note that the series $F_{-\nu} ( k \eta) = {}_0 F_{1} \left( 1 - \nu ; - \frac{1}{4} k^2 \eta^2    \right)$ can also be written in terms of a hypergeometric function, which makes the analytic structure manifest---we have used a Bessel $J$ function in \eqref{eqn:Fnu} to make contact with the Hankel mode functions of \eqref{eqn:modefalpha}.
} of both $\eta$ and $\bfk$,  
\begin{align}
F_{-\nu} ( k \eta) = \sum_{r=0}^\infty \frac{ (-4)^{-r} }{ \Gamma \left(  1 + r - \nu  \right) }  \frac{ ( - k \eta )^{2r} }{r!} = 2^{-\nu} \Gamma ( 1- \nu ) \frac{ J_{- \nu} ( - k \eta ) }{ ( - k \eta )^{-\nu} } \, . 
\label{eqn:Fnu}
\end{align}
The other boundary behaviour corresponds to the solution $(-\eta)^{d - \Delta} F_{+\nu} (- k \eta)$, and so the general solution for the mode function is given by,
\begin{align}
 f_{\bfk} ( \eta)  = N' (k)  (- \eta)^{\Delta} \left(  
  F_{-\nu} ( k\eta) -  \alpha_{\bfk}^* (-\eta)^{2\nu} F_{+\nu} (  k \eta )    
  \right)   \; , 
 \label{eqn:fa}
\end{align}
where $N'(k)$ is an overall normalisation that does not affect any observable, and $\alpha_{\bfk}$ is a constant function of the momentum which must be fixed using the initial condition for $\Psi [ \phi]$. 
Near the boundary, $f_{\bfk} (\eta) \propto (-\eta)^{\Delta} + \alpha_{\bfk} ( - \eta )^{d - \Delta}$, and so physically $\alpha_{\bfk}$ is capturing the subleading $\eta$ dependence at late times\footnote{
One can therefore relate $\alpha_{\bfk}$ to the conjugate momentum for $\phi$, since it is well-known holographically that the coefficient of the $(-\eta)^{d-\Delta}$ mode is conjugate to the coefficient of the $(-\eta)^{\Delta}$ mode. 
}. 
 
\eqref{eqn:fa} already exhibits a very general feature: when written in terms of the data $\alpha_n$ at the conformal boundary\footnote{
Neglecting the overall normalisation, \eqref{eqn:fa} obeys, $ f_{\bfk} \overset{\leftrightarrow}{\partial}_{\eta} f_{\bfk}^* \propto 2 i \nu \text{Im} \left( \alpha_{\bfk} \right) \Omega^{1-d} $, and so when expanding around $\alpha_{\bfk} = 0$ the commutator $[ \hat{\phi} , \hat{\Pi} ]$ vanishes and fluctuations behave classically. 
}, the coefficients in this expansion are analytic functions of the momenta at small $\eta$ (for $|k \eta|  < 1$) and only develop non-analyticities once $| k \eta| > 1$ and the mode crosses the horizon. This behaviour is not manifest if the initial condition for $\Psi [\phi]$ is supplied in the bulk, for instance comparing \eqref{eqn:fa} with the de Sitter mode function given in \eqref{eqn:modefalpha}, we see that $\alpha_{\bfk}$ is related to $\alpha_{\bfk}^{\rm in}$ by\footnote{
Note that the overall normalisations are also related, $N' (k) = - N(k) \frac{i}{\pi} 2^n  e^{i \nu \pi }   \Gamma (\nu) (-k)^{-\nu} \left(1 + (\alpha_{\bfk}^{\rm in} )^* \right) $.
},
\begin{align}
\alpha_{\bfk} = \frac{i k^{2\nu}}{\left( 1 - e^{2
   i \pi  \nu} \right)} \, \frac{  1+ \alpha_{\bfk}^{\rm in} e^{-2 i \pi  \nu} }{ 1 + \alpha_{\bfk}^{\rm in}  }  \frac{\pi  2^{1-2 \nu}}{\Gamma (\nu) \Gamma (\nu+1)}\; .
   \label{eqn:aToain}
\end{align}
The Bunch-Davies initial condition, for instance, sets $\alpha_{\bfk} \propto i k^{2\nu}$, and this introduces non-analyticity\footnote{
Fixing $\alpha_{\bfk}$ is essentially integrating out the conjugate momentum, which leaves behind an effective description of $\phi$ dynamics which contains non-analyticities---much the same way as integrating out a heavy field in a Wilsonian QFT produces non-local interactions for the remaining light fields. 
} into $f_{\bfk} ( \eta )$ (since $k = \sqrt{ \bfk \cdot \bfk}$ is not analytic in $\bfk$).

\paragraph{Quadratic Coefficient:}
Analogously, \eqref{eqn:HJfree} fixes the small $\eta$ limit of $c_{\bfk - \bfk} (\eta)$ to be either $\Delta / (-\eta)^d$ or $(d-\Delta) / (-\eta)^d$. Choosing $\Delta/(-\eta)^d$ results in a series solution $c_{\bfk-\bfk} (\eta) = (-\eta)^{-d} c_{\bfk -\bfk}^{\rm series}$ in which $c_{\bfk -\bfk}^{\rm series}$ is analytic in both $\eta$ and $\bfk$,
\begin{align}
 c_{\bfk - \bfk}^{\rm series} ( \eta ) &=  \Delta   + \frac{ ( - k \eta)^2 }{2 ( \nu -1 )} + \frac{ (- k \eta)^4 }{ 8 ( \nu -2 ) ( \nu -1 )^2 } + ...    \nonumber \\ 
 &=   \Delta  +  k \eta  \frac{ J_{1-\nu} ( - k \eta) }{ J_{-\nu} ( - k \eta ) }     \; . 
 \label{eqn:c2series}
\end{align}
that coincides with $\Omega^d \eta \partial_\eta F_{-\nu} / F_{-\nu}$ of \eqref{eqn:Fnu}. 

\begin{figure}
\centering
\includegraphics[width=0.9\textwidth]{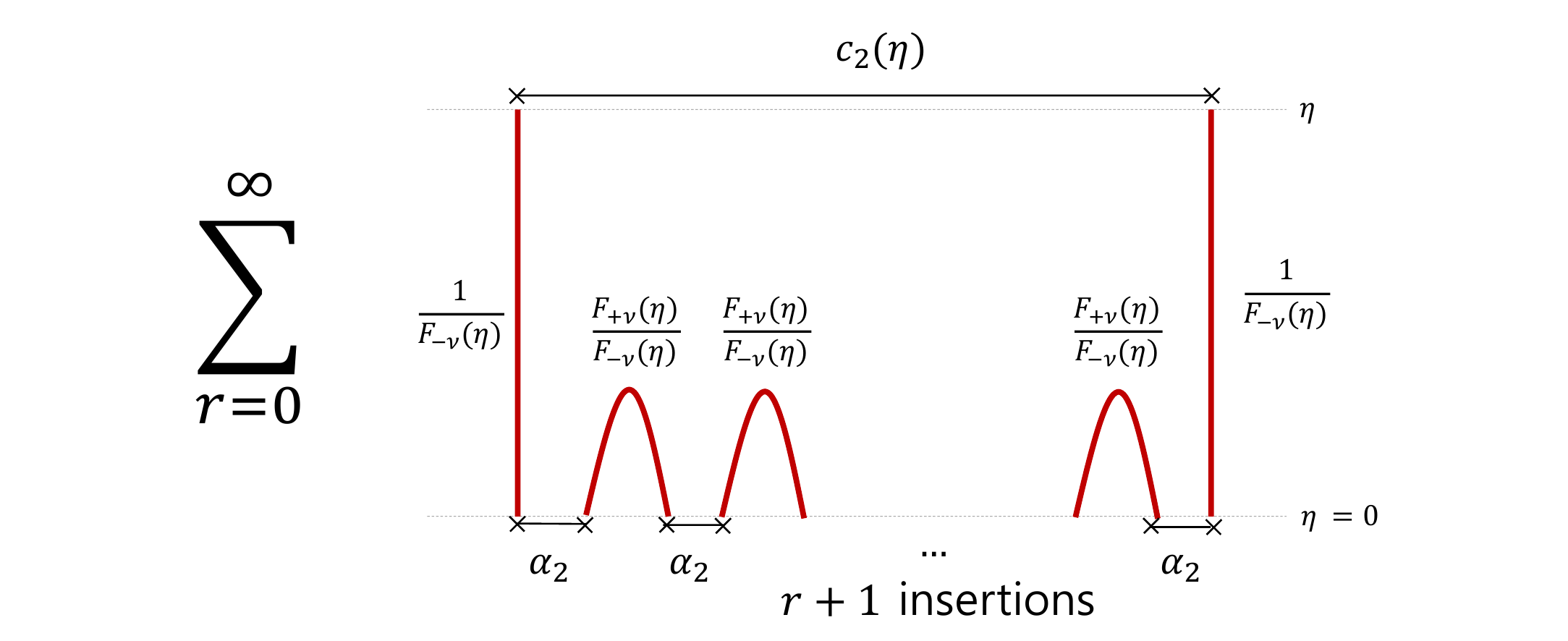}
\caption{Adding a quadratic $\alpha_2 \phi^2$ to the boundary wavefunction (at time $\eta = 0$) affects the two-point coefficient $c_2 (\eta) \phi^2$ at a bulk time $\eta$ via the resummation shown above. This is analogous to a mass insertion, $m^2 \phi^2$, shifting the propagator from $1/p^2$ to $1/(p^2-m^2)$ in a standard Lorentzian QFT.}
\label{fig:a2Resum}
\end{figure}

While $c_{\bfk - \bfk}^{\rm series} / (-\eta)^d$ is a particular solution to the Hamilton-Jacobi equation,
\begin{align}
 - (-\eta)^d  \eta \partial_\eta \left(  \frac{ c_{\bfk - \bfk}^{\rm series} }{ (-\eta)^d } \right) =  \left( c_{\bfk - \bfk}^{\rm series} \right)^2   -   \left( m^2 +  k^2 \eta^2  \right) \; , 
\end{align}
there is also the freedom to add to $c_{\bfk - \bfk}$ any solution of
\begin{align}
 - \eta \partial_\eta c_{\bfk - \bfk}  = 2  c_{\bfk - \bfk}^{\rm series} c_{\bfk - \bfk} +  (-\eta)^d ( c_{\bfk -\bfk}  )^2 
\end{align}
which depends on the Hamiltonian only implicitly (via $c_{\bfk - \bfk}^{\rm series}$). 
This corresponds to the freedom to add $ (-\eta)^{d-\Delta} F_{+\nu}$ to our solution to the classical equations of motion. In particular, for the mode function \eqref{eqn:fa}, the corresponding wavefunction coefficient is,
\begin{align}
 c_{\bfk - \bfk} (\eta) = \frac{ c_{\bfk -\bfk}^{\rm series} (\eta) }{ (-\eta)^d}  + \frac{ \alpha_{\bfk} }{ (- \eta )^{2 \Delta} } c_{\bfk -\bfk}^{\rm initial} (\eta)  \; , 
\end{align} 
where $\alpha_{\bfk}$ can be determined by specifying an initial condition for $c_{\bfk -\bfk}$, and the function $c_{\bfk - \bfk}^{\rm initial} (\eta)$ can also be written in terms of analytic functions as,
\begin{align}
c_{\bfk-\bfk}^{\rm initial} (\eta) = \sum_{r = 0} \alpha_{\bfk}^r ( - \eta )^{2 r \nu}  I^{(r)}_k ( \eta )
\label{eqn:c2initial}
\end{align}
where the functions $I^{(r)}_k (\eta)$ are analytic in both $\eta$ and $\bfk$,
\begin{align}
 I^{(r)}_k = \frac{1}{ F_{-\nu} F_{-\nu} }  \left(   \frac{ F_{+\nu} }{ F_{-\nu} }  \right)^r    \; . 
\end{align}
This leads to the resummed, $c_{\bfk -\bfk}^{\rm initial} = 1/ F_{-\nu} / ( F_{-\nu} - \alpha_{\bfk} (-\eta)^{2\nu} F_{+\nu} )$. 
This resummation can be represented graphically\footnote{
The reason that it is factors of $F_{+\nu}/F_{-\nu}$ which appear on the red internal lines of figure~\ref{fig:a2Resum} is that,
\begin{align}
(- \eta)^{2 \nu} \frac{ F_{+\nu} (k \eta)  }{ F_{-\nu} (k \eta ) } =  \int \frac{d \eta}{\eta}  \, \frac{ 2 \nu  (- \eta)^{2 \nu} }{ F_{-\nu} ( k \eta) F_{-\nu} (k \eta) }   \; . 
\end{align}
where the right-hand side is an integral over all time of the $\alpha_{\bfk} = 0$ propagator, $1/F_{-\nu}F_{-\nu}$.
}, as shown in figure~\ref{fig:a2Resum}.
 
\paragraph{Cubic Coefficient:}
We can represent the transition from boundary $\alpha_n$ to bulk $c_n$ graphically for any non-Gaussian wavefunction coefficient. The simplest of these is the cubic coefficient.
Anticipating the resummation of the $\alpha_{\bfk}$ series for the external lines, we first define a new coefficient\footnote{
Note that this is simply $(-\eta)^{-3 \Delta} c^I_{\bfk_1 \bfk_2 \bfk_3}$ for the mode function \eqref{eqn:fa}. 
}, 
\begin{align}
\tilde{c}_{\bfk_1 \bfk_2 \bfk_3} =  c_{\bfk_1 \bfk_2 \bfk_3}  \prod_{j=1}^3 \left( F_{-\nu} (k_j \eta) - \alpha_{k_j} (-\eta )^{2\nu} F_{+\nu} ( k_j \eta)   \right)
\end{align}
such that $\tilde{c}_{\bfk_1 \bfk_2 \bfk_3}$ coincides with $c_{\bfk_1 \bfk_2 \bfk_3}$ near the boundary and depends at most linearly on any single $\alpha_{\bfk_J}$,
\begin{align}
 \tilde{c}_{\bfk_1 \bfk_2 \bfk_3} (\eta) &=  \frac{\alpha_{\bfk_1 \bfk_2 \bfk_3} }{ (-\eta)^{3 \Delta}} + \tilde{c}_{\bfk_1 \bfk_2 \bfk_3} (\eta) |_{\alpha = 0}   
 + \sum_{\rm perm.}^3 \frac{ \delta  \tilde{c}_{\bfk_1 \bfk_2 \bfk_3} }{\delta \alpha_{\bfk_1}} \Big|_{\alpha = 0} \alpha_{\bfk_1}   \nonumber  \\
& + \sum_{\rm perm.}^3 \frac{ \delta^2  \tilde{c}_{\bfk_1 \bfk_2 \bfk_3} }{\delta \alpha_{\bfk_1} \delta \alpha_{\bfk_2} } \Big|_{\alpha = 0} \alpha_{\bfk_1} \alpha_{\bfk_2} +  \frac{ \delta^3  \tilde{c}_{\bfk_1 \bfk_2 \bfk_3} }{\delta \alpha_{\bfk_1} \delta \alpha_{\bfk_2} \alpha_{\bfk_3}} \Big|_{\alpha = 0} \alpha_{\bfk_1} \alpha_{\bfk_2} \alpha_{\bfk_3} \; .   
\label{eqn:c3exp}
\end{align}
where $|_{\alpha=0}$ denotes the function evaluated with all $\alpha_n = 0$. The $\eta$ dependence of $\tilde{c}_{3}$ is now encoded in the various \emph{transfer coefficients} $\delta^n \tilde{c}_i/\delta \alpha_j^n\Big|_{\alpha=0}$. 
For instance, for the cubic interaction $\mathcal{H}_{\rm int} = \Omega^{d+1} \, \frac{1}{3!} v_3 \phi^3$, the corresponding coefficients are,
\begin{align}
(-\eta)^d \tilde{c}_{\bfk_1 \bfk_2 \bfk_3 } (\eta) |_{\alpha = 0} &= v_3 \mathcal{I}_{\bfk_1 \bfk_2 \bfk_3}^{---} ( \eta ) \;   \label{eqn:c3transfer} \\
\frac{ (-\eta)^{d} }{ (-\eta)^{2 \nu_1} } \frac{ \delta \tilde{c}_{\bfk_1 \bfk_2 \bfk_3} (\eta)  }{ \delta \alpha_{\bfk_1} } \Big|_{\alpha = 0} &=   v_3 \mathcal{I}_{\bfk_1 \bfk_2 \bfk_3 }^{+--} ( \eta )    \nonumber \\
\frac{ (-\eta)^{d} }{ (-\eta)^{2 \nu_1 + 2 \nu_2} } \frac{ \delta^2 \tilde{c}_{\bfk_1 \bfk_2 \bfk_3} (\eta)  }{ \delta \alpha_{\bfk_1} \delta \alpha_{\bfk_2} } \Big|_{\alpha = 0} &= v_3 \mathcal{I}_{\bfk_1 \bfk_2 \bfk_3}^{++-} ( \eta )  \nonumber  \\
\frac{ (-\eta)^{d} }{ (-\eta)^{2 \nu_1+2 \nu_2 + 2\nu_3} } \frac{ \delta^3 \tilde{c}_{\bfk_1 \bfk_2 \bfk_3 \bfk_4} (\eta)  }{ \delta \alpha_{\bfk_1} \delta \alpha_{\bfk_2} \delta \alpha_{\bfk_3}  } \Big|_{\alpha = 0} &=   v_3 \mathcal{I}_{\bfk_1 \bfk_2 \bfk_3}^{+++} ( \eta ) \nonumber 
\end{align}
where the functions $\mathcal{I}_{\bfk_1 \bfk_2 \bfk_3}^{\nu_1 \nu_2 \nu_3}$ are analytic for $|k_j \eta| < 1$, and straightforward to write down by solving \eqref{eqn:HJc3} with mode function \eqref{eqn:Fnu},
\begin{align}
\mathcal{I}_{\bfk_1 \bfk_2 \bfk_3}^{\sigma_1 \sigma_2 \sigma_3} (\eta )  &=
\frac{1}{ \sum_i \Delta_i^{\sigma_i} - d }  +  \frac{ (-\eta)^2}{ \sum_i \Delta_i^{\sigma_i} - d + 2} \sum_{j}  \frac{ k_j^2 }{  4 ( \sigma_j \nu_j - 1 ) }  \nonumber \\
&\;\;+   \frac{ (-\eta)^4}{\sum_i \Delta_i^{\sigma_i} - d + 4} \left( \sum_{j}   \frac{ k_j^4 }{ 32 ( \sigma_j \nu_j - 1 ) ( \sigma_j \nu_j - 2 ) }  +
\sum_{j j'}   \frac{ k_j^2 k_{j'}^2 }{ 16 ( \sigma_j \nu_j - 1 ) ( \sigma_{j'} \nu_{j'} - 1 ) } 
\right)  \nonumber \\
&\;\;+ \mathcal{O}( k^6 \eta^6) \;.  \label{eqn:I3}  
\end{align}
where $\Delta^{\pm}_j = \frac{d}{2} \pm \nu_j$ (so $\Delta_j^- = \Delta_j$ and $\Delta_j^+ = d - \Delta_j$). 
The main virtue of the expansion \eqref{eqn:c3exp} is that, since the $\mathcal{I}_{\bfk_1 \bfk_2 \bfk_3}^{\sigma_1 \sigma_2 \sigma_3} (\eta)$ are analytic in the momentum, \emph{locality} of the interaction Hamiltonian (i.e. that $v_3$ is an analytic function\footnote{
Of course, for a local interaction which is also \textit{de Sitter invariant}, $v_3$ will depend on $\eta$ through the combination $\Omega^{-2} \bfk_i \cdot \bfk_j$. This does not affect our argument, since any polynomial dependence on $\eta$ can be incorporated into an analogous $\mathcal{I}_{\bfk_1 ... \bfk_n}^{\sigma_1... \sigma_n} (\eta)$ object by shifting the location of the poles in \eqref{eqn:I3} (but this will not change the analyticity of $\mathcal{I}$ in the momentum).
} of the $\bfk_j$) translates directly into \emph{analyticity of the transfer coefficients} given by \eqref{eqn:c3transfer}. 

\begin{figure}
\centering
\includegraphics[width=0.85\textwidth]{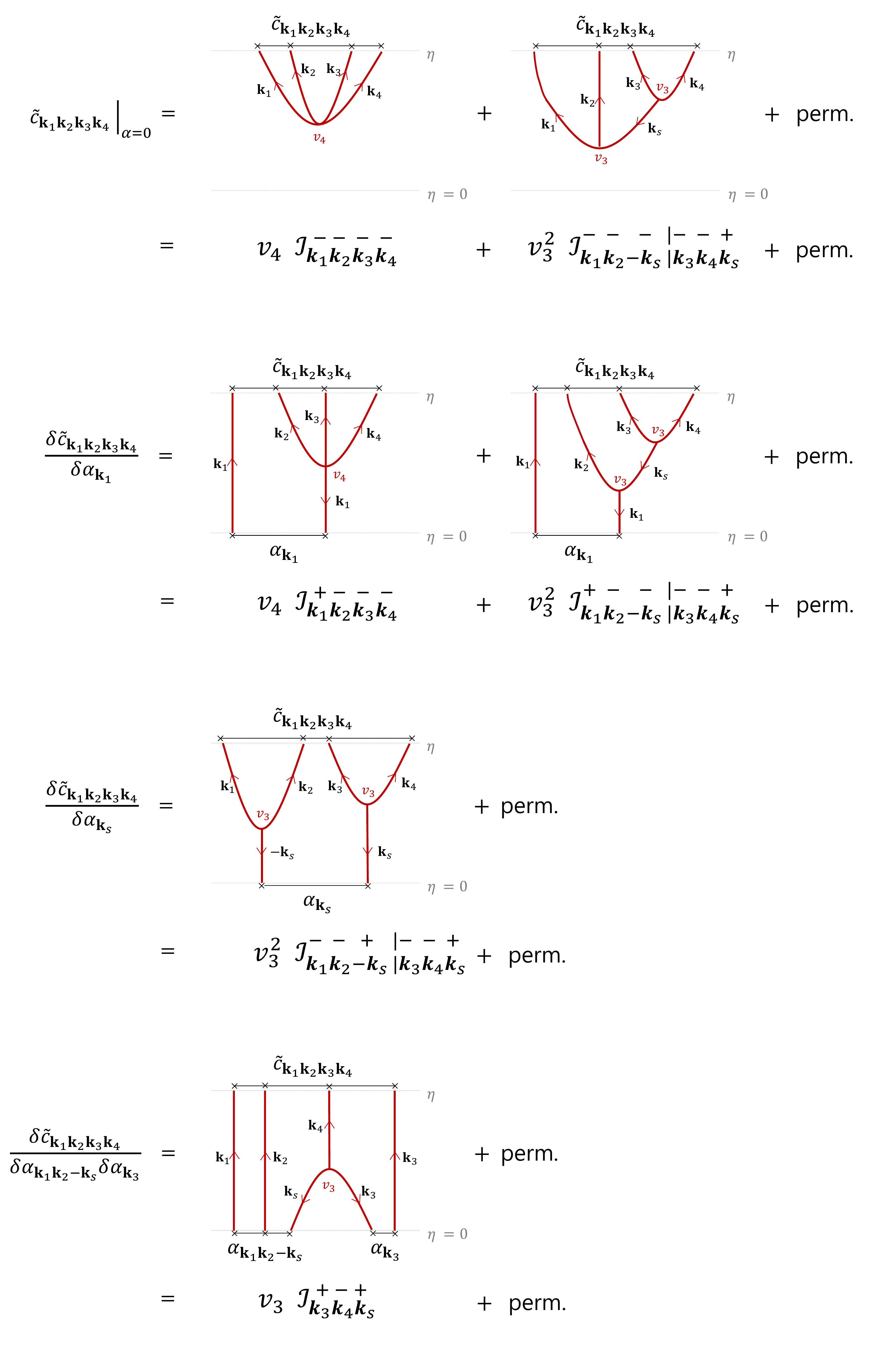}
\caption{Diagrammatic representation of the wavefunction coefficient $\tilde{c}_{\bfk_1 \bfk_2 \bfk_3 \bfk_4} (\eta)$ and its dependence on the boundary wavefunction coefficients $\alpha_{\bfk_1 ... \bfk_n}$ at $\eta = 0$.}
\label{fig:C4diagrams}
\end{figure}

\paragraph{Quartic Coefficient:}
Next, we move on to the quartic coefficient, $c_4$. 
Again anticipating the resummation of the $\alpha_{\bfk}$ series for the external lines, we can define\footnote{
Note that this is the analogue of $(-\eta)^{-4 \Delta} C_{\bfk_1 \bfk_2 \bfk_3 \bfk_4}$ for the mode function \eqref{eqn:fa}. 
}, 
\begin{align}
\tilde{c}_{\bfk_1 \bfk_2 \bfk_3 \bfk_4} (\eta)   &=     c_{\bfk_1 \bfk_2 \bfk_3 \bfk_4}   \prod_{j=1}^4 \left( F_{-\nu} (k_j \eta) - \alpha_{k_j} (-\eta )^{2\nu} F_{+\nu} ( k_j \eta)   \right)  \\
&- (-\eta)^d c_{\bfk_1 \bfk_2 - \bfk_s} c_{\bfk_3 \bfk_4 \bfk_s}  F_{+\nu} ( k_s \eta) \left( F_{-\nu} (k_s \eta) - \alpha_{k_s} (-\eta )^{2\nu} F_{+\nu} ( k_s \eta)   \right)  \nonumber 
\end{align}
which coincides with $c_{\bfk_1 \bfk_2 \bfk_3 \bfk_4} (\eta)$ at small $\eta$, and has an analogous expansion to \eqref{eqn:c3exp} which is at most linear in any particular $\alpha_{\bfk_J}$. A few of the relevant transfer functions are shown graphically in figure~\ref{fig:C4diagrams}, and are listed below. As with the cubic coefficient, the quartic transfer functions can be written in terms of manifestly analytic functions $\mathcal{I}_{\bfk_1 \bfk_2 \bfk_3 \bfk_4}^{\sigma_1 \sigma_2 \sigma_3 \sigma_4} (\eta)$ (which describes the contact contribution) and $\mathcal{I}_{\bfk_1 \bfk_2 -\bfk_s | \bfk_3 \bfk_4 \bfk_s}^{\sigma_1 \sigma_2 \sigma_s | \sigma_3 \sigma_4 +} (\eta) $ (which describes the exchange contribution). While $\mathcal{I}_{\bfk_1 \bfk_2 \bfk_3 \bfk_4}^{\sigma_1 \sigma_2 \sigma_3 \sigma_4} (\eta)$ is given by \eqref{eqn:I3} (where the sums now run from 1 to 4), the exchange contribution is captured by\footnote{
This follows from solving \eqref{eqn:HJtilde} at small $\eta$, using the relation,
\begin{align}
 \int \frac{d \eta}{\eta}  \, \frac{ 2 \nu  (- \eta)^{2 \nu} }{ \left( F_{-\nu} ( k \eta) - \alpha_k (-\eta)^{2 \nu} F_{+\nu} (k \eta) \right)^2  }  =  \frac{  (- \eta)^{2 \nu} F_{+\nu} (k \eta)  }{ F_{-\nu} (k \eta ) - \alpha_k (-\eta)^{2\nu} F_{+\nu} ( k \eta)  } \; . 
\end{align}
in place of $\Omega^{1-d}/f_{\bfk}^*f_{\bfk}^* = i \partial_\eta \left[ f_{\bfk} / f_{\bfk}^* \right]$. 
},
\begin{align}
\mathcal{I}_{\bfk_1 \bfk_2 -\bfk_s| \bfk_3 \bfk_4 \bfk_s}^{ \sigma_1 \sigma_2 \sigma_s | \sigma_4 \sigma_4 +} (\eta) 
&=  \frac{1}{ \left( \sum_{i=1}^4 \Delta_i^{\sigma_i} - d - (\sigma_s + 1) \nu_s  \right) \left(  \Delta_1^{\sigma_1} + \Delta_2^{\sigma_2} + \Delta_s^{\sigma_s} - d  \right) }    \nonumber \\
&+ \frac{  \frac{ ( - k_3 \eta )^2 }{ 4 ( \sigma_3 \nu_3 -1) } + \frac{ ( - k_4 \eta )^2 }{ 4 ( \sigma_4 \nu_4 -1 )} + \frac{ ( - k_s \eta )^2 }{ 4 ( \nu_s -1 )}   }{ \left( \sum_{i=1}^4 \Delta_i^{\sigma_i} - d - (\sigma_s + 1) \nu_s  + 2 \right) \left(  \Delta_1^{\sigma_1} + \Delta_2^{\sigma_2} + \Delta_s^{\sigma_s} - d  \right) }       \nonumber \\
&+ \frac{  \frac{ ( - k_1 \eta )^2 }{ 4 ( \sigma_1 \nu_1 -1) } + \frac{ ( - k_2 \eta )^2 }{ 4 ( \sigma_2 \nu_2 -1 )} + \frac{ ( - k_s \eta )^2 }{ 4 ( \sigma_s \nu_s -1 )}   }{ \left( \sum_{i=1}^4 \Delta_i^{\sigma_i} - d - (\sigma_s + 1) \nu_s + 2  \right) \left(  \Delta_1^{\sigma_1} + \Delta_2^{\sigma_2} + \Delta_s^{\sigma_s} - d + 2  \right) }       \nonumber \\
&+ \mathcal{O} ( k^4 \eta^4)   \label{eqn:I3exch}
\end{align} 
where $\Delta_s = \frac{d}{2} - \nu_s$ is the weight of the exchanged particle. 

In terms of the analytic $\mathcal{I}$ functions \eqref{eqn:In} and \eqref{eqn:I3exch}, the transfer functions for $\tilde{c}_4$ under the interaction Hamiltonian $\mathcal{H}_{\rm int} = \Omega^{d+1} \left( \frac{1}{3!} v_3 \phi^3 + \frac{1}{4!} v_4 \phi^4 \right)$ are given by:

\begin{itemize}

\item The term which is independent of the $\alpha_n$ boundary conditions is,
\begin{align}
(-\eta)^d \tilde{c}_{\bfk_1 \bfk_2 \bfk_3 \bfk_4} (\eta) |_{\alpha = 0} &= v_4 \mathcal{I}_{\bfk_1 \bfk_2 \bfk_3 \bfk_4}^{----} ( \eta )  \nonumber  \\
&+ v_3^2 \, \mathcal{I}_{\bfk_1 \bfk_2 -\bfk_s | \bfk_3 \bfk_4 \bfk_s}^{--- | --+} (\eta)  +  5 \text{ perm.}
\label{eqn:c4exp1}
\end{align}

\item The dependence on the quadratic initial condition, $\alpha_{\bfk}$, is given by,
\begin{align}
\frac{ (-\eta)^d }{ (-\eta)^{2 \nu_1}} \frac{ \delta \tilde{c}_{\bfk_1 \bfk_2 \bfk_3 \bfk_4} (\eta)  }{ \delta \alpha_{\bfk_1} } \Big|_{\alpha = 0} &=   v_4 \mathcal{I}_{\bfk_1 \bfk_2 \bfk_3 \bfk_4}^{+---} ( \eta ) 
+ v_3^2 \, \mathcal{I}_{\bfk_1 \bfk_2 -\bfk_s | \bfk_3 \bfk_4 \bfk_s}^{+-- | --+} (\eta)  +  5 \text{ perm.}    \nonumber \\
\frac{ (-\eta)^d }{ (-\eta)^{2 \nu_s}} \frac{ \delta \tilde{c}_{\bfk_1 \bfk_2 \bfk_3 \bfk_4} (\eta)  }{ \delta \alpha_{\bfk_s} } \Big|_{\alpha = 0} &=  v_3^2 \, \mathcal{I}_{\bfk_1 \bfk_2 -\bfk_s | \bfk_3 \bfk_4 \bfk_s}^{--+ | --+} (\eta)  +  5 \text{ perm.}     \nonumber  \\
\frac{ (-\eta)^d }{ (-\eta)^{2 \nu_1+2\nu_2}} \frac{ \delta^2 \tilde{c}_{\bfk_1 \bfk_2 \bfk_3 \bfk_4} (\eta)  }{ \delta \alpha_{\bfk_1} \delta \alpha_{\bfk_2} } \Big|_{\alpha = 0} &= v_4 \mathcal{I}_{\bfk_1 \bfk_2 \bfk_3 \bfk_4}^{++--} ( \eta )  + v_3^2 \, \mathcal{I}_{\bfk_1 \bfk_2 -\bfk_s | \bfk_3 \bfk_4 \bfk_s}^{++- | --+} (\eta)  +  5 \text{ perm.}   \nonumber  \\
\frac{ (-\eta)^d }{ (-\eta)^{2 \nu_1+2\nu_s}} \frac{ \delta^2 \tilde{c}_{\bfk_1 \bfk_2 \bfk_3 \bfk_4} (\eta)  }{ \delta \alpha_{\bfk_s} \delta \alpha_{\bfk_1} } \Big|_{\alpha = 0} &= v_3^2 \, \mathcal{I}_{\bfk_1 \bfk_2 -\bfk_s | \bfk_3 \bfk_4 \bfk_s}^{+-+ | --+} (\eta)  +  5 \text{ perm.}      
\label{eqn:c4exp2}
\end{align}
and so forth---just as for the cubic coefficient, each of the $\alpha_{\bfk_1} , ... , \alpha_{\bfk_4}$ may only appear once in any term, and now there may further be an additional $\alpha_{\bfk_s}$ or $\alpha_{\bfk_t}$ or $\alpha_{\bfk_u}$. 

\item The cubic initial condition, $\alpha_{\bfk_1 \bfk_2 \bfk_3}$, contributes in only three ways,
\begin{align}
(-\eta)^{\Delta_1 + \Delta_2 + \Delta_s} \frac{ \delta \tilde{c}_{\bfk_1 \bfk_2 \bfk_3 \bfk_4} (\eta)  }{ \delta \alpha_{\bfk_1 \bfk_2 -\bfk_s} } \Big|_{\alpha = 0} &= v_3 \mathcal{I}^{--+}_{\bfk_3 \bfk_4 \bfk_s} (\eta) \nonumber  \\
\frac{ (-\eta)^{\Delta_1 + \Delta_2 + \Delta_s}}{ (-\eta)^{2\nu_3} } \frac{ \delta \tilde{c}_{\bfk_1 \bfk_2 \bfk_3 \bfk_4} (\eta)  }{ \delta \alpha_{\bfk_1 \bfk_2 -\bfk_s} \delta \alpha_{\bfk_3} } \Big|_{\alpha = 0} &= v_3 \mathcal{I}^{+-+}_{\bfk_3 \bfk_4 \bfk_s} (\eta)  \nonumber  \\
\frac{ (-\eta)^{\Delta_1 + \Delta_2 + \Delta_s}}{ (-\eta)^{2\nu_3+2\nu_4} } \frac{ \delta \tilde{c}_{\bfk_1 \bfk_2 \bfk_3 \bfk_4} (\eta)  }{ \delta \alpha_{\bfk_1 \bfk_2 -\bfk_s} \delta \alpha_{\bfk_3} \delta \alpha_{\bfk_4} } \Big|_{\alpha = 0} &=  v_3  \mathcal{I}^{+++}_{\bfk_3 \bfk_4 \bfk_s}  (\eta)  \; , 
\label{eqn:c4exp3}
\end{align}
plus terms related by permuting the external legs. 

\item Finally, we have a new initial condition, $\alpha_{\bfk_1 \bfk_2 \bfk_3 \bfk_4}$, which appears as a single term in $\tilde{c}_{\bfk_1 \bfk_2 \bfk_3 \bfk_4}$, 
\begin{align}
(- \eta)^{4 \Delta} \frac{ \delta \tilde{c}_{\bfk_1 \bfk_2 \bfk_3 \bfk_4} (\eta)  }{ \delta \alpha_{\bfk_1 \bfk_2 \bfk_3 \bfk_4} } \Big|_{\alpha = 0} &=  1   \; , 
\label{eqn:c4exp4}
\end{align}
just like the first term in \eqref{eqn:c3exp}.

\end{itemize} 

Since all of the transfer coefficients have the form $v_j \times \mathcal{I}$, where the $\mathcal{I}$ functions are analytic outside the horizon, we can again conclude that locality of the interaction Hamiltonian (in this case analyticity of both $v_3$ and $v_4$) is translated into analyticity of the $\tilde{c}_4$ transfer coefficients. The same procedure can be carried out for any $\tilde{c}_n$ wavefunction coefficient, and although we have focused on interactions of the form $\phi^n$ the same conclusion can be reached for more general interactions which also contain $\Pi$ (this requires defining a new $\mathcal{I}$ object in which a $\partial_\mu F_{-\nu}$ is used in place of $F_{-\nu}$, but otherwise proceeds as above). Locality of the bulk interactions guarantees that the transfer coefficients which relate $\tilde{c}_n (\eta)$ to $\alpha_n$ on superhorizon scales are analytic functions of the momenta. 
 
\paragraph{Horizon Crossing:}
As $|k \eta|$ approaches unity, the series solutions for the transfer coefficients must be resummed. Formally, this can be done using integrals over products of the Bessel mode function, resumming \eqref{eqn:I3} and \eqref{eqn:I3exch} into,
\begin{align}
\mathcal{I}_{\bfk_1 ... \bfk_n}^{\sigma_1 ... \sigma_n} (\eta )  &=   \int^{\eta}_{0} \frac{d \eta'}{\eta'} \left( \frac{\eta'}{\eta} \right)^{-d + \sum_i \Delta_i^{\sigma_i} }  F_{\sigma_1 \nu_1} ( k_1 \eta' ) ... F_{\sigma_n \nu_n} (k_n \eta' )  \;  \, ,  \label{eqn:In} \\
\mathcal{I}_{\bfk_1 \bfk_2 -\bfk_s| \bfk_3 \bfk_4 \bfk_s}^{ \sigma_1 \sigma_2 \sigma_s | \sigma_4 \sigma_4 +} (\eta)  &= 
  \int^{\eta}_0 \frac{d \eta''}{\eta''} \left( \frac{\eta''}{\eta}  \right)^{ \Delta_{3}^{\sigma_3} + \Delta_{4}^{\sigma_4} - \Delta_{s}  }  F_{\sigma_3 \nu_3} ( k_3 \eta'' )  F_{\sigma_4 \nu_4} ( k_4 \eta'' ) F_{+\nu_s} ( k_s \eta'' )   \nonumber \\
 &\qquad \times
  \int^{\eta''}_0 \frac{d \eta'}{\eta'} \left( \frac{\eta'}{\eta}  \right)^{-d + \Delta_{1}^{\sigma_1} + \Delta_{2}^{\sigma_2} + \Delta_{s}^{\sigma_s}  } F_{\sigma_1 \nu_1} ( k_1 \eta' ) F_{\sigma_2 \nu_2} ( k_2 \eta' )F_{\sigma_s \nu_s} ( k_s \eta' )  \; . 
  \label{eqn:Inexch}
\end{align}
where again we have used $\Delta^{\pm}_j = \frac{d}{2} \pm \nu_j$ (so $\Delta_j^- = \Delta_j$ and $\Delta_j^+ = d - \Delta_j$). 
While these expressions are not particularly enlightening for general $\nu$, in section~\ref{sec:examples} we will see particular cases in which these integrals can be performed and the transfer functions extended to $| k \eta | >1$ and into the horizon. 

\paragraph{Method of Regions:}
One useful feature of \eqref{eqn:In} and \eqref{eqn:Inexch}, even if they cannot always be evaluated explicitly, is that they allow us to infer properties of our transfer functions in the deep subhorizon limit, $| k \eta | \gg 1$. We do this by dividing the integration over $\eta$ into a number of regions, determined by the momenta. For example, consider the quartic coefficients $\tilde{c}_{\bfk_1 \bfk_2 \bfk_3 \bfk_4}$, where the arguments have been ordered so that $k_1 \geq k_2 \geq k_3 \geq k_4$. The $\eta$ integrals in the transfer function can be written as,
\begin{align}
 \int_0^{\eta} d \eta' =  \left( \int_0^{-1/k_1}  + \int_{-1/k_1}^{-1/k_2} + \int_{-1/k_2}^{-1/k_3} + \int_{-1/k_3}^{-1/k_4}    + \int_{-1/k_4}^{\eta}  \right) d \eta' \; ,
 \label{eqn:etaRegions}
\end{align}
 and in each region a different approximation for the mode functions may be used. If $|\eta'|$ is smaller than $1/k_j$, then the analytic expansion \eqref{eqn:Fnu} is used, while if $|\eta'|$ is larger than $1/k_j$ then the asymptotic expansion\footnote{
Strictly speaking, one requires $|k \eta | \gg | \nu^2 - 1/4 |$ in $d=3$ dimensions to use this expansion, but for light fields we can safely treat the right-hand side as an order one number.   
 },
\begin{align}
(-\eta)^{\Delta} F_{-\nu} ( k \eta ) \propto \eta \left( e^{i k \eta} -i e^{ i \nu \pi}  e^{- i k \eta} \right) \; , 
\label{eqn:FLargeEta}
 \end{align}
 is used. 
 
 Focussing on the final region in \eqref{eqn:etaRegions}, in which $|\eta|$ is much larger than any momenta and we can use \eqref{eqn:FLargeEta}, we see that each $\mathcal{I}_{\bfk_1 \bfk_2 \bfk_3 \bfk_4}^{\sigma_1 \sigma_2 \sigma_3 \sigma_4} (\eta)$ will contain,
\begin{align}
 \mathcal{I}_{\bfk_1 \bfk_2 \bfk_3 \bfk_4}^{\sigma_1 \sigma_2 \sigma_3 \sigma_4} (\eta) \supset \int_{-1/k_4}^{\eta} d \eta  \, (-\eta)^{-d+3} \, \sum_{\sigma'_j = \pm} e^{i ( \sigma_1' k_1 + \sigma_2' k_2 + \sigma_3' k_3 + \sigma_4' k_4 ) \eta}
 \end{align}
and therefore in $d=3$ will generically develop poles in $1/\sum_j k_j$, as well as all other folded configurations ($k_1 + k_2 + k_3 - k_4$, $k_1 + k_2 - k_3 - k_4$, etc.), when we transfer from the boundary to deep inside the horizon. 
While the transfer functions are analytic outside the horizon for local interactions, non-analyticities inevitably develop once $| k \eta | > 1$. 
This argument is somewhat heuristic, since we have neglected the other regions in \eqref{eqn:etaRegions}, but nonetheless we will observe precisely this behaviour in explicit examples in section~\ref{sec:examples}. 

The above discussion has assumed generic values of the conformal weights $\Delta_j$. However, 
note that for particular values of $\sum_j \Delta_j$ and $d$ there are simple-pole divergences in both $\mathcal{I}_{\bfk_1 .. \bfk_n}^{\nu_1 ... \nu_n}$ and $\mathcal{I}_{\bfk_1 ... \bfk_{n} | \bfp_1 ... \bfp_{n'} }^{\nu_1 ... \nu_n | \nu'_1 ... \nu'_{n'}}$ (the integrals which determine the transfer coefficients). 
We will now show how these can be systematically renormalised.

\subsection{Renormalisation of the Boundary Wavefunction}
\label{sec:33}

In addition to the manifest analyticity, another advantage of expressing the wavefunction coefficients in terms of the transfer functions is that the structure of boundary divergences becomes clear. The integrals \eqref{eqn:In} and \eqref{eqn:Inexch} have simple poles at particular values of the scalar field masses and spatial dimension $d$. We will now show how these can be removed in a systematic way by renormalising the wavefunction's boundary condition.  

\paragraph{Types of Divergence:}
Let us begin by listing the various kinds of divergence we may encounter at the boundary. Beginning with the contact integral \eqref{eqn:I3}, we see that this leads to two qualitatively different kinds of divergence in the transfer functions.
Firstly, the $\mathcal{I}_{\bfk_1 ... \bfk_n}^{-... -}$ contain divergences when $\sum_{j=1}^n \Delta_j = d - 2 \ell$ (for any positive integer $\ell$), as can be seen from \eqref{eqn:In}. 
Adopting the language of \cite{Bzowski:2015pba}, we will refer to this as an \textit{ultralocal} divergence. 
Secondly, there is an analogous divergence in $\mathcal{I}_{\bfk_1 ... \bfk_n}^{+ -... -}$ when $d - \Delta_1 + \sum_{j=2}^{n} \Delta_j = d - 2 \ell$ approaches zero. 
Again following \cite{Bzowski:2015pba}, we will refer to this as a \textit{semilocal} divergence. 
$\mathcal{I}^{++-...-}_{\bfk_1 \bfk_2 ... \bfk_n}$ and the other functions in \eqref{eqn:In} are finite for light fields\footnote{
By `light', we mean that they belong to the ``complementary series'' of de Sitter representations. 
} in any $d$ since $0 \leq \Delta < d/2$, so $2d - \Delta_1 - \Delta_2 + \sum_{j=3}^n \Delta_j$ is always greater than $d$ (and can never approach a pole at $d-2\ell$). 

Moving on to the simplest exchange integral \eqref{eqn:I3exch}, we find the same ultralocal and semilocal divergences, plus a new kind of divergence which appears only in $\mathcal{I}_{\bfk_1 \bfk_2 -\bfk_s | \bfk_3 \bfk_4 \bfk_s}^{\sigma_1 \sigma_2 + | \sigma_3 \sigma_4 +}$, when $\sum_{j=1}^4 \Delta_j^{\sigma_j} = 2 \Delta_s - 2 \ell$, where $\Delta_s$ is the conformal weight of the exchanged field. Since this kind of divergence can appear only in 4-point correlators and higher, it does not appear in the 3-point analysis of \cite{Bzowski:2015pba}. We dub these \textit{exchange} divergences. In higher $n$-point correlators, there are triple integrals, quadruple integrals, etc., and it seems that at each order new exchange divergences are introduced. 

We will not attempt a systematic classification of all such divergences here. Rather, we will focus on the ultralocal and semilocal divergences stemming from the contact integrals \eqref{eqn:In}.  
We will first show that both types of divergence can be dealt with by renormalising the boundary condition at the conformal boundary, and then further show that this is equivalent to performing a Boundary Operator Expansion to replace the (singular) bulk operators with (finite) boundary operators.

\paragraph{Ultralocal Divergences:}
An ultralocal divergence appears as a $1/\delta_{n \ell}$ pole, where $\delta_{n \ell} = \sum_{j=1}^n \Delta_j - d + 2 \ell$. This kind of divergence can be removed by renormalising a single wavefunction coefficient,
\begin{align}
 \alpha_n^{\rm ren} = \alpha_n - \frac{  \mu^{- \delta_{n \ell} }  }{ \delta_{n \ell} } P_{2\ell} ( \bfk_J ) \; ,
 \label{eqn:localrenorm}
\end{align}
where $P_{2\ell}$ is an analytic polynomial of degree $2\ell$ in the momenta, and $\mu$ is an arbitrary scale introduced on dimensional grounds. For instance, the example of three conformally coupled scalars from \eqref{eqn:egaren} corresponds to $\sum_J \Delta_J = 3$ in $d=3$, which gives $\ell = 0$ and so $P_{0} = 1$ is a simple constant. More generally, from our expression \eqref{eqn:In} for the integral $\mathcal{I}_{\bfk_1 ... \bfk_n}^{\sigma_1 ... \sigma_n}$ (which determines the transfer function for any $v_n \phi^n$ contact interaction), we see that an ultralocal divergence in $\tilde{c}_n$ can indeed always be removed by \eqref{eqn:localrenorm},
\begin{align}
\tilde{c}_n &= \frac{\alpha_n}{ (-\eta)^{\sum_j \Delta_j} } + \frac{v_n}{(-\eta)^d} \mathcal{I}_{\bfk_1 ... \bfk_n}^{-...-} + \mathcal{O} \left(  \alpha_{j<n} \right)  \nonumber  \\
&= \lim_{\delta_{\ell} \to 0} \left[  \frac{\alpha_n}{ (-\eta)^{ \sum_j \Delta_j } }  +  \frac{v_n}{(-\eta)^{\sum_j \Delta_j + 2 \ell - \delta_{n\ell}}} \, \frac{ (- k \eta)^{2\ell} }{ \delta_{n \ell} } + \text{finite} \right]   \nonumber \\
&= \frac{\alpha_n^{\rm ren} }{ (-\eta)^{ \sum_j \Delta_j } }  + v_n (- k \eta)^{2 \ell} \log ( - \eta \mu ) + \text{finite}  \; .   \label{eqn:localrenorm2}
\end{align}
where we have used $(-k\eta)^{2\ell}$ to represent a polynomial of degree $2\ell$ in the $k_J$ (i.e. the coefficient at order $(-\eta)^{2\ell}$ in \eqref{eqn:I3}). In fact, any interaction (not only $\phi^n$) can be renormalised in the same way, by choosing the function $P_{2\ell}$ in \eqref{eqn:localrenorm} appropriately. 

\paragraph{Semilocal Divergences:}
A semilocal divergence appears as a $1/\delta'_{n \ell}$ pole, where $\delta'_{n \ell} = \sum_{j=2}^{n} \Delta_j - \Delta_1  + 2 \ell$. This kind of divergence can only be removed by renormalising an infinite number of wavefunction coefficients, starting at order $n$, 
\begin{align}
 \alpha_{\bfk_1 ... \bfk_n}^{\rm ren} &= \alpha_{\bfk_1 ... \bfk_n} +  \sum_{\rm perm.} \frac{ \alpha_{\bfp -\bfp} }{ \delta_{n \ell}' }  P_{2\ell} ( \bfp ; \bfk_2 , ... , \bfk_n )   \;\;\;, \;\;\;   \label{eqn:asemilocal} \\
  \alpha_{\bfk_1 ... \bfk_{n+1} }^{\rm ren} &= \alpha_{\bfk_1 ... \bfk_{n+1} } +  \sum_{\rm perm.} \frac{ \alpha_{\bfk_1 \bfk_2 -\bfp} }{ \delta_{n \ell}' }  P_{2\ell} ( \bfp  ; \bfk_3, ... , \bfk_{n+1} )  \;\;\; , \;\;\;  \nonumber   \\
  \vdots   \nonumber \\
  \alpha_{\bfk_1 ... \bfk_{2n-2} }^{\rm ren} &= \alpha_{\bfk_1 ... \bfk_{2n-2}}  + \sum_{\rm perm.}  \frac{\alpha_{\bfk_1  ... \bfk_{n-1}  \bfp } }{\delta_{n \ell}' }  P_{2\ell} \left(  -\bfp ,  \bfk_n , ... , \bfk_{2n-2}   \right)    \nonumber   \\
  &\qquad+ \sum_{\rm perm.} \frac{ \alpha_{\bfp -\bfp }  }{ \delta_{n \ell}'^2} P_{2\ell} ( \bfp ; \bfk_1, ... , \bfk_{n-1} ) P_{2\ell} ( -\bfp ; \bfk_n , ... , \bfk_{2n-2} ) \nonumber   \\
  &\vdots  \nonumber 
\end{align}
where the $P_{2 \ell} ( \bfp ; \bfk_1 ... , \bfk_{n-1})$ are again polynomials of degree $2\ell$ in the momenta, and we adopted the convention that $\bfp = \sum_{j=1}^{n-1} \bfk_j$ is always set to be the sum of the remaining $n-1$ arguments. Taking $n=3$ and focussing on the $\delta_{30}'$ divergences found in our $\tilde{c}_{4}$ transfer coefficients above, we can see explicitly that \eqref{eqn:asemilocal} (with $P_{0}$ a constant) has the effect of removing the $1/\delta_{30}'$ divergence from $\mathcal{I}_{\bfk_1 \bfk_2 \bfk_3 \bfk_4}^{+---}$ and from every $\mathcal{I}_{\bfk_1 \bfk_2 - \bfk_s | \bfk_3 \bfk_4 \bfk_s}^{\sigma_1 \sigma_2 \sigma_3 | --+}$, and also the double pole from $ \mathcal{I}^{--+|--+}_{\bfk_1 \bfk_2 - \bfk_s | \bfk_3 \bfk_4 \bfk_s} $.
In fact, this tower of redefinitions \eqref{eqn:asemilocal} coincides with a particular redefinition of the field $\phi_1$ (with conformal weight $\Delta_1$),
\begin{align}
 \phi_1 \to \phi_1  + \frac{ P_{2\ell} (\bfk_1 ; \bfk_2,  ... ,\bfk_n)  }{\delta_{n \ell}' } \phi_2 ... \phi_{n} \; . 
 \label{eqn:phiredef}
\end{align}

In section~\ref{sec:examples} we will study an example of this kind of divergence, namely the three-point coefficient of a massless field (which requires performing \eqref{eqn:phiredef} with $n=3$ and $\ell=0$). 
Another example of this kind of divergence is the two-point coefficient, $\alpha_2$, which always exhibits an $\ell = 0$ divergence since $ (d-\Delta) + \Delta = d$. In this case, $n=2$ in \eqref{eqn:phiredef} and all that is required is a rescaling $\phi \to Z \phi$. We will show below that this is most easily done using a hard cutoff, namely $\phi \to (-\eta_{\delta} )^{\Delta} \phi$ (which is the familiar renormalisation of the boundary value routinely used in holography).  

When the initial condition for $\Psi$ is provided at the boundary, $\eta = 0$, then renormalisation can be carried out straightforwardly by shifting the $\alpha_n$ coefficients as in \eqref{eqn:localrenorm} or \eqref{eqn:asemilocal}. On the other hand, when the initial condition is provided in the bulk (e.g. Bunch-Davies vacuum in the past), then the renormalisation must be carried out at the level of the operators $\hat{\phi}$ and $\hat{\Pi}$. We will now describe how this is done.

\paragraph{Operator Mixing:}
On the boundary, we have a set of local operators, namely $\hat{\varphi}$, its momentum $\hat{\pi}$ and their descendents, $k^2 \hat{\varphi} , k^2 \hat{\pi}, ...$. When the bulk operator $\hat{\phi} (\eta)$ (and its canonical momentum $\hat{\Pi} (\eta)$) approach the boundary, we must specify how it is mapped onto the boundary operators\footnote{
Note that we have switched to the Heisenberg picture for the field operators for notational convenience. 
}. This mapping is known in the CFT literature as the ``Boundary Operator Expansion" (BOE), and in general takes the form, 
\begin{align}
\lim_{\eta \to 0} \hat{\mathcal{O}}_i^{\rm bulk} ( \eta ) = \sum_j  Z_{ij}  \; \hat{ \mathcal{O}}^{\rm boundary}_j \; .
\end{align}
Since this limit is singular, it requires a small regulator, $\delta$. The BOE coefficients $Z_{ij}$ then depend on this regulator in a way which is fixed by the isometries. For instance, in the free theory with a hard cutoff at $\eta_{\delta}$, the BOE is,
\begin{align}
 \lim_{\eta \to 0} \left[  \frac{ \Phi (\eta) }{ (-\eta)^{\Delta_-}}  \right] &= Z_{\phi \varphi}  \hat{\varphi} + Z_{\phi \pi}  ( - \eta_{\delta} )^{2 \nu} \hat{\pi}   \\
 \lim_{\eta \to 0} \left[  (-\eta)^{\Delta_-} \Pi (\eta)   \right] &= Z_{\Pi \pi} \hat{\pi} + Z_{\Pi \varphi} (-\eta_{\delta} )^{-2 \nu} \hat{\varphi}    
\end{align}
where the scaling weights are $[\hat{\varphi}] = \Delta$ and $[ \hat{\pi} ] = d-\Delta$, and we have used scale-invariance to write the $Z_{ij}$ as constants (multiplying the appropriate power of $\delta$).  
The commutation relation,
\begin{align}
 [ \hat{\varphi} , \hat{\pi} ] = \frac{ \lim_{\eta \to 0} [ \hat{ \phi } (\eta) ,  \hat{\Pi} (\eta) ]] }{ Z_{\phi \varphi} Z_{\Pi \pi} - Z_{\phi \pi} Z_{\Pi \varphi}}
\end{align}
is canonically normalised providing $Z_{\phi \varphi} Z_{\Pi \pi} - Z_{\phi \pi} Z_{\Pi \varphi} = 1 $ . 
The quadratic correlators are given by,
\begin{align}
\langle \hat{\varphi} \hat{\varphi} \rangle &=  \frac{1}{2 (-\eta_\delta)^{2 \Delta} \text{Im} \, c_2 (\eta_\delta)} \left( Z_{\Pi \pi} - Z_{\phi \pi} \, (-\eta_\delta )^d \text{Re} \, c_2 ( \eta_\delta ) \right)    \\
\langle \hat{\varphi} \hat{\pi} \rangle &= - \frac{i}{2} + (-\eta_\delta)^{-2\nu}  \frac{  (Z_{\phi \varphi} Z_{\Pi \pi} + Z_{\phi \varphi} Z_{\Pi \varphi} )  (-\eta_\delta)^d  \text{Re}\,c_2 ( \eta_\delta)  }{2 (-\eta_\delta)^{2\Delta} \text{Im} \, c_2 (\eta_\delta) }    \nonumber \\
&\qquad\quad- (- \eta_\delta )^{-2\nu} \frac{  
 Z_{\Pi \pi} Z_{\Pi \varphi}  + Z_{\phi \varphi} Z_{\phi \pi}  \left( (-\eta_\delta)^d \text{Re} \, c_2 ( \eta_\delta) \right)^2 }{2 (-\eta_\delta)^{2\Delta} \text{Im} \, c_2 (\eta_\delta) }    \nonumber \\ 
\langle \hat{\pi} \hat{\pi} \rangle &=     (-\eta)^{-2 \nu}  \frac{  \left( Z_{\Pi \varphi}  - Z_{\phi \varphi} (\eta_\delta)^{d} \text{Re} \, c_2 ( \eta_\delta)  \right)^2  }{2  (-\eta_\delta )^{2 \Delta} \text{Im} \, c_2 ( \eta_\delta) }  + Z_{\phi \varphi}^2 \frac{ (-\eta_\delta )^{2 \Delta} \text{Im} \, c_2 ( \eta_\delta )  }{2} \, . 
\nonumber 
\end{align}
Note that since $\text{Re} \, c_2 \sim \Delta / \eta^d$ and $\text{Im} \, c_2 \sim 1/\eta^{2\Delta}$, these are finite providing we fix  $Z_{\Pi \varphi} = \Delta Z_{\phi \varphi}$. The subleading $Z_{\phi \pi}$ parameter may take any value\footnote{
It is related to the Reparametrization Invariance (RPI) that one inevitably introduces when splitting up degrees of freedom, see e.g. \cite{Cohen:2020php}.
}, and we choose,
\begin{align}
 Z_{\phi \varphi} = 1 \;\;, \;\; Z_{\phi \pi} = 0 \;\; , \;\; Z_{\Pi \pi} = 1 \;\; ,\;\; Z_{\Pi \varphi} = \Delta \; ,
\end{align}
which corresponds to the definition $\hat{\phi} = (-\eta_{\delta} )^{\Delta} \hat{\varphi}$ in the free theory. This is the reason that the rescaling \eqref{eqn:andef} is necessary to translate $c_n (\eta)$ (the wavefunction coefficients of $\Psi_{\eta} [ \phi] $) to $\alpha_n$ (the wavefunction coefficients of $\Psi_{0} [ \varphi ]$). 

In an interacting theory, there can be further non-zero $Z_{ij}$ coefficients. For instance, whenever $\sum_{j=1}^n \Delta_j = d$, or equivalently $d-\Delta_1 = \sum^{n}_{j=2} \Delta_j$, we can have mixing between $\hat{\pi}$ and $\hat{\varphi}^{n-1}$, 
\begin{align}
 \lim_{\eta \to 0} \left(  \eta^{\Delta_-} \Pi_{\bfk_1} (\eta)   \right) &=  \Delta (-\eta_\delta)^{-2 \nu} \hat{\varphi}_{\bfk_1}   + \hat{\pi}_{\bfk_1}  + \frac{ Z_{\Pi \varphi^{n-1} } }{n!} \int_{\bfk_2 ... \bfk_{n}} \hat{\varphi}_{\bfk_2} ... \hat{\varphi}_{\bfk_n} \delta^d ( \sum_{j=1}^n \bfk_j )   \; . 
\end{align}
because $Z_{\Pi \varphi^{n-1}}$ is no longer constrained to be zero by scale invariance. This occurs precisely when $\delta_{n0} = 0$, corresponding to the ultralocal type of divergence.
The effect of this mixing is to introduce an additive counterterm in the boundary wavefunction coefficient \eqref{eqn:cnEqualTime},
\begin{align}
\lim_{\eta \to 0} \left[   (- \eta )^{\sum_j \Delta_j} c_n (\eta)   \right] =  \alpha_n + Z_{\Pi \varphi^{n-1}}  
\end{align}
which can be used to renormalise the ultralocal divergence, as shown in \eqref{eqn:localrenorm2}. The higher $\ell$ divergences, when $\delta_{n\ell} = \sum_{j=1}^n \Delta_j -d + 2\ell \to 0$, correspond to the mixing of $\partial^{2\ell} \hat{\varphi}^{n-1}$ into the BOE of $\hat{\Pi}$, and can be similarly renormalised using the mixing coefficients $Z_{\Pi \partial^{2\ell} \varphi^{n-1}}$.

Similarly, there is a second kind of mixing that takes place when $\Delta_1 = \sum_{j=2}^n \Delta_j$, which allows $\hat{\varphi}$ to mix with $\hat{\varphi}^{n-1}$, 
\begin{align}
 \lim_{\eta \to 0} \left(  \frac{ \Phi_{\bfk_1} (\eta) }{\eta^{\Delta_-}}  \right) &=  \hat{\varphi}_{\bfk_1} +  \frac{Z_{ \phi \varphi^{n-1} }  }{n!} \int_{\bfk_2 ... \bfk_{n}} \hat{\varphi}_{\bfk_2} ... \hat{\varphi}_{\bfk_n} \delta^3 ( \sum_{j=1}^n \bfk_j)   \; ,
\end{align}
since now a non-zero $Z_{\phi \varphi^{n-1}}$ is permitted by scale invariance. 
This has the effect of mixing wavefunction coefficients of different order, e.g. when $n=3$ the boundary coefficient \eqref{eqn:andef} becomes,
\begin{align}
\lim_{\eta \to 0} \left[   (- \eta )^{\sum_j \Delta_j} c_3 (\eta)   \right] =  \alpha_3 +  Z_{\phi \varphi^{n-1}} \,\sum_{\rm perm.} \alpha_{2}  \, ,
\end{align}
while the $c_4$, $c_5$, etc. coefficients are also shifted into each other, as shown in \eqref{eqn:asemilocal}. 
This shift can be used to renormalise the semilocal divergences when $\delta_{n0}' \to 0$.
The higher $\ell$ divergences, when $\delta_{n\ell}' = \sum_{j=2}^n \Delta_j - \Delta_1 + 2\ell \to 0$, correspond to the mixing of $\partial^{2\ell} \hat{\varphi}^{n-1}$ into the BOE of $\hat{\phi}$, and can be similarly renormalised using the mixing coefficients $Z_{\phi \partial^{2\ell} \varphi^{n-1}}$.

Finally, let us close this section with a conjecture. The new kind of exchange divergence which we encountered in \eqref{eqn:Inexch} is a result of $2 \Delta_s = \sum_{j=1}^4 \Delta_j + 2\ell$, which would be the condition for a composite operator like $: \varphi_{s} \varphi_s :$ to mix with $\partial^{2\ell}: \varphi_1 \varphi_2 \varphi_3 \varphi_4 :$. Although contact divergences can always be removed by the Boundary Operator Expansion of $\hat{\phi}$ and $\hat{\Pi}$, we speculate that to remove exchange divergences may also require an analogous expansion of composite bulk operators. Understanding how to renormalise exchange divergences is important, since they appear in even simple examples like the exchange of a massless field (for which $\Delta = 0$). We will not pursue this further here, and instead turn now to three concrete examples to which the above discussion in sections~\ref{sec:timeEvolution}, \ref{sec:isometries} and \ref{sec:superhorizon} can be applied.

\section{Some Examples}
\label{sec:examples}

In the preceding sections, we have discussed the impact of unitarity, de Sitter invariance and locality of the bulk interactions on the wavefunction coefficients (and equal-time correlation functions). These basic properties allowed us to draw very general conclusions about the properties that we should expect from cosmological correlators, regardless of the specific details of the interactions.  

We will now focus on a small number of specific models, primarily as a sanity check---to confirm that our general conclusions are realised in practice---but also to make many of our main ideas more concrete and intuitive. Throughout this section we will work in $d=3$ spatial dimensions (except for the purposes of dimensional regularisation).

 
\subsection{Conformally Coupled Scalar}

As our first example, consider a conformally coupled scalar field on a fixed de Sitter background,
\begin{align}
 \mathcal{H} [ \sigma , \Sigma ] =  \int_{\bfk}   \left[   \frac{ 1 }{2 \Omega^{d-1}} \Sigma_{\bfk} \Sigma_{-\bfk} +  \frac{ \Omega^{d-1} }{2} \left( k^2 + \frac{2}{\eta^2}   \right) \sigma_{\bfk} \sigma_{-\bfk}   \right] +  \mathcal{H}_{\rm int}    \, , 
 \label{eqn:Sconf}
\end{align}
where $\Sigma$ is the momentum conjugate to $\sigma$, and $\Omega = 1/(-H \eta)$.

\subsubsection*{Free Theory}

We begin by solving the free theory, with $\mathcal{H}_{\rm int} = 0$.

\paragraph{Gaussian Coefficient:}
The two-point wavefunction coefficient is given by solving the nonlinear first-order differential equation \eqref{eqn:HJfree}, with $\nu = 1/2$ (i.e. $\Delta = 1$), which gives,
\begin{align} 
 c_{\bfk , - \bfk} (\eta) =  \frac{ 1 }{ H^2 \eta^3 }   + \frac{ i k }{ H^2 \eta^2 }  \left(  1  -  \frac{2 \alpha_\bfk^{\rm in} }{ \alpha_\bfk^{\rm in} - e^{i 2 k \eta}  }   \right)    \, ,
 \label{eqn:c2conf}
\end{align}
up to an undetermined constant of integration $\alpha_\bfk^{\rm in}$. 
Solving \eqref{eqn:HJfreeMode} for the mode function introduces a second undetermined constant, $N$, 
\begin{align}
 f_{\bfk} ( \eta ) =  \frac{ - H \eta }{ \sqrt{2 i k} }  N (k) \left(  e^{-i k \eta} + \left( \alpha_{\bfk}^{\rm in} \right)^* \, e^{ + i k \eta } \right) \, . 
\end{align}
The choice of $N$ does not affect any wavefunction coefficient, and it is conventional to set $N(k) =1/\sqrt{1-| \alpha_{\bfk}^{\rm in} |^2 }$ in order to normalise the Wronskian, $f_{\bfk}^* \overset{\leftrightarrow}{\partial}_\eta f_{\bfk} = - i H^2 \eta^2$ (which gives canonical commutation relations).
Choosing the Bunch-Davies vacuum in the past corresponds to setting the integration constant $\alpha_{\bfk}^{\rm in} = 0$ (which ensures that $f_{\bfk} (\eta) \sim e^{- i k\eta}$ at early times). 
Note that we have chosen the constant phase of $f_{\bfk} (\eta)$ so that, when restricted to the Bunch-Davies state, $f_{\bar{\bfk}} (\eta) = f_{\bfk}^* (\eta)$ implies simply $|\bar{\bfk}| = - | \bfk |$. 

\paragraph{de Sitter Isometries:}
\eqref{eqn:phiphic2} and \eqref{eqn:quadratic2} give the following two-point correlation functions, 
\begin{align}
\langle \sigma_{\bfk} \sigma_{-\bfk}  \rangle' (\eta) &= -  \frac{H^2 \eta^2}{2 k}   \;  \frac{ 1 + |\alpha^{\rm in}_{\bfk}|^2 + 2 \text{Re} \left[  \alpha^{\rm in}_{\bfk} e^{i 2 k \eta } \right]   }{1 - |\alpha^{\rm in}_{\bfk}|^2} \; , 
\label{eqn:sigsigc2}  \\
\langle \Sigma_{\bfk} \sigma_{-\bfk}  \rangle' (\eta) &= \frac{i}{2} - \frac{1}{2 k \eta}   \frac{1 + |\alpha^{\rm in}_{\bfk} |^2 + 2\,  \text{Re} \left[ \alpha^{\rm in}_{\bfk} (1 + i k \eta) e^{i 2 k \eta} \right] }{1- |\alpha^{\rm in}_{\bfk}|^2}   \; ,  \\
\langle \Sigma_{\bfk} \Sigma_{-\bfk}  \rangle' (\eta) &=  - \frac{
k\left(  
1 - | \alpha_\bfk^{\rm in} |^2 +  \frac{ \left(  1 + |\alpha^{\rm in}_{\bfk} |^2 + 2\,  \text{Re} \left[ \alpha^{\rm in}_{\bfk} (1 + i k \eta) e^{i 2 k \eta} \right]  \right)^2 }{k^2 \eta^2 ( 1 - | \alpha_\bfk^{\rm in} |^2  )} 
\right)
}{
2 H^2 \eta^2 \left( 1 + | \alpha_\bfk^{\rm in} |^2 + 2 \text{Re} \left[ \alpha_\bfk^{\rm in} e^{i 2 k \eta}  \right] \right)  
}   \; ,  
\end{align}
where $ \Sigma $ is the momentum conjugate to $\sigma$.
For the corresponding unequal-time correlator to be invariant under dilations, $\sum_{J} D_J  \langle \sigma_{\bfk_1} (\eta_1) \sigma_{\bfk_2} (\eta_2) \rangle  = 0$, the equal-time correlator (with $\delta^3 ( \bfk_1 + \bfk_2)$ removed), must satisfy\footnote{
Note that $\partial_\eta \langle \phi_{\bfk_1} ...  \phi_{\bfk_n} \rangle (\eta) = \sum_J \partial_{\eta_J} \langle \phi_{\bfk_1} (\eta_1) ... \phi_{\bfk_n} (\eta_n) \rangle |_{\eta}$, and for the two-point function in particular $k \partial_k \langle \phi_{\bfk} \phi_{-\bfk} \rangle' = \sum_J  \bfk_J \cdot \partial_{\bfk_J} \langle \phi_{\bfk_1} \phi_{\bfk_2} \rangle'$. 
},
\begin{align}
\left( 3  - \eta \partial_\eta  + k \partial_k    \right)  \langle \sigma_{\bfk} \sigma_{-\bfk}  \rangle' (\eta) &= 0  \,.
\label{eqn:c2D}
\end{align}
which in the case of \eqref{eqn:sigsigc2} sets $\partial_k \alpha_\bfk^{\rm in} =0$ (and we have assumed rotational invariance, so that $\alpha_{\bfk}$ is a function of $k$ only). Note that invariance under the other three de Sitter isometries, $\sum_J K_J \langle  \sigma_{\bfk} \sigma_{-\bfk} \rangle'=0$, is automatic since $K_J$ is odd in $\bfk_J$ (and $\bfk_1 = - \bfk_2$). The same conclusion is reached by instead applying \eqref{eqn:Dilc} and \eqref{eqn:Kc} directly to \eqref{eqn:c2conf}. 

\paragraph{Boundary Coefficient:}
If we instead write $c_2$ in terms of the late time boundary condition $\alpha_k$, so that,
\begin{align}
  c_{\bfk - \bfk} (\eta) = \frac{1}{H^2 \eta^3}  + \frac{k}{H^2 \eta^2} \frac{ \alpha_k \cos (k \eta) - k \sin (k \eta)}{ k \cos (k \eta) + \alpha_k \sin (k\eta ) } \; , 
\end{align}
which corresponds to mode functions,
\begin{align}
 f_{\bfk} (\eta) \propto \eta \cos ( k \eta ) +  \alpha^*_k \frac{ \eta  }{k} \sin ( k \eta )  = \eta + \alpha^*_k \eta^2 + \mathcal{O} ( \eta^3 ) \; ,
\end{align}
then we see that,
\begin{align}
 \alpha_k = i k  \;  \frac{ 1 + \alpha_{\bfk}^{\rm in} }{1- \alpha_{\bfk}^{\rm in} } \; , 
\end{align}
consistent with the general relation \eqref{eqn:aToain}. The Bunch-Davies condition in the past corresponds to the non-analytic boundary condition $\alpha_k = i k$ in the future, and more generally any de Sitter invariant state is characterised by an $\alpha_k =  i k \times \text{const.}$ 

\subsubsection*{Quartic Interaction, $\sigma^4$}

Now consider turning on a single quartic interaction, $\mathcal{H}_{\rm int} =  \Omega^{d+1} \lambda \frac{1}{4!} \sigma_{\bfx}^4$. 

\paragraph{Quartic Coefficient:}
Solving \eqref{eqn:HJint}, assuming for the moment a Bunch-Davies boundary condition for $c_2$, gives a simple quartic wavefunction coefficient,
\begin{align}
 c_{\bfk_1 \bfk_2 \bfk_3 \bfk_4} ( \eta ) =  \frac{1}{ H^4 \eta^4} \left[ \frac{i \lambda}{ k_T }  +  \alpha^{\rm in}_{\bfk_1 \bfk_2 \bfk_3 \bfk_4}  e^{-i  k_T  \eta }  \right]   \, . 
 \label{eqn:c4conf}
\end{align}
in terms of a single undetermined coefficient, $\alpha_{\bfk_1 \bfk_2 \bfk_3 \bfk_4}^{\rm in}$ (recall that $k_T = k_1 + k_2 + k_3 + k_4$ is the total $|\bfk_J|$). 

Note that in this case there is no exchange contribution to $c_4$, and consequently the constant of motion $\beta_4$ \eqref{eqn:constants} is given solely by the discontinuity \eqref{eqn:Disc} of \eqref{eqn:c4conf}, 
\begin{align}
\beta_4 =   \text{Disc} \, c_4^I = \frac{ \text{Disc} \, \alpha_4^{\rm in}  }{ \sqrt{i k_1 ik_2 ik_3 ik_4} }
\end{align}
and $\partial_\eta \beta_4 = 0$ is indeed conserved, since it is determined by the initial condition $\alpha_4^{\rm in}$.

\paragraph{de Sitter Isometries:}
The corresponding equal-time correlators from \eqref{eqn:phiphiphic4} and \eqref{eqn:Piphiphiphi} are,
\begin{align} 
\langle \sigma_{\bfk_1} \sigma_{\bfk_2} \sigma_{\bfk_3} \sigma_{\bfk_4}\rangle' (\eta) &=  \frac{\lambda H^4 \eta^4}{  \prod_j k_j } \left[ \frac{1}{ k_T} +  \text{Im} \left[  \alpha_4^{\rm in}  e^{-i k_T \eta} \right]  \right] \, , \\
\langle \Sigma_{\bfk_1} \sigma_{\bfk_2} \sigma_{\bfk_3} \sigma_{\bfk_4}\rangle' (\eta) &=  \frac{\lambda H^2 \eta}{  \prod_j k_j } \left[ \frac{1}{ k_T  }  + \text{Im} \left[  \alpha_4^{\rm in}  e^{ -i k_T \eta} \right] - \text{Re} \left[ \alpha_4^{\rm in}  i  k_T \eta  e^{-i k_T \eta} \right]   \right] \, .
\end{align}
For the corresponding unequal-time correlator to be invariant under the de Sitter isometries, the equal-time correlator (with $\delta^3 ( \bfk_1 + \bfk_2)$ removed) must satisfy \eqref{eqn:Dilc} and \eqref{eqn:Kc}, 
\begin{align}
\left( +9  - \eta \partial_\eta  + \sum_J  k_J  \partial_{k_J}    \right)  \langle \sigma_{\bfk_1} \sigma_{\bfk_2} \sigma_{\bfk_3} \sigma_{\bfk_4}  \rangle'  &= 0   \label{eqn:Dconf} \\
\sum_J \frac{\bfk_J}{k_J} \left[  \left( 4 \partial_{k_J}   + k_J  \partial_{k_J}^2   \right)  \langle \sigma_{\bfk_1} \sigma_{\bfk_2} \sigma_{\bfk_3} \sigma_{\bfk_4}  \rangle'  -  \frac{2 \eta}{\Omega^2}  \partial_{k_J} \langle \sigma_{\bfk_1} ... \Sigma_{\bfk_J} ... \sigma_{\bfk_4}  \rangle' \right] &= 0  \label{eqn:Kconf}  \, . 
\end{align}
where we assumed that the correlator is a function of the $| \bfk_j |$ only (for a simple $\phi^4$ contact interaction, this is guaranteed providing $\alpha_4^{\rm in}$ depends only on the $|\bfk_j|$). Dilations require that $\alpha_4^{\rm in} \sim 1/k$, while de Sitter boosts require that $(\partial_{k_I}^2 - \partial_{k_J}^2) \alpha_4^{\rm in} $ for every pair of $I$ and $J$ . Obvious solutions are $\alpha^{\rm in}_4 \propto 1/k_T$, and the Bunch-Davies condition, $\alpha_4^{\rm in} = 0$, which both satisfy these conditions\footnote{
We also note that when $\alpha_4^{\rm in} = 0$, choosing any constant $\alpha_2^{\rm in}$ also gives a $\langle \sigma_{\bfk_1} \sigma_{\bfk_2} \sigma_{\bfk_3} \sigma_{\bfk_4} \rangle'$ which satisfies \eqref{eqn:Dconf} and \eqref{eqn:Kconf}, and therefore any $\alpha$-vacuum is also consistent with the de Sitter isometries. 
}. This coincides with the conformal constraints \eqref{eqn:Dila} and \eqref{eqn:Ka} placed on $\alpha_{4}^{\rm in}$ by the isometries $\delta_D \Gamma$ \eqref{eqn:DGamma} and $\delta_{\mathbf{K}} \Gamma$ \eqref{eqn:KGamma} acting directly on the wavefunction.

\paragraph{Boundary Coefficient:}
Near the boundary at $\eta = 0$, there is no divergence in $\eta^4 c_4 (\eta)$, and so we can immediately write down the boundary coefficient $\alpha_4$ in terms of the initial condition $\alpha_4^{\rm in}$ (assuming $\alpha_2^{\rm in} = 0 $),
\begin{align}
\alpha_4 =  \frac{ i \lambda }{ k_T}  + \alpha_4^{\rm in}  . 
\label{eqn:O4conf}
\end{align}
Providing $\alpha_4^{\rm in}$ satisfies the above conditions, this satisfies the Ward identities of a primary correlator in a 3d CFT. Also note that the Bunch Davies initial condition, $\alpha_4^{\rm in} = 0$, corresponds to the non-analytical boundary state, $\alpha_4 = i \lambda / k_T$ in the future.  

Despite $\sigma_{\bfx}^4$ being a simple local interaction, the expressions \eqref{eqn:c4conf} and \eqref{eqn:O4conf} are not analytic in the momenta since we have enforced a condition on $\alpha_{\bfk}^{\rm in}$ in the far past (effectively integrating out the canonical momentum, fixing $\partial_\eta \phi$ in terms of a non-analytic function of $\bfk$). Instead, we can write a more general expression for $c_4$ by leaving the boundary condition for $c_2$ arbitrary, and utilizing the transfer functions (\ref{eqn:c4exp1}-\ref{eqn:c4exp4}). For this simple contact interaction, only the $\mathcal{I}_{\bfk_1 \bfk_2 \bfk_3 \bfk_4}^{\sigma_1 \sigma_2 \sigma_3 \sigma_4} (\eta )$ integrals \eqref{eqn:In} are needed, and they are given by,
\begin{align}
 \mathcal{I}_{\bfk_1 \bfk_2 \bfk_3 \bfk_4}^{----} (\eta') &= \frac{1}{\eta'} \int^{\eta'}_0 d\eta  \,  \cos ( k_1 \eta)  \cos ( k_2 \eta ) \cos (k_3 \eta) \cos (k_4 \eta) \\
 \mathcal{I}_{\bfk_1 \bfk_2 \bfk_3 \bfk_4}^{+---} (\eta') &= \frac{1}{\eta'} \int^{\eta'}_0 d\eta  \, \sin ( k_1 \eta)  \cos ( k_2 \eta ) \cos (k_3 \eta) \cos (k_4 \eta)   \;\;\; ... \;\;\; \text{etc.}  
\end{align}
These are manifestly analytic outside the horizon (i.e. for $| k \eta | <1$), for instance,
\begin{align}
 \mathcal{I}_{\bfk_1 \bfk_2 \bfk_3 \bfk_4}^{----} (\eta) = 1 - \frac{1}{6} \sum_{\rm perm.}^4 k_1^2 \eta^2 + \frac{1}{120} \sum_{\rm perm.}^4 k_1^4 \eta^4 + \frac{1}{20} \sum_{\rm perm.}^6 k_1^2 k_2^2 \eta^4 + \mathcal{O} ( | k \eta |^6  ) \; , 
\end{align}
but then as $|k \eta | \sim 1$ it is necessary to resum this series, which in this simple example can be done exactly, 
\begin{align}
 \mathcal{I}_{\bfk_1 \bfk_2 \bfk_3 \bfk_4}^{----} (\eta) = \frac{1}{16} \sum_{s_J = \pm}^{16} \; \frac{ e^{i (s_1 k_1 + s_2 k_2 + s_3 k_3 + s_4 k_4)\eta} }{s_1 k_1 + s_2 k_2 + s_3 k_3 + s_4 k_4}
\end{align}
which produces the familiar $k_T$ pole, as well as the various folded singularities, as anticipated by our previous Method of Regions argument in section~\ref{sec:smallEta}. 
Choosing $\alpha_k = i k$ (the Bunch-Davies condition) leads to a cancellation of all folded singularities, producing \eqref{eqn:O4conf}, but while this has simplified the analytic structure at $| k \eta | >1$ it comes at the cost of manifest analyticity at $| k \eta | < 1$ scales.

\subsubsection*{General Contact Contribution, $\sigma^n$}

For a general $\frac{g}{n!} \Omega^{d+1} \sigma^n$ interaction, the contact contribution to $c_n$ (assuming a Bunch-Davies initial condition for $c_2$) can be written simply as,
\begin{align}
 c_n (\eta)  =  \frac{ \alpha_n^{\rm in}  }{ \eta^{n} } e^{- i k_T \eta}  - \frac{g}{\eta^d} \, e^{- i k_T \eta} E_{n-d +1} \left( - i k_T \eta \right) \; .
 \label{eqn:cnConf}
\end{align}
where we have used $d=3$ dimensional mode functions but allowed for a general $d$-dimensional volume element $\Omega^{d+1}$ in the interaction (since this will allow us to compare with the divergent case $n=3$ below). We have also momentarily set $H=1$ (since it can always be restored by inspecting the mass dimension of each term).

\paragraph{Constants of Motion:}
Since this is purely a contact contribution to $c_n$, the discontinuity \eqref{eqn:Disc} must be conserved. Indeed, we see explicitly that,
\begin{align}
\beta_n =  \text{Disc} \, c_n^I (\eta) =  \frac{ \text{Disc} \, \alpha_n^{\rm in} }{ \sqrt{ ik_1 ... ik_n } }   - \frac{ \text{Im} \, g}{ \eta^{d-n} \sqrt{ ik_1 ... ik_n } } E_{n-d +1} \left( - i k_T \eta \right)
\end{align} 
and so for unitary dynamics (i.e. a real potential, so that $\text{Im} \, g = 0$), $\beta_n$ is indeed a constant of motion, and is set by the initial condition for $\alpha_n^{\rm in}$. 

\paragraph{de Sitter Isometries:}
A dilation acting on this $c_n (\eta)$ gives,
\begin{align}
 \delta_D \, c_n (\eta) = \frac{e^{- i k_T \eta} }{\eta^n}  \left[ - d  + n +  \sum_J  \bfk_J \cdot \partial_{\bfk_J}    \right]  \alpha_{n}^{\rm in}
\end{align}
using \eqref{eqn:Dilc1} and \eqref{eqn:Dila}. For boosts, notice that since $\partial_{k_1}^2$ and $\frac{1}{k_1} \left( 1 -  \eta \partial_\eta f_{\bfk_1}^*/ f^*_{\bfk_1}  \right)  \partial_{k_1}$ acting on the interaction contribution both yield results which depend only on $k_T$, the interaction contribution to \eqref{eqn:KcSph} vanishes by momentum conservation, leaving only a special conformal transformation \eqref{eqn:Ka} acting on $\alpha_n^{\rm in}$,
\begin{align}
 \delta_{\mathbf{K}} c_n (\eta) = \frac{e^{-i k_T \eta}}{\eta^n} \sum_J \left[  2  \left( \Delta^-_J   + \bfk_J \cdot \partial_{\bfk_J}   \right)  \frac{\partial}{\partial \bfk_J} - \bfk_J \partial_{\bfk_J}^2    \right]  \alpha_{n}^{\rm in}   \; .
\end{align} 
The result \eqref{eqn:cnConf} is therefore de Sitter invariant providing that the initial state $\alpha_n^{\rm in}$ is conformally invariant, with the scaling of a correlator of $n$ fields each of weight 2 ($=d - \Delta$ for a conformally coupled mass).  

\paragraph{Boundary Coefficient:}
At late times, assuming $n>d$, 
\begin{align}
 c_n (\eta)  =   \frac{1}{\eta^n} \left[   \alpha_n^{\rm in}  + \frac{g}{ ( - i k_T )^{n-d} }   \Gamma ( n- d )  + \mathcal{O} (\eta ) \right]  \; , 
\end{align}
we see that this has an order $(n-d)$ pole in $k_T$. 

So the general picture for contact interactions is quite simple. In terms of the $\alpha_n$ near $\eta = 0$, the wavefunction coefficient $c_n$ can be written via transfer functions which are analytic providing every $| k_J \eta | < 1$, e.g.,
\begin{align}
 \mathcal{I}_{\bfk_1 ... \bfk_n}^{-...-} (\eta) = \frac{1}{n-d} -  \frac{1}{2 (n-d+2)} \sum_{\rm perm.}^n k_1^2 \eta^2 + \mathcal{O} ( | k \eta |^4 ) \; , 
 \label{eqn:Inconf}
\end{align}
but once $| k \eta |$ exceeds 1 then this series resums into $k_T$ poles up to order\footnote{
Note that if we had considered an interaction with derivatives, then this would also increase the order of the $k_T$ pole (at fixed $n-d$).
} $(n-d)$. 

If $n = d$, then there are divergences in the transfer functions which must be regulated---we will now discuss this case in more detail.

\subsubsection*{Cubic Interaction, $\sigma^3$}

Consider the single interaction $\mathcal{H}_{\rm int} = \Omega^{d+1}  \frac{H}{3!}  \lambda \sigma^3_\bfx$ (the factor of $H$ ensures that $\lambda$, our expansion parameter, is dimensionless). 

\paragraph{Bunch-Davies State:}
Solving the Hamilton-Jacobi equation gives,
\begin{align}
 c_{\bfk_1 \bfk_2 \bfk_3} (\eta) = \frac{e^{ i \sum_J k_J \eta } }{ H^3 \eta^3} \left( \lambda \, \text{Ei} \left[ -i k_T \eta \right] + \alpha_{\bfk_1 \bfk_2 \bfk_3}^{\rm in}   \right)
 \label{eqn:c3conf}
\end{align}
where $\alpha_3^{\rm in}$ is an undetermined constant of integration.

\paragraph{de Sitter Isometries:}
Assuming $\alpha_3^{\rm in}$ is real, the equal-time correlators are,
\begin{align} 
\langle \sigma_{\bfk_1} \sigma_{\bfk_2} \sigma_{\bfk_3} \rangle (\eta) &=  \frac{\lambda H^3 \eta^3 }{ \prod_J k_J}  \left[  - (  \text{Ci} + \alpha_3^{\rm in} ) \, \sin   +  \text{Si}  \, \cos          \right]  \, , \label{eqn:phi3conf} \\
\langle \Sigma_{\bfk_1} \sigma_{\bfk_2} \sigma_{\bfk_3}  \rangle (\eta) &=  \frac{\lambda H}{ \prod_J k_J} \Bigg[ - \left( k_1 \eta \cos  + \sin  \right) \left( \text{Ci} + \alpha_3^{\rm in} \right) + \left( \cos - k_1 \eta \sin  \right) \text{Si}    \Bigg] \, . \nonumber
\end{align}
where we have suppressed the $k_T \eta$ argument of $\sin$ and $\cos$, as well as of the sine integral, $\text{Si}$, and the cosine integral $\text{Ci}$. Invariance of $\langle \sigma_{\bfk_1} (\eta_1) \sigma_{\bfk_2} (\eta_2) \sigma_{\bfk_3} (\eta_3) \rangle$ under dilations and de Sitter boosts places the constraints \eqref{eqn:equalTimeIsometries} on \eqref{eqn:phi3conf}, requiring that $\sum_J k_J \partial_{k_J} \alpha_3^{\rm in} = 0$ and that $( \partial_{k_I}^2 - \partial_{k_J}^2 ) \alpha_3^{\rm in} = 0$ for every pair of $I$ and $J$ (where we have used that $\alpha_3^{\rm in}$ can always be written in terms of the three $|\bfk_j|$ only, using $\bfk_1 + \bfk_2+ \bfk_3 = 0$). $\alpha_3^{\rm in}$ is therefore constrained to be dimensionless and simultaneously a function of $k_1 \pm k_2$, $k_1 \pm k_3$ and $k_2 \pm k_3$ only---the final requirement of crossing symmetry then fixes $\alpha_3^{\rm in}$ to be a constant.

\paragraph{Boundary Divergence:}
Near the boundary, there is a logarithmic divergence,
\begin{align}
 \text{Ei} \left[  - i \sum_J k_J \eta  \right] = \log ( i k_T \eta  ) + \gamma_E  + \mathcal{O} ( \eta ) 
\end{align}
and so we must renormalise $\alpha_3 = \lim_{\eta \to 0} H^3 \eta^3 c_{\bfk_1 \bfk_2 \bfk_3}$.

In $d = 3 - \delta$ dimensions, the boundary coefficient $\alpha_3$ can be computed directly in terms of $\alpha^{\rm in}_3$ using a triple-$H$ integral (evaluated in the Appendix), 
\begin{align}
\alpha_{\bfk_1 \bfk_2 \bfk_3} &= N_1 N_2 N_3 \int_{-\infty}^0 d\eta \;   \partial_\eta c^I_{\bfk_1 \bfk_2 \bfk_3}   \nonumber \\
 &= \alpha^{\rm in}_{\bfk_1 \bfk_2 \bfk_3} +  \frac{\lambda}{2} \left[   \frac{1}{\delta} +  \log \left( k_T  \right) 
 + \text{finite}  \right] \, ,  
\end{align}
where $N_J = i \pi ( k_J/2)^{\nu_J} / \Gamma ( \tilde{\nu}_J ) N (k_J)$ is the finite $\lim_{\eta \to 0} H \eta^{\Delta}/f_{\bfk} (\eta)$, and $\partial_\eta c^I_{\bfk_1 \bfk_2 \bfk_3} = \lambda H  \Omega^{d+1} f_{\bfk_1} f_{\bfk_2} f_{\bfk_3}$ from the equation of motion \eqref{eqn:HJc3}. 
In the minimal subtraction scheme \eqref{eqn:localrenorm}, the coefficient is rendered finite by an additive renormalisation of the boundary condition,
\begin{align}
 \alpha_{\bfk_1 \bfk_2 \bfk_3} = \alpha_{\bfk_1 \bfk_2 \bfk_3}^{\rm ren} - \frac{\lambda \mu^{-\delta}}{2 \delta}
\end{align}
as discussed in \eqref{eqn:egaren}. This results in a finite $\alpha_{3}^{\rm ren}$, which depends logarithmically on the momenta, and satisfies an anomalous conformal Ward identity \eqref{eqn:anomaly}. 

Rather than start in the Bunch-Davies state and evolve forward until we encounter this divergence at the endpoint of the evolution, we can instead describe the bulk $c_3 (\eta)$ using the transfer functions \eqref{eqn:c3exp}.
 
\paragraph{Transfer Functions:}
Since $\sum_i \Delta_i = d$, it is $\mathcal{I}_{\bfk_1 \bfk_2 \bfk_3}^{---}$ which is responsible for the late time divergence in the transfer function,
\begin{align}
 \mathcal{I}_{\bfk_1 \bfk_2 \bfk_3}^{---} (\eta) = \frac{1}{d-3} - \log ( \eta )  - \frac{1}{4} \sum_J k_J^2 \eta^2  + \mathcal{O} (  | k \eta |^4 ) 
\end{align}
we see that the transfer function is analytic in momenta for $|k \eta | < 1$, but due to the divergence has developed a logarithmic dependence on $\eta$. As we transfer from the boundary at $\eta = 0$ into the bulk, eventually we reach horizon crossing at $|k \eta| \sim 1$ and it is necessary to resum the series. In this case we can perform the resummation exactly,
\begin{align}
 \mathcal{I}_{\bfk_1 \bfk_2 \bfk_3}^{---} &= \frac{1}{3-d} - \log ( \eta ) \nonumber  \\
 &- \frac{1}{8} \sum_{s_J = \pm}^8  \left(  \text{Ci} \left( (s_1 k_1 + s_2 k_3 + s_3 k_3) \eta \right) -  \log \left( (s_1 k_1 + s_2 k_3 + s_3 k_3) \eta  \right) - \gamma_E  \right) \; , 
 \label{eqn:I3conf}
\end{align}
and the series in $k^{2j}$ becomes a logarithmic singularity at large $\eta$ (in both $k_T$ and its folded counterparts), consistent with our previous expectation of an order $n-3$ pole from a general contact contribution to $c_n$. 

To remove this divergence from $\mathcal{I}^{\rm ---}_{\bfk_1 \bfk_2 \bfk_3}$, we must redefine the boundary condition on $\alpha_3$,
\begin{align}
\alpha^{\rm ren}_{\bfk_1 \bfk_2 \bfk_3} = \alpha_{\bfk_1 \bfk_2 \bfk_3} - \frac{ \lambda \mu^{d-3} }{d-3}
\end{align}
where we have introduced a scale $\mu$ required on dimensional grounds. When written in terms of this $\alpha_3^{\rm ren}$, the coefficient $c_3 (\eta)$ is now finite as $d \to 3$ for any value of $\alpha_2$. In particular, choosing the Bunch-Davies $\alpha_2 = i k$ removes the folded singularities at early times and produces a renormalised coefficient,
\begin{align}
 c_{\bfk_1 \bfk_2 \bfk_3}^I ( \eta ) = \alpha^{\rm ren}_{\bfk_1 \bfk_2 \bfk_3}  + \lambda \left(  \text{Ei} [ - i k_T \eta ] 
  - \log \left( \frac{- i k_T}{ \mu} \right) - \gamma_E  \right) \; .
  \label{eqn:c3renconf}
\end{align}
Taking the limit $\eta \to -\infty (1-i \epsilon)$ to connect with the state in the far past, we find that,
\begin{align}
\alpha^{\rm in}_3 = \alpha^{\rm ren}_3 - \lambda \log \left( \frac{-i k_T}{\mu} \right) - \lambda \gamma_E  \; . 
\end{align}
Setting the scale $\mu$ close to $k_T$ ensures that the log is small and allows a constant $\alpha_3^{\rm ren}$ to reproduce the Bunch-Davies correlators in the bulk.



\subsection{Massless Scalar}

Now we will consider a massless scalar, $\phi$, on de Sitter, with free Hamiltonian given by \eqref{eqn:Hfree} with $m = 0$ and $\Omega = 1/(-H\eta)$. 

\subsubsection*{Free Theory}

The Gaussian width is given by,
\begin{equation}
c_{\bfk - \bfk} (\eta)  =   \frac{  k^2 }{ H^2 \eta ( 1 - i  k \eta ) }  \;  ,
\end{equation}
where we have chosen the Bunch-Davies boundary condition $\alpha_2^{\rm in} = 0$ so that the corresponding mode function,
\begin{align}
 f_{\bfk} (\eta) = \frac{H ( 1 + i k \eta ) }{i k} \frac{e^{-i k \eta }}{\sqrt{2 i k}} \; ,
\end{align}
vanishes as $\eta \to -\infty (1- i \epsilon)$. Note that we have chosen the overall phase of $f_{\bfk} (\eta)$ so that sending $k \to -k$ gives $f_\bfk (\eta) \to f_{\bfk}^* (\eta)$. 

This corresponds to a field correlator,
\begin{align}
 \langle \phi_{\bfk} \phi_{-\bfk} \rangle = \frac{H^2}{2 k^3}  (1 + k^2 \eta^2) \; . 
\end{align}

\paragraph{Boundary Coefficient:}
Note that while $c_2$ diverges as we approach the boundary at $\eta = 0$, since $\Delta = 0$ for massless fields the divergence is softened from $\Delta \eta^{-d}$ to $k^2 \eta^{2-d} / (2 \nu-2)$. In dimensional regularisation this vanishes, but with a hard cutoff at $\eta_\delta$ this requires an additional renormalisation of the $\alpha_2$ boundary condition,
\begin{align}
 \alpha_k = \alpha_k^{\rm in} - \frac{k^2}{3 \eta_\delta} \; , 
\end{align} 
which can be thought of as adding a counterterm $(\partial_\mu \phi )^2$ to the effective action. The Bunch-Davies condition in the past corresponds to setting this $\alpha_2^{\rm ren} = i$ in the future (and not the bare $\alpha_2$).

This gives a renormalised correlator of boundary sources,
\begin{align}
\alpha_2^{\rm ren} =  \frac{k^3}{i H^2}
\end{align}
which satisfies the Ward identity for a $3$-dimensional two-point function of primary operators each with weight $3$.

\subsubsection*{$\phi^3$ Interaction}

\paragraph{Cubic Coefficient:}
The cubic wavefunction coefficient (with $\alpha_2^{\rm in} = 0$) is given by,
\begin{align}
 c_{\bfk_1 \bfk_2 \bfk_3}^I (\eta) = \alpha^{\rm in}_{\bfk_1 \bfk_2 \bfk_3} + \frac{g}{ 6 \sqrt{2} H ( -i k_1 k_2 k_3)^{3/2} } \Bigg(&   -  \text{Ei} [ i k_T \eta ] \sum_{\rm perm.}^3 i k_1^3  \nonumber \\ 
 &\qquad e^{i k_T \eta} \sum_{\rm perm.}^3 \left(  \frac{1}{3\eta^3}  - \frac{i k_1}{\eta^2} + \frac{k_1^2 - k_1 k_2}{\eta}    \right) 
  \Bigg) \; . 
  \label{eqn:c3massless}
\end{align}
where $k_T = k_1 + k_2 + k_3$. 
Note that since the exponential integral obeys $ \text{Ei} [ z ]^* = \text{Ei} [ z^* ]$, we have that,
\begin{align}
 \text{Disc} \, c_{\bfk_1 \bfk_2 \bfk_3}^I (\eta) =  \text{Disc} \, \alpha^{\rm in}_{\bfk_1 \bfk_2 \bfk_3} 
\end{align}
whenever $g \phi^3$ is a unitary interaction (i.e. $\text{Im} \, g = 0$), and so the $\beta_{\bfk_1 \bfk_2 \bfk_3} = \text{Disc} \, \alpha_{\bfk_1 \bfk_2 \bfk_3}^{\rm in}$ of the initial state is conserved throughout the time evolution. 

\paragraph{de Sitter Isometries:}
This corresponds to an equal-time correlator,
\begin{align}
 \langle \phi_{\bfk_1} \phi_{\bfk_2} \phi_{\bfk_3} \rangle' &= 
 \text{Im} \, \left(  \alpha^{\rm in}_{\bfk_1 \bfk_2 \bfk_3}  \prod_{J} f_{\bfk_J} (\eta ) \right)  \nonumber \\
 &+ \frac{g H^2 \sum_J k_J^3}{24 k_1^3 k_2^3 k_3^3 } 
 \Bigg( 
 \left( 1 - \sum_{\rm perm.}^3 k_1 k_2 \eta^2  \right)  \text{Re} \left(  e^{-i k_T \eta}  \text{Ei} \left[ i k_T \eta \right]   \right)  \nonumber  \\
&\qquad\qquad\qquad\qquad-  \left(  k_T \eta  + \eta^3 k_1 k_2 k_3  \right)  \text{Im} \left(  e^{-i k_T \eta}  \text{Ei} \left[ i k_T \eta \right]   \right)  -1
  \Bigg)   \nonumber \\
&+ \frac{g H^2}{24 k_1^3 k_2^3 k_3^3 }  \Bigg( 
k_1 k_2 k_3 \left( 1 + \sum_J k_J^2 \eta^2  - \sum_{\rm perm.}^3 k_1 k_2 \eta^2  \right) - \sum_{\rm perm.}^6 k_1 k_2^2 
 \Bigg)
\end{align}
which indeed obeys  \eqref{eqn:equalTimeIsometries}, providing $\alpha_3^{\rm in}$ satisfies the conformal Ward identities \eqref{eqn:Dila} and \eqref{eqn:Ka}. 

\paragraph{Boundary Renormalisation:}
As $\eta \to 0$, the contribution from interactions to \eqref{eqn:c3massless} diverges---the interactions do not turn off fast enough at late times, and lead to a formally infinite change in the wavefunction between $\alpha_3^{\rm in}$ and $\alpha_3$ on the boundary.  
To renormalise the divergence in $c_3$, we will now write it in terms of the boundary value of $\Psi$ at $\eta \to 0$. Using a hard-cutoff, this is given by,
\begin{align}
 c_{\bfk_1 \bfk_2 \bfk_3}^I ( \eta ) &= \lim_{\eta_{\delta} \to 0} \left( f_{\bfk_1} (\eta_\delta) f_{\bfk_2} (\eta_\delta)f_{\bfk_3} (\eta_\delta)f_{\bfk_4} (\eta_\delta)  \alpha_{\bfk_1 \bfk_2 \bfk_3} (\eta_\delta ) + \int_{\eta_{\delta} }^\eta d \eta'  \, \partial_\eta c_{\bfk_1 \bfk_2 \bfk_3} (\eta')   \right) \label{eqn:c3a3} \\
&=  \alpha_{\bfk_1 \bfk_2 \bfk_3}^{\rm ren} (\mu) - \frac{g}{H^4} \sum_{\rm perm.}^3 \left( (-2+2\gamma_E + i \pi ) k_1^3 - 2 k_1^2 (k_2 + k_3 )  +  \frac{i k_1^3}{3} \log ( - k_T \mu )      \right)   \nonumber \\
&+ \frac{g}{ 6 \sqrt{2} H ( -i k_1 k_2 k_3)^{3/2} }          \Bigg(   -  \text{Ei} [ i k_T \eta ] \sum_{\rm perm.}^3 i k_1^3 + e^{i k_T \eta} \sum_{\rm perm.}^3 \left(  \frac{1}{3\eta^3}  - \frac{i k_1}{\eta^2} + \frac{k_1^2 - k_1 k_2}{\eta}    \right)  \nonumber 
\Bigg) 
\end{align}
where we have renormalised the boundary condition (adopting a minimal subtraction scheme)\footnote{
In order to remove the logarithmic $\log (\eta_\delta)$ divergence, it is necessary to introduce a scale $\mu$ in this redefinition.  
},
\begin{align}
 \alpha_{\bfk_1 \bfk_2 \bfk_3}^{\rm ren} (\mu) = \alpha_{\bfk_1 \bfk_2 \bfk_3}  - \frac{2g}{H^4} \left[
  \sum_{\rm perm.}^3 \left(  \frac{1}{3\eta_\delta^3} + \frac{k_1^2 - k_1 k_2}{\eta_\delta}  - \frac{i k_1^3}{3} \log ( - \mu \eta_\delta  \right) 
  \right] \, .
\end{align}
This produces a $c_3 (\eta)$ which is now finite for any $\eta$ (assuming $|\eta| > |\eta_\delta|$ before $\eta_\delta$ is taken to zero), and which asymptotes to $\alpha_{\bfk_1 \bfk_2 \bfk_3}^{\rm ren}$ at the boundary.  Choosing Bunch-Davies in the past, $\alpha_3^{\rm in} = 0$ in \eqref{eqn:c3massless}, corresponds to choosing,
\begin{align}
 \alpha^{\rm ren}_{\bfk_1 \bfk_2 \bfk_3} =  \frac{g}{H^4} \sum_{\rm perm.}^3 \left( (-2+2\gamma_E + i \pi ) k_1^3 - 2 k_1^2 (k_2 + k_3 )       \right) \; ,
\end{align}
in \eqref{eqn:c3a3}. At the conformal boundary, the Bunch-Davies vacuum corresponds to a non-local state.

The coefficient of the logarithmic running is scheme-independent. For instance, using instead dimensional regularisation,
\begin{align}
 c_{\bfk_1 \bfk_2 \bfk_3}^I ( \eta ) &= \lim_{\delta \to 0} \left( \alpha_{\bfk_1 \bfk_2 \bfk_3} + \int_{0}^\eta d \eta'  \, \partial_\eta c_{\bfk_1 \bfk_2 \bfk_3} (\eta')   \right)   \nonumber \\
&= \lim_{\delta \to 0} \left(   - \frac{i}{H^4 } \; \frac{ k_1^3 + k_2^3 + k_3^3 }{3 \delta}  \right)  \; . 
\end{align}
requires a different redefinition of the boundary condition\footnote{
The mass dimension of $\alpha_{\bfk_1 \bfk_2 \bfk_3}$ requires the addition of a scale $\mu$ in this redefinition.  
},
\begin{align}
 \alpha_{\bfk_1 \bfk_2 \bfk_3} = \alpha_{\bfk_1 \bfk_2 \bfk_3}^{\rm ren} - \mu^{-\epsilon} \frac{k^3}{\delta } 
 \label{eqn:masslessa3redef}
\end{align}
but produces a renormalised coefficient with the same $\log (k \mu )$ dependence on the RG scale $\mu$.

\paragraph{Boundary Operator Mixing:}
For massless  scalars, $\Delta^- = 0$ and so $2 \Delta^- + \Delta^+ - d = 0$ and the three-point function has a logarithmic semi-local divergence.
The renormalisation of $\alpha_3$ can be understood as the operator redefinition,
\begin{align}
 \phi \to  \phi - \frac{g}{\delta} \phi^2 \, ,
\end{align}
which implements the following redefinition of the boundary condition,
\begin{align}
 \alpha_{\bfk_1 \bfk_2 \bfk_3} = \alpha^{\rm ren}_{\bfk_1 \bfk_2 \bfk_3} - \frac{2 g}{\epsilon} \alpha_2 \;\;\;\; , \;\;\;\; \alpha_{\bfk_1 \bfk_2 \bfk_3 \bfk_4} = \alpha^{\rm ren}_{\bfk_1 \bfk_2 \bfk_3 \bfk_4} - \frac{g}{\delta} \alpha_{\bfk_1 \bfk_2 \bfk_3}  - \frac{g}{\delta^2}  \alpha_2 \;\; , \;\; ... 
 \label{eqn:masslessaredef}
\end{align}
Note that since we are working with the Bunch-Davies initial condition, $\alpha_2^{\rm in} = 0$, this corresponds to $\alpha_2 = i k^3$, and so \eqref{eqn:masslessaredef} indeed corresponds to \eqref{eqn:masslessa3redef} for $\alpha_3$.

\subsection{EFT of Inflation}

Let us move from the pure de Sitter spacetime  to  discuss  inflation. An inflationary  epoch can be thought of as a de Sitter spacetime where time translations are spontaneusly  broken.  This broken symmetry will imply the existence of a Goldstone mode $\pi(x)$. From an effective field theory point of view we can write the most general action for the scalar field $\pi$ which is consistent with the rest of the symmetries of de Sitter. The advantge of this approach is that we can be rather  agnostic about the field content of de Sitter, and for example include the case where there is a non canonical kinetic term for the inflaton.

 There is a general procedure for implementing this~\cite{Cheung:2007st}, where one starts by writing down the most general action consistent with softly broken time traslations. This implies that, for example, the first terms will be $S\sim\int\d^4x\sqrt{-g}( c(t)R+g^{00}+\Lambda(t)+...)$, where the coeffiecients $c(t)$ and $\Lambda(t)$ parametrise the new time dependence, and the dots are higher order terms in derivatives of the metric.  To obtain the Goldstone mode one reintroduces time reparametrisations via the Stuckelberg trick, $t\to t-\pi(x)$. Doing so, one gets an action for the graviton and the scalar field $\pi$. This action contains the kinetic term and  also the self interactions of the Goldston mode, but also all allowed   interactions  with the graviton, so  solving the system in full generality is not possible analytically. 
 
 In order to proceed let us notice following~\cite{Cheung:2007st}, that  the leading order mixing term of  the scalar field $\pi$ with the graviton is given by $2\mpl^2\dot H\dot\pi\delta g^{00}\sim -6 \mpl^2\dot H (\epsilon H^2\pi^2)$. This has to be compared with the kinetic term for $\pi$, $-\mpl^2\dot H\dot \pi^2$. We have that,
 \bea
 \frac{\mathrm{mixing}}{\mathrm{kinetic}}=\frac{3\epsilon H^2\pi^2}{\dot\pi^2}=\frac{3\epsilon H^2}{\omega^2}.
 \eea
 The mixing becomes negligible in the limit when wavelengths are $\omega\gg \omega_{\mathrm{mix}}\equiv\sqrt{\epsilon} H$. So large wavelenghts are effectively decoupled from gravitational fluctuations, and we can  write the action for Goldstone mode on a fixed curved background neglecting couplings with the graviton modes. The decoupled action is,
\begin{align}
  S=\int d^4 x\sqrt{-g}\left[-\frac{\mpl^2\dot H}{c_s^2}\left(\dot\pi^2-c_s^2\frac{(\partial_i\pi)^2}{a^2}\right)+\mpl^2\dot H \frac{(1-c_s^2)}{c_s^2}\left(\dot \pi^3-\dot\pi\frac{(\partial_i\pi)^2}{a^2}\right)-\frac{4}{3}M_3^4\dot\pi^3\right]+..\nonumber\\
  \label{EFT:pi_action}
  \end{align}
First let us note that modes  leave the horizon at $\omega= H$, so we can relate the Goldstone to observations by considering this action.
Let us discuss the  action (\ref{EFT:pi_action}). Besides the usual kinetic term we have introduced the speed of sound for the peturbations $c_s^2$. This parametrises the effect of a non-canonical kinetic term (and also as the effect of integrated out heavy degree of freedoms~\cite{Tolley:2009fg,Achucarro:2012sm,Cespedes:2012hu}). Note that when $c_s\neq 1$ scale invariance is broken even at $k\eta\to 0$ and so fixing the correlation functions by using conformal symmetry at the future infinity is no longer possible. The second term $M_3^4$, parameterises self interactions of the Goldstone mode and is present even in single field inflation where it is of order $\epsilon^2$.

It is convenient to introduce the symmetry breaking scale $f_\pi^4\equiv 2\mpl^2\vert\dot H\vert c_s$, which corresponds to the scale when time traslations are broken~\cite{Baumann:2011su}. For the case of single field inflation $f_\pi^4=\dot\phi$, which matches the intuition that a background spontaneously break  the de Sitter symmetry.

\paragraph{Equations of Motion:}
To calculate the Hamiltonian associated to (\ref{EFT:pi_action}) we first need the conjugate momentum to $\pi$, $P_{\pi}=\frac{\partial\calL}{\partial_{\dot\pi}}$, which  is a non linear  equation in $\dot \pi$. We can invert this relation if we consider that $f^{-4}_\pi\ll 1$. Using the solution for $\dot\pi$ we can write down the Hamiltonian, which at leading order is,
\bea
H=-\frac{1}{\sqrt{-g}}\frac{c_s^3}{8f_\pi^4}P_\pi^2+2\sqrt{-g}\frac{f_\pi^4}{a^2c_s}\nabla_\pi^2-\frac{(c_s^2-1)}{2a^2})P_\pi(\nabla\pi)^2+\frac{1}{g}\left(\frac{c_s^6(c_s^2-1)}{32 f_\pi^8}+\frac{1}{48}\frac{M_3^4 c_s^9}{f_\pi^{12}}\right)P_\pi^3+...\nonumber
\label{eqn:EFTHamiltonian}
\eea
For energies much below the symmetry breaking scale  $f_\pi^4$, the wavefunction evolution is given by the  Schr\"{o}dinger equation $H=\partial_\eta S$. 
Now given an observed state $\vert\Psi\rangle$ we can constrain the coefficients of the Hamiltonian by assuming that the initial state was Bunch Davis. The equation for the two point function is given by
\bea
3H  c_{\bfk -\bfk} +\frac{1}{a^2} \partial_\eta c_{\bfk -\bfk} = \frac{c_s^3}{f_\pi^4} c_{\bfk -\bfk}^2+\frac{f_\pi^4}{c_s}\frac{1}{a^2}k^2 \label{EFT:HJ}.
\eea
This equation looks similar to a massless scalar field on FLRW but now the coefficients are time dependent. The above equation can be solved analytically when we consider de Sitter evolution and that  $c_s$ and $f_\pi^4$ are time and scale independent. We have that,
\bea
c_{\bfk -\bfk} (\eta) =\frac{f^4_\pi H^2}{c_s^3}\frac{k^2\eta^2 c_s^2}{1+i k\eta c_s},
\eea
where to fix the initial conditions we have assumed  a Bunch-Davies initial state. In general inflation breaks the de Sitter isometries and in particular the spacetime is no longer conformally invariant at $k\eta\to 0$. 
 
The Hamiltonian \eqref{eqn:EFTHamiltonian}  contains interactions which are higher order in momentum and that we did not take into account before. We can still use \eqref{eqn:HJint} to obtain the wavefunction coefficients.  For example the equation for the three point function is given by,
\begin{align}
&\left(3H+\frac{1}{a^2}\frac{\partial }{\partial \eta}\right)  c_{\bfk_1 \bfk_2 \bfk_3}^I =\nonumber\\
&\left(\frac{c_s^6(c_s^2-1)}{32 f_\pi^8}+\frac{1}{48}\frac{M_3^4 c_s^9}{f_\pi^{12}}\right)\prod_{i=1}^3\frac{ c_{\bfk -\bfk}^I  }{f_{\bfk_i} }-\frac{(c_s^2-1)}{a^2}f_{\bfk_1}f_{\bfk_2}f_{\bfk_3}\sum_{\mathrm{perms}(1,2,3)}\frac{ c_{\bfk -\bfk}^I }{f_{\bfk_i}f_{-\bfk_i}} (\bfk_j\cdot \bfk_k) \nonumber\\
\label{HJ_G:EFT}
\end{align}

\paragraph{Boundary Coefficients:}
This differential equation can be brought to an integral form.  Although it is possible to specify an arbitrary initial condition, for now let us assume  a Bunch-Davies initial state. This allows to specify the contour at  $\eta\to -\infty$ so the integral vanishes there. Then to fix the Hamiltonian one will have to rely on a potentially observed bispectrum. This means that one, in principle,  can constrain all the coefficients appearing in the Hamiltonian \eqref{eqn:EFTHamiltonian}.
The first term appears in slow roll inflation, in which case $c_s=1$ and $\frac{1}{48}\frac{M_3^4 c_s^9}{f_\pi^{12}}=-\frac{1}{\epsilon}$, where $\epsilon$ is the slow roll parameter.  The three point function coefficient is given by,
\bea
\left(  \alpha_{\bfk_1 \bfk_2 \bfk_3}   \right)_{\dot\pi^3}&\sim& \frac{ (H \eta_\delta )^3 }{ \prod_{i=1}^3 f_{\k_i}^*(\eta_{\delta} ) } \int^{ \eta_\delta }_{-\infty}\frac{\d\eta'}{(H\eta')^4}\prod_{i=1}^3 f_{\k_i}^*(\eta') c_{\bfk_i -\bfk_i} (\eta')  \nonumber\\
&=&\frac{1}{a( \eta_\delta )^3}\left(\frac{k_1^2k_2^2k_3^2}{(k_1+k_2+k_3)^3}+\mathcal{O}( \eta_\delta )\right)
\eea
When $c_s\neq 1$ the second term also contributes to the three point function. After a straightforward calculation we get,
\begin{align}
&( \alpha_{\bfk_1 \bfk_2 \bfk_3} )_{\dot\pi\nabla\pi^2} = \frac{c_s^2-1}{a( \eta_\delta )^3 \prod_{i=1}^3 f_{\k_i}^*(\eta_{\delta} ) }\int^{\eta_\delta}_{-\infty}\frac{\d\eta'}{(H\eta')^2}\prod_{i=1}^3  f_{\k_i}^*(\eta')  \sum_{\rm perm.} c_{\bfk_1 -\bfk_1} (\eta') \bfk_2\cdot\bfk_3 \nonumber\\ 
&\quad =\frac{c_s^2-1}{a( \eta_\delta )^3} \sum_{\rm perm.} \left(
\frac{k_1^2-k_2^2-k_3^2}{(k_1+k_2+k_3)^3}k_1^2(k_1^2+3k_1(k_2+k_3)+2(k_2^2+3k_2k_3+k_3^2)) +\mathcal{O}( \eta_\delta )
\right)
\end{align}
Both results match known expressions at $\eta\to 0$ for the three point function \citep{Cheung:2007sv}.

\paragraph{Bulk Coefficients:}
We close this section with explicit formulae for the wavefunction in the EFT of inflation at finite $\eta$, 
\bea
( c_{\bfk_1\bfk_2 \bfk_3} ) ^S_{\dot\pi^3}&\sim&(H\eta)^3\int^{\eta}_{-\infty}\frac{\d\eta'}{(H\eta')^4}\prod_{i=1}^3\left( K_{\k_i}(\eta') c_{\bfk_i -\bfk_i}^S \right)\nonumber\\
&=&\frac{i}{   k_T^3}\prod_{i=1}^3{\left(\frac{k_i}{k_i\eta-i}\right)}\left(\sum_{i=1}^{3} k_i\eta\left(k_T\eta-2i\right)-2\right) 
\eea

When $c_s\neq 1$ the second term also contributes to the three point function,
\begin{align}
&( c_{\bfk_1 \bfk_2 \bfk_3})^S_{\dot\pi\nabla\pi^2}  
= \frac{c_s^2-1}{a(\eta)^3}\int^{\eta}_{-\infty}\frac{\d\eta'}{(H\eta')^2}\prod_{i=1}^3 K_{\k_i}(\eta') \sum_{\mathrm{perms}(1,2,3)}\left(\langle \mathcal{O}_{\bfk_i} \mathcal{O}_{-\bfk_i}\rangle^S(\bfk_j\cdot\bfk_k)\right)\nonumber\\ 
&=-\frac{(c_s^2-1)a(\eta)^{-3}}{(k_1
  \eta-i) (k_2 \eta-i) (k_3 \eta-i) k_T^3} \nonumber\\
 &\times\sum_{\mathrm{perm}}^3 k_1^2\left(-k_1^2+k_2^2+k_3^2)\left(\right(2k_T-k_1)k_T+2k_2k_3+i k_T(k_1k_T+2k_2k_3)\eta-k_T^2 k_2k_3 \eta^2\right)
\end{align}

\section{Discussion}
\label{sec:discussion}

In this work, we have studied the time evolution of scalar field correlators on a fixed de Sitter background using the wavefunction in the Schr\"{o}dinger picture. 
This is a concrete arena in which to explore properties of cosmological correlators, since the dominant signals produced during inflation originate in the correlations of a scalar field on a quasi-de Sitter background. 
Rather than focus on any particular model of inflation, we have assumed only that the interaction Hamiltonian has certain foundational properties---namely unitary, de Sitter invariance and locality. 
We have shown that,

\begin{itemize}[leftmargin=5em]

\item[Unitarity $\Rightarrow$] One constant of motion, $\beta_n$, for every wavefunction coefficient $c_n (\eta)$.

\item[de Sitter $\Rightarrow$] Constraints on equal-time correlators (of both $\phi$ and $\Pi$) and wavefunction coefficients (which depend explicitly on $\mathcal{H}_{\rm int}$) at any time in the bulk---the latter reduce to the familiar conformal Ward identities at $\eta \to 0$. 

\item[Locality $\Rightarrow$] Analyticity of transfer functions outside the horizon---as a result, non-analyticities in the bulk $c_n (\eta)$ only arise when $| k \eta | >1$ and fields enter the horizon and begin to oscillate (or from the boundary condition itself).

\end{itemize}

These results have a number of interesting consequences. 
Firstly, the absence of a conserved invariant energy with which to label states on de Sitter has long made describing interactions more difficult than on Minkowski space. Since our $\beta_n$ are conserved in any unitary theory, and can be computed from the initial data (assuming the free theory can be solved exactly so that $\bar{\bfk}$ can be identified from $f_{\bar{\bfk}} (\eta) = f^*_{\bfk} (\eta)$), they could be used as good quantum numbers to label the different states in the Hilbert space. Furthermore, since the constraints from de Sitter isometries in the bulk may depend explicitly on the interaction Hamiltonian, it seems that performing a model-independent bootstrap (in which $c_n$ is determined by symmetries alone) is only possible in the limits where the interactions turn off---namely $\eta \to - \infty$ in the far past, and $\eta \to 0$ at late times (modulo IR divergences). Finally, although the Bunch-Davies initial condition (fixing $\alpha_{\bfk}^{\rm in} = 0$ in the past) often results in a simpler analytic structure for the correlators (e.g. removing the folded non-analyticities), this corresponds to a non-analytic boundary condition at late times ($\alpha_{\bfk} = k$) and can obscure the otherwise analytic behaviour of $c_n (\eta)$ near the conformal boundary. Rather than studying $c_n (\eta)$ itself, whose analytic structure is sensitive to the initial conditions, one might investigate the properties of objects like $\delta c_n / \delta \alpha_{n'} |_{\alpha_j = 0}$, which are always analytic outside the horizon, and more closely resemble a scattering amplitude (since it is the change in the (final) bulk state in response to a change in the (initial) boundary state). 

Furthermore, there are interesting connections between our results and other recent progress from a purely boundary perspective. 
The asymptotic expansion of our equations of motion for the wavefunction coefficients is similar to  the Hamilton-Jacobi approach to Holographic Renormalisation \cite{Bianchi:2001kw, Heemskerk:2010hk,Bzowski:2015pba}, in which the equations of motions are solved perturbatively in a derivative expansion (in order to identify the required local counterterms), 
which also underpins recent holographic approaches to inflationary cosmology \cite{Larsen:2003pf, vanderSchaar:2003sz, McFadden:2009fg,Anninos:2011ui, Bzowski:2011ab, Mata:2012bx}.
The boundary divergences which we have encountered are entirely distinct from the bulk divergences from loops, which have a separate degree of divergence and must be renormalised at any $\eta$, not just at the boundary (they are related to the divergences that appear in flat space, and can be resummed using dynamical RG \cite{Green:2020txs}).
Recently,  \cite{Cohen:2020php}  has proposed a soft de Sitter EFT for superhorizon modes---it would be interesting to explore further the connection between the Hamiltonian approach used here (in particular our transfer functions and boundary renormalisation) and the Lagrangian approach of \cite{Cohen:2020php}. 

In the future, it is imperative that we continue to exploit the connections between cosmological correlators and amplitudes. 
For instance, our equation of motion for $c_4$ takes the form of a factorisation condition, $\partial_\eta c_4 \sim c_3 c_3$, and one could explore whether there is some general factorisation theorem for cosmological correlators (akin to the factorisation of flat space amplitude)---see for instance \cite{Hertzberg:2020ird, Hertzberg:2020yzl, Pajer:2020wnj} for recent discussions of the consistency conditions imposed by factorisation when Lorentz boosts are broken. 
Further, combining such axiomatic properties of the $S$-matrix (namely unitarity, causality, locality and Lorentz invariance) gives rise to powerful UV/IR relations known as ``positivity bounds'' \cite{Adams:2006sv}. 
One key goal should be to develop analogous relations for cosmological correlators, so that our late-time (low-energy) observables can be translated into statements about UV properties of the fields/interactions present during inflation\footnote{See \cite{Baumann:2015nta} for an application of these ideas to subhorizon scattering during inflation and \cite{Baumann:2019ghk} for an anti-de Sitter analogue.}.

Cosmological correlators offer an exiting new window into the early Universe and fundamental physics in the high-energy regime. As a new generation of cosmological surveys searches the skies for signs of primordial non-Gaussianity, we must be ready with a developed theoretical framework for translating these observations into concrete features of the high-energy inflationary Universe. 
In this work, we have made a step forward in deriving model-independent constraints on the evolution of the wavefunction---from fundamental properties like unitarity, de Sitter invariance and locality---and hereby advanced this ambitious cosmological correlator programme.

\subsubsection*{Acknowledgements}
SM is supported by an Emmanuel College Research Fellowship.
SC is supported through MCIU (Spain) through contract PGC2018-096646-A-I00 and by the IFT UAM-CSIC Centro de Excelencia Severo Ochoa SEV-2016-0597 grant. 
This work has been partially supported by STFC consolidated grant
ST/P000681/1.

\appendix
\section{Comparison with Wavefunctions on Minkowski Spacetime}

In this short appendix we compare some of our above (de Sitter) discussion with the Minkowski case.

\subsection*{Equations of Motion}
Since $c_2 = i k$ and time-translation invariance sets $\partial_\eta c_n = 0$, the equations of motion \eqref{eqn:HJint} for the wavefunction coefficients become purely algebraic, 
\begin{align}
0 =  i k_T c_{\bfk_1 ... \bfk_n} +   \frac{\delta^n \mathcal{H}}{\delta \phi_{\bfk_1} ... \delta \phi_{\bfk_n} } + \text{exchange}
\end{align}
and the $n$-point coefficient is related to lower order coefficients (as well as bulk sources) by a simple factor of $i/k_T$. 

Since the mode functions are simply $f_{\bfk} (t) = e^{-i k t}/\sqrt{2i k}$, they have the property that $f_{\bar{\bfk}} (t) = f_{\bfk}^* (t)$ for $|\bar{\bfk}| = - | \bfk|$. The constants of motion \eqref{eqn:constants} are also conserved on Minkowski space (and indeed on any conformally flat spacetime), but with the simple relation for $\bar{\bfk}$ the \text{Disc} in \eqref{eqn:Disc} can now be interpreted as a genuine discontinuity of the function as one crosses $|\bfk| = 0$. 
See \cite{Goodhew:2020hob} for a detailed discussion of this important point.

\subsection*{Minkowski Isometries}
The isometries of Minkowski spacetime are,
\begin{align}
 P_\alpha &= \partial_\alpha   \;\;\;\;  \text{     and     } \;\;\;\;
 M_{\alpha \beta} =  x_{\alpha} \partial_{\beta} - x_{\beta} \partial_{\alpha}   \; .
\end{align}
Invariance of the scalar field action $S = \int d^4 x \, \mathcal{L}$ under these symmetries leads to 10 conserved Noether currents,
\begin{align}
J_{P_\alpha}^{\; \mu} &= \frac{\partial \mathcal{L}}{ \partial_\mu \phi} \partial_\alpha \phi - \delta_\alpha^\mu \mathcal{L}  \\
J_{ M_{\alpha\beta} }^{\; \mu} &= x_\alpha J_{P_\beta}^{\; \mu}  - x_\beta J_{P_\alpha}^{\; \mu}    \; ,
\end{align}
whose corresponding charges may be written in terms of the Hamiltonian density and canonical momenta,
\begin{align}
Q_{P_0}   &= \int d^3 \bfx \,  \mathcal{H}_{\bfx}  \;\;\;\; , \;\;\;\; Q_{P_i}  = \int d^3 \bfx \;  \Pi_{\bfx} \partial_i  \phi_{\bfx}    \\
Q_{ M_{0i} }  &= \int d^3 \bfx \left[ - t \,  \Pi_{\bfx} \partial_i  \phi_{\bfx}   -  \bfx_i \, \mathcal{H}_{\bfx}    \right]
\end{align}
In the quantum theory, these must be normally ordered. For instance, the charge associated with translations may be written
\begin{align}
 \hat{Q}_{P_i} = \int_{\bfk}  i \bfk_i \, \frac{ \hat{\Pi}_{-\bfk} \hat{\phi}_{\bfk} + \hat{\phi}_{\bfk} \hat{\Pi}_{-\bfk}  }{2}  =  \int_{\bfk}  \bfk_i  \, \hat{a}_{\bfk}^\dagger \hat{a}_{\bfk}
\end{align}
so that a one-particle state, $\hat{a}^\dagger_{\bfk} | 0 \rangle$ has momentum $+\bfk$.
The charge associated with boosts may be written\footnote{
Note that since the charge is time-independent we can define the Schr\"{o}dinger-picture operator using the classical $Q_{M_{0i}}$ at $t=0$. 
},
\begin{align}
\hat{Q}_{M_{0i}} =  \frac{\partial}{\partial \bfq^i} \mathcal{H}_{\bfq} \big|_{\bfq = 0} \; ,
\end{align}
where $\mathcal{H}_{\bfq} = \int_{\bfx} e^{i \bfq \cdot \bfx} \mathcal{H}_{\bfx}$ is the momentum space Hamiltonian density, such that $\mathcal{H}_{\bfq = 0}$ is the usual Hamiltonian and generator of time translations. In the free theory,
\begin{align}
\mathcal{H}_{\bfq} &= \int_{\bfk_1 \bfk_2} \frac{ k_1 k_2 - \bfk_1 \cdot \bfk_2 }{2 \sqrt{k_1 k_2}}  \, \hat{a}_{\bfk_1}^{\dagger} \hat{a}_{\bfk_2} \, \delta ( \bfq + \bfk_1 - \bfk_2 ) \;   \\
&= \mathcal{H}_{\bfq = 0}  + \bfq  \cdot \left(  \int_{\bfk}  k \hat{a}_{\bfk}^\dagger   \partial_{\bfk} \hat{a}_{\bfk}  +  \int_{\bfk} \frac{ \bfk }{2 k} \hat{a}^\dagger_{\bfk} \hat{a}_{\bfk}  \right) + \mathcal{O} ( q^2  ) \, .  
\end{align}
Note that $\hat{\mathcal{H}_{\bfq}}^\dagger = \mathcal{H}_{-\bfq}$ since the Hamiltonian is Hermitian. This generates the expected transformation of $\hat{\phi}_{\bfk}$,
\begin{align}
 [ \hat{Q}_{M_{0i}} , \hat{\phi}_{\bfk} ] = - i t \bfk \hat{\phi}_{\bfk} - i \partial_\bfk \hat{\Pi}_{\bfk} \, .  
\end{align}
which is indeed $x_0 \partial_i - \bfx_i \partial_0$ when written in position-space. 

Under an infinitesimal boost, the wavefunction $\Psi_i [\phi] = \langle \phi | \Psi_t \rangle$ changes by,
\begin{align}
 \delta \Psi_t [\phi ]  = \langle \phi | i  \hat{Q}_{M_{0i}} | \Psi_t \rangle \, . 
\end{align}
In terms of the effective action, $\Psi_t [ \phi] = e^{i \Gamma_t [ \phi]}$, 
\begin{align}
 \delta_{M_{0i}} \Gamma_t =  \int_{\bfp_1 \bfp_2} \left(  \frac{1}{2} \frac{\delta \Gamma_t}{\delta \phi_{\bfp_1}}  \frac{ \delta \Gamma_t}{ \delta \phi_{\bfp_2} }  -  \frac{ \bfp_1 \cdot \bfp_2 }{2} \phi_{\bfp_1}  \phi_{\bfp_2}  \right)  \partial_i \delta^3 \left(  \bfp_1 + \bfp_2  \right) + \partial_{\bfq^i} \mathcal{H}_{\bfq}^{\rm int} \big|_{\bfq=0}
 \label{eqn:delG}
\end{align}
where we have split $\mathcal{H}_{\bfx} = \frac{1}{2} \Pi_{\bfx}^2 + \frac{1}{2} (\partial_i \phi_{\bfx} )^2 + \mathcal{H}_{\bfx}^{\rm int}$ into a free and interacting part. 

For example, while there are many possible vacuum states, $c_{\bfk_1 \bfk_2} (t)$, for Minowski, only one is both time-translation and boost invariant, namely $c_{\bfk_1 \bfk_2} (t) = i | \bfk_1| \delta^3 ( \bfk_1+ \bfk_2 )$. In fact, suppose that we demand this particular $\hat{Q}_{M_{0i}}$ as a symmetry (i.e. invariance under boosts with speed $c=1$), then if we consider the time evolution generated by,
\begin{align}
 \mathcal{L} =  \frac{1}{2} \left(  \dot \phi^2  - c'^2 (\partial_i \phi )^2      \right)
\end{align}
we find that \eqref{eqn:delG} requires that $c' =1$. 

As another example, take the cubic interactions $\lambda \phi^3$ and $\lambda' \phi (\nabla \phi )^2$. The corresponding wavefunction coefficients are,
\begin{align}
 c_{\bfk_1 \bfk_2 \bfk_3} (t) = \lambda  \left(  \frac{i}{k_T} + \alpha (k_1, k_2, k_3) e^{- i k_T t}  \right) \delta^3 ( \bfk_1 + \bfk_2 + \bfk_3)
\end{align}
and
\begin{align}
 c_{\bfk_1 \bfk_2 \bfk_3} (t) =  \lambda' \left(  i  k_T + \alpha (k_1, k_2, k_3) e^{- i k_T t}  \right) \delta^3 ( \bfk_1 + \bfk_2 + \bfk_3)
\end{align}
respectively, where $k_T = k_1 +k_2 + k_3$. Lorentz invariance in this case requires,
\begin{align}
\delta  c_{\bfk_1 \bfk_2 \bfk_3} = \sum_j  k_j \partial_{\bfk_j}  c_{\bfk_1 \bfk_2 \bfk_3}  = 0  \, 
\label{eqn:dc3}.
\end{align}
This is satisfied by the $\alpha$-independent terms (which were generated by the Lorentz-invariant interactions), but for our initial state to preserve Lorentz-invariance we must have that $\alpha$ is a function of $k_T$ only\footnote{
By crossing symmetry, $\alpha$ can depend only on the combinations $k_T$, $k_1^2 + k_2^2 + k_3^2$ and $k_1 k_2 k_3$. Only the first of these satisfies \eqref{eqn:dc3}. 
}. Further demanding invariance under time-translations requires that $\alpha$ contains a $\delta (k_T)$, and so Lorentz invariance uniquely fixes the three-point coefficients in terms of a single constant, 
\begin{align}
 c_{\bfk_1 \bfk_2 \bfk_3} (t) = \lambda  \left(  \frac{i}{k_T} +  \text{const.} \times \delta (k_T)  \right) \delta^3 ( \bfk_1 + \bfk_2 + \bfk_3)
\end{align}
and 
\begin{align}
 c_{\bfk_1 \bfk_2 \bfk_3} (t) = \lambda'  \left(  i k_T +\text{const.} \times \delta (k_T)   \right) \delta^3 ( \bfk_1 + \bfk_2 + \bfk_3)
\end{align}

For higher derivative interactions, e.g. $\nabla_\mu \phi \nabla^\mu ( \nabla \phi )^2$ has a simple wavefunction coefficient,
\begin{align}
 c_{\bfk_1 \bfk_2 \bfk_3}  &= \frac{ i \lambda}{ k_1 + k_2 + k_3} \left[ \left(  k_1 (k_2 + k_3)  - \bfk_1 \cdot \bfk_{23}  \right) \left( k_2 k_3 - \bfk_2 
 \cdot \bfk_3  \right) + \text{2 perms.}  \right]  \, . 
\end{align}
Since,
\begin{align}
\sum_J k_J \frac{\partial}{ \partial \bfk_J} c_{\bfk_1 \bfk_2 \bfk_3}  &= - \frac{ c_{\bfk_1 \bfk_2 \bfk_3 } }{ k_T} \sum_J \bfk_J  \, . 
\end{align}
the boost constraint is satisfied. 
Note that using momentum conservation we may also write this coefficient more simply as,
\begin{align}
 c_{\bfk_1 \bfk_2 \bfk_3} =  i \lambda  \left(  k_1 \bfk_2 \cdot \bfk_3 + \text{2 perms.}     \right)
\end{align}
which is consistent with integrating the interaction by parts, contributing only a boundary term $\dot \phi (\nabla \phi)^2$ to the on-shell action. 

We will now show that the boost constraint \eqref{eqn:delG} is also satisfied by in a number of simple examples

\subsection*{Some Examples}

\subsubsection*{$\lambda \phi^4$ Interaction}

Consider a $\lambda \phi^4$ interaction. The corresponding wavefunction coefficient is,
\begin{align}
 c_{\bfk_1 \bfk_2 \bfk_3 \bfk_n}  &= \frac{ i \lambda}{ k_T  }     \; .
\end{align}
Since,
\begin{align}
 \sum_J k_J \frac{\partial}{\partial \bfk_J } c_{\bfk_1 \bfk_2 \bfk_3 \bfk_4} &= \frac{ i \lambda }{ k_T^2}  \sum_J \bfk_J  
\end{align}
the boost constraint is satisfied. 

\subsubsection*{$\lambda \phi^3$ Interaction}

Consider a $\lambda \phi^3$ interaction. The corresponding wavefunction coefficients are,
\begin{align}
 c_{\bfk_1 \bfk_2 \bfk_3}  &= \frac{  i \lambda}{ k_1 + k_2 + k_3}  \\
 c_{\bfk_1 \bfk_2 \bfk_3 \bfk_n}  &= \frac{ i \lambda}{ k_T ( k_1 + k_2 + k_{12}  ) ( k_3 + k_4 + k_{34} ) } + \text{2 perms.}   
\end{align}
and so on. Note that the following $\bfk$ derivatives vanish when the total momentum is conserved,
\begin{align}
 \sum_J k_J \frac{\partial}{\partial \bfk_J } c_{\bfk_1 \bfk_2 \bfk_3} &= \frac{ i \lambda }{ k_T^2}  \sum_J \bfk_J  \\ 
 \sum_J k_J \frac{\partial}{\partial \bfk_J} c_{\bfk_1 \bfk_2 \bfk_3 \bfk_4} &=  \left(  \frac{\lambda^2}{ (k_1 + k_2+  k_{12}) ( k_3 + k_4 + k_{34} )  } + \text{2 perms.}  \right) \frac{1+k_T}{k_T^2} \sum_J \bfk_J   \\ 
 c_{\bfk_1 \bfk_2 -\bfk_{12} } \frac{\partial}{\partial \bfk_{12}} c_{\bfk_3 \bfk_4 \bfk_{12}} + \text{2 perms} &=  \left(   \frac{ \lambda^2}{ k_{12}  ( k_1 + k_2 + k_{12} )^2   ( k_3 + k_4 + k_{34}) } + \text{2 perms.}    \right) \sum_J \bfk_J    
\end{align}
and so the boost constraint is satisfied for any $\bfk_J$.

\subsubsection*{$\lambda \phi (\nabla \phi )^2$ Interaction}

Consider a $\lambda \phi (\nabla \phi)^2$ interaction. The corresponding wavefunction coefficient is,
\begin{align}
 c_{\bfk_1 \bfk_2 \bfk_3}  &= \frac{ i \lambda}{ k_1 + k_2 + k_3} \left(  k_1 k_2  - \bfk_1 \cdot \bfk_2  + \text{2 perms.}  \right)  \, . 
\end{align}
Since,
\begin{align}
\sum_J k_J \frac{\partial}{ \partial \bfk_J} c_{\bfk_1 \bfk_2 \bfk_3}  &= - \frac{ c_{\bfk_1 \bfk_2 \bfk_3 } }{ k_T} \sum_J \bfk_J  \, . 
\end{align}
the boost constraint is satisfied. 
Note that using momentum conservation we may also write this coefficient more simply as,
\begin{align}
 c_{\bfk_1 \bfk_2 \bfk_3} =  i \lambda k_T
\end{align}
which is consistent with writing the interaction as $\nabla_\mu \phi \nabla^\mu ( \phi^2)/2$ and integrating by parts, contributing only a boundary term $\dot \phi \phi^2$ to the on-shell action.

\subsubsection*{$\lambda \nabla_\mu \phi \nabla^\mu (\nabla \phi )^2$  Interaction}

Consider a $\lambda \nabla_\mu \phi \nabla^\mu (\nabla \phi)^2$ interaction. The corresponding wavefunction coefficient is,
\begin{align}
 c_{\bfk_1 \bfk_2 \bfk_3}  &= \frac{ i \lambda}{ k_1 + k_2 + k_3} \left[ \left(  k_1 (k_2 + k_3)  - \bfk_1 \cdot \bfk_{23}  \right) \left( k_2 k_3 - \bfk_2 
 \cdot \bfk_3  \right) + \text{2 perms.}  \right]  \, . 
\end{align}
Since,
\begin{align}
\sum_J k_J \frac{\partial}{ \partial \bfk_J} c_{\bfk_1 \bfk_2 \bfk_3}  &= - \frac{ c_{\bfk_1 \bfk_2 \bfk_3 } }{ k_T} \sum_J \bfk_J  \, . 
\end{align}
the boost constraint is satisfied. 
Note that using momentum conservation we may also write this coefficient more simply as,
\begin{align}
 c_{\bfk_1 \bfk_2 \bfk_3} =  i \lambda  \left(  k_1 \bfk_2 \cdot \bfk_3 + \text{2 perms.}     \right)
\end{align}
which is consistent with integrating the interaction by parts, contributing only a boundary term $\dot \phi (\nabla \phi)^2$ to the on-shell action.

\subsection*{Locality and Analyticity}
Finally, we remark that it would be interesting to study the analogue of our ``transfer functions'' from section~\ref{sec:smallEta} on Minkowski space. Although there is no horizon, one could nonetheless imagine specifying a boundary condition for the wavefunction at $t=0$, and then write the future time evolution of the state in terms of this boundary condition. By analogy with section~\ref{sec:smallEta}, the coefficients in this expansion should be analytic functions of the momenta for times $|t| < 1/k$, i.e. before the field begins to oscillate appreciably.

\section{Some Properties of Bessel Functions}

\subsection*{Dimensional Regularisation}

In Minkowski, analytically continuing to $d$ dimensions does not change the form of the momentum-space mode function, $e^{i k_\mu x^\mu}$. 

In $d$ dimensions, the free vacuum state is given by,
\begin{align}
- \partial_\eta c_{\bfk, -\bfk}  = \Omega^{1-d} c_{\bfk, - \bfk}^2 + \Omega^{1-d} \left( k^2 + \Omega^2 m^2 \right)
\end{align}
Choosing the Bunch-Davies vacuum\footnote{
In principle one could mix in $e^{+ i k \eta}$ by an amount which vanishes as $d \to 3$, but we will not explore such schemes. 
},
\begin{align}
 c_{\bfk, -\bfk} (\eta) =  H  \Omega^d \left( -  \frac{ d - 2 \tilde{\nu} }{2} + \frac{  k \eta H_{\tilde{\nu}-1} ( - k \eta) }{ H_{\tilde{\nu}} ( - k \eta ) }  \right)
\end{align}  
where $H_\nu$ is a Hankel function of the first kind, with order given by,
\begin{align}
 \tilde{\nu} = \sqrt{ \frac{d^2}{4} - \frac{m^2}{H^2} } = \nu + \frac{3 (d-3) }{4 \nu} + \mathcal{O} ( (d-3)^2 ) \, ,  
\end{align}
in the limit $d \to 3$, where $\nu = \sqrt{ 9/4 - m^2/H^2}$.
The mode functions, given by $\Omega^{d-1} \partial_\eta \tilde{f}_{\bfk} = c_{\bfk, -\bfk} \tilde{f}_\bfk$, become,
\begin{align}
\tilde{f}_{\bfk} (\eta) = N (k) \,  \Omega^{-d/2} \, H_{\tilde{\nu}} ( - k \eta ) 
\end{align}
and its complex conjugate, where $N (k)$ is an overall normalisation which does not affect correlations. 

When regulating divergent integrals, the limit $\eta \to 0$ does not commute with taking $d \to 3$. In particular, this means that integrals of the form $\int_{\infty}^\eta d \eta/\eta^4 \partial_\eta c^I_{\bfk_1 ... \bfk_n}$ must be first carried out up to $\eta =0$ at general $d$, and only after the integral is performed can we expand in powers of $(d-3)$ (prematurely expanding the integrand in powers of $(d-3)$ would give spurious results). This scheme therefore requires integrals over products of Hankel functions\footnote{
\cite{Bzowski:2015pba} also discuss alternative schemes in which the mode function order is not analytically continued with the spacetime dimension---although computationally simpler, this results in wavefunction coefficients which do not respect the equations of motion (just like a hard cutoff). 
}. For convenience, we collect relevant identities below.

\subsection*{Triple-$H$ Integrals}

In general, an integral over $n$ Hankel functions can be expressed in closed form as a generalised hypergeometric function of $n-1$ variables. For example, we will show below that,
\begin{align}
&\int_0^\infty \frac{dt}{t} \, t^{\lambda} \; \prod_{J=1}^3 H^{(1)}_{\nu_J} ( k_J t )   \label{eqn:tripleH} \\
&=
 \frac{i 2^\lambda }{2 \pi^3 }  e^{ - i   \pi/ 2 \sum_J \nu_J  } \left(  k_3 e^{- i \pi/2}  \right)^{-\lambda} \left(  A_{\nu_1 \nu_2} + A_{\nu_1 - \nu_2} + A_{-\nu_1 \nu_2}  + A_{-\nu_1 -\nu_2}   \right)
   \; ,   \nonumber
\end{align}
\begin{align}
 A_{\nu_1 \nu_2}  &=   \left( \frac{ k_1 }{ k_3} \right)^{\nu_1} \left(  \frac{k_2}{k_3}  \right)^{\nu_2}    \Gamma ( - \nu_1) \Gamma ( - \nu_2 ) \Gamma (a_1) \Gamma (a_2) \, F_4 \left( a_1 ,a_2 \; ; \; 1+\nu_1 , 1+\nu_2 \; ; \;  \frac{k_1^2}{k_3^2} , \frac{k_2^2}{k_3^2}  \right)  \, ,  \nonumber 
\end{align}
where $F_4$ is the fourth Appell function (hypergeometric series in two variables), $a_1 = (\lambda + \sum_J \nu_J )/2$ and $a_2 = a_1 - \nu_3$, and we have assumed that $\text{Re} \left( \lambda - \sum_J \nu_J \right) > 0$ and that the $t$ contour can be deformed to $t \to \infty (1 + i \epsilon)$. 
While these special functions are not particularly enlightening (particularly when $n>3$, not much more is known about them than the original Hankel integrals), they can be simplified for particular values of the $\nu_j$.  
For example, we see from the triple-$H$ integral \eqref{eqn:tripleH} with $\lambda = d/2$ the simple pole arising whenever $\Delta_1^- + \Delta_2^- + \Delta_3^- = d$ (i.e. $a_1 = 0$) or $\Delta_1^- + \Delta_2^- + \Delta_3^+ = d$ (i.e. $a_2 =0$), whose residue is straightforwardly read off using the fact that,
\begin{align}
 F_4 ( \epsilon,  a_2 \; ; \; b_1  , b_2 \; ; \; x , y     ) = 1  + \mathcal{O} ( \epsilon )  \, . 
\end{align}
In particular, for the three-point coefficient of conformally coupled scalars ($\lambda = d/2, \; \nu_j = 1/2$ for all $j$) discussed in section~\ref{sec:examples}, the first subleading correction is,
\begin{align}
& F_4 ( a_1,  a_2 \; ; \; 1 -\nu_1  , 1 -\nu_2 \; ; \;  \frac{k_1^2}{k_3^2}  ,  \frac{k_2^2}{k_3^2}     )  \nonumber \\
&= 1  - 2 (d-3) \frac{\partial}{\partial a_1}  F_4 ( a_1,  \frac{1}{2} \; ; \; \frac{1}{2}  , \frac{1}{2} \; ; \; \frac{k_1^2}{k_3^2}  , \frac{k_2^2}{k_3^2}     ) \big|_{a_1 = 0}  + \mathcal{O} ( (d-3)^2 )  \,  \nonumber \\
&=  1  + (d-3)  \log \left(  \frac{ (  k_{23} - k_1 ) ( k_{12} - k_3) ( k_{13} - k_2 ) k_{123}   }{ k_3^4 }    \right)  + \mathcal{O} ( (d-3)^2 ) 
 \label{eqn:F4expconf}
\end{align}
when $d \to 3$. Inside the log, the numerator is a manifestly crossing-symmetric function of the $k_J$, while the denominator cancels against the $\log$ coming from the factor of $k_3^{-2a_1} = k_3^{4 (d-3)+...}$ in \eqref{eqn:tripleH}.

In order to derive \eqref{eqn:tripleH}, we will make use of the Mellin-Barnes representation of the Bessel functions, 
\begin{align}
 J_\nu (z) &= - \frac{1}{2\pi} \int_{c - i \infty}^{c + i \infty} ds  \Gamma (s) \Gamma (s-\nu)  \left(  \frac{z}{2}  \right)^{\nu - 2 s}   \; \frac{i \sin \left( \pi (s - \nu )  \right)}{\pi}   \\
i Y_\nu (z) &= - \frac{1}{2\pi} \int_{c - i \infty}^{c + i \infty} ds  \Gamma (s) \Gamma (s-\nu)  \left(  \frac{z}{2}  \right)^{\nu - 2 s}  \; \frac{ \cos \left( \pi (s - \nu )  \right)}{\pi}   \\ 
 H_\nu^{(1)} ( z) &=  - \frac{1}{2 \pi} \int_{c - i \infty}^{c + i \infty} ds \, \Gamma ( s ) \Gamma ( s - \nu ) \left(  \frac{z}{2}   \right)^{ \nu - 2 s}  \;  \frac{ e^{i \pi ( s - \nu ) } }{ \pi}
\end{align}
where $c$ is a real positive constant $> \text{Re} ( \nu )$. 
Using the relation $\pi / \sin (\nu \pi) = \Gamma (\nu) \Gamma ( 1- \nu)$, we can express\footnote{
Note that while the $J_\nu$ integral representation is usually defined with $\nu >0$, we can use that $J_{-\nu} = \cos ( \nu \pi ) J_\nu - \sin (\nu \pi ) Y_\nu$ to derive the analogous representation also for negative order. 
}
\begin{align}
 J_{\pm \nu} (z) &=  \frac{1}{2\pi i} \int_{ - i \infty}^{i \infty} ds  \frac{ \Gamma (-s) }{ \Gamma (s + 1 \pm \nu) } \left(  \frac{z}{2}  \right)^{2 s \pm \nu}     \\
  H_\nu^{(1)} (z) &= \frac{i}{ \sin (\nu \pi ) } \left(  e^{-i \pi \nu} J_\nu - J_{-\nu}   \right)  
\end{align}
and so the triple-$H$ integral is given by,
\begin{align}
 \frac{-1}{\sin (\nu_1 \pi) \sin ( \nu_2 \pi )} \left[  \mathcal{I}_{-\nu_1 -\nu_2} - e^{-i \pi \nu_1} \mathcal{I}_{-\nu_1 \nu_2} - e^{-i \pi \nu_2} \mathcal{I}_{\nu_1 - \nu_2} + e^{-i \pi (\nu_1 + \nu_2 )} \mathcal{I}_{\nu_1 \nu_2}    \right]
\end{align}
where $\mathcal{I}_{\nu_1 \nu_2}$ is the integral,
\begin{align}
\mathcal{I}_{\nu_1 \nu_2}  &= \int_0^\infty \frac{dt}{t} \, t^{\lambda} \; J_{\nu_1} ( k_1 t ) J_{\nu_2} ( k_2 t ) H_{\nu_3} ( k_3 t )  \\
&= \frac{1}{(2 \pi)^3} \int d^3 s \, 2 \pi i \delta ( \lambda + \nu_T + 2 ( s_1 + s_2 - s_3)  ) \, \frac{ \Gamma (-s_1) \Gamma (-s_2 ) \Gamma (s_3 ) \Gamma (s_3 - \nu_3 ) }{ \Gamma (s_1 + 1 + \nu_1 ) \Gamma (s_2 + 1 + \nu_2 ) }   \nonumber  \\
&\qquad \times  \left(  \frac{k_1}{2}  \right)^{2 s_1 + \nu_1}  \left(  \frac{k_2}{2}  \right)^{2 s_2 + \nu_2} \left(  \frac{k_3}{2}  \right)^{\nu_3 - 2 s_3}   \frac{ e^{i \pi (s_3 - \nu_3 )} }{\pi }   
\end{align}
where $\nu_T = \sum_J \nu_J$ and the integration is $\int d^3 s = \prod_J \int_{c_J-i \infty}^{c_J + i \infty} ds_J$ (with $c_1>0$, $c_2 >0$ and $c_3 > \text{Re} (\nu_3)$), and we have used the identity,
\begin{align}
\int_0^{\infty} \frac{dt}{t} \, t^n =  2 \pi i \delta (n) \, ,
\end{align}
to perform the $t$ integration. 
We can perform the $s_3$ integration using the delta function,
\begin{align}
&\mathcal{I}_{\nu_1 \nu_2}   \nonumber \\
&=  -  \frac{ i 2^\lambda  e^{i \pi a_2 }  }{ 2 \pi }  \frac{ k_1^{\nu_1} k_2^{\nu_2} k_3^{\nu_3}  }{ k_3^{2 a_1} }   \nonumber  \\
&\qquad \times  \int \frac{ d^2 s }{(2\pi i)^2}  \, \frac{ \Gamma (-s_1) \Gamma (-s_2 ) \Gamma (s_1 + s_2 + a_1 ) \Gamma ( s_1 + s_2 + a_2 ) }{ \Gamma (s_1 + 1 + \nu_1 ) \Gamma (s_2 + 1 + \nu_2 ) }   \left( - \frac{k_1^2}{k_3^2}  \right)^{ s_1}  \left( -  \frac{k_2^2}{k_3^2}  \right)^{s_2}     \\
&=  -  \frac{ i 2^\lambda  e^{i \pi a_2 }  }{ 2 \pi }  \frac{ k_1^{\nu_1} k_2^{\nu_2} k_3^{\nu_3}  }{ k_3^{2 a_1} }   \frac{\Gamma (a_1) \Gamma (a_2) }{ \Gamma (1+\nu_1) \Gamma (1 + \nu_2 ) } \, F_4 \left( a_1 ,a_2 \; ; \; 1+\nu_1 , 1+\nu_2 \; ; \;  \frac{k_1^2}{k_3^2} , \frac{k_2^2}{k_3^2}  \right)  \, ,  
\end{align}
where the $\delta$ function has been used to set $s_3 =  s_1 + s_2 + (\lambda + \nu_T )/2$, and $\lambda + \nu_T$ is assumed positive.  
So defining,
\begin{align}
 A_{\nu_1 \nu_2}  &=   \left( \frac{ k_1 }{ k_3} \right)^{\nu_1} \left(  \frac{k_2}{k_3}  \right)^{\nu_2}    \Gamma ( - \nu_1) \Gamma ( - \nu_2 ) \Gamma (a_1) \Gamma (a_2) \, F_4 \left( a_1 ,a_2 \; ; \; 1+\nu_1 , 1+\nu_2 \; ; \;  \frac{k_1^2}{k_3^2} , \frac{k_2^2}{k_3^2}  \right)  \, ,
\end{align}
the triple-$H$ integral is,
\begin{align}
 \frac{i 2^\lambda }{2 \pi^3 }  e^{ - i \nu_T  \pi/ 2  } \left(  k_3 e^{- i \pi/2}  \right)^{-\lambda} \left(  A_{\nu_1 \nu_2} + A_{\nu_1 - \nu_2} + A_{-\nu_1 \nu_2}  + A_{-\nu_1 -\nu_2}   \right)
\label{eqn:tripleHans}
\end{align}
(note a sign which comes from $\pi / \sin ( \nu \pi) = \Gamma (\nu ) \Gamma (1- \nu) = - \Gamma (-\nu) \Gamma ( 1+ \nu )$). 
The sum over four such functions is clearly required by the property $H_\nu (z) = e^{i \pi \nu} H_{-\nu} (z)$.
The factor of $( k_3 e^{-i \pi /2} )^{- \lambda}$ out front can be recognized as the analytic continuation from the triple-$K$ integral \cite{Bzowski:2013sza}, since $K_\nu (z) = \frac{i \pi}{2} e^{i \nu \pi/2} H_\nu \left( z e^{i \pi/2}  \right) $.

\paragraph{Appel $F_4$ Identities:}


The Appel $F_4$ function is a hypergeometric series of two variables,
\begin{align}
F_4 ( a_1 , a_2 ; b_1 , b_2  ; x , y  )  = \sum_{m,n = 0}^{\infty}  \frac{ (a_1)_{m+n} (a_2)_{m+n} }{ (b_1)_m (b_2)_n } x^m y^n \; , 
\label{eqn:F4sum}
\end{align}
which can be represented as an integral,
\begin{align}
&F_4 ( a_1 , a_2 ; b_1 , b_2  ; x , y  ) \\
 &= \frac{ \Gamma (b_1 ) \Gamma (b_2) }{ \Gamma (a_1) \Gamma (a_2) } \int_{-i \infty}^{+ i \infty} \frac{d s_1 ds_2}{(2 \pi i )^2 }  \frac{ \Gamma (s_1 + s_2 + a_1 ) \Gamma (s_1 + s_2 + a_2 )  }{ \Gamma (s_1 + b_1 ) \Gamma (s_2 + b_2)  } \Gamma (-s_1) \Gamma (-s_2) ( -x)^{s_1} (-y)^{s_2} \, .
 \nonumber 
\end{align}
These representations make manifest the symmetry properties,
\begin{align}
 F_4 ( a_1 , a_2 \; ; \; b_1 , b_2 \; ; \; x ,y ) = F_4 ( a_2 , a_1 \; ; \; b_1, b_2 \; ; \; x , y) = F_4 ( a_1 , a_2 \; ; \; b_2, b_1 \; ; \; y , x)
\end{align}
as well as various recurrence relations such as,
\begin{align}
&a_2 \left(  F_4 ( a_1 , a_2 + 1 \; ; \; b_1 , b_2 \; ; \; x ,y )  - F_4 ( a_1 , a_2 \; ; \; b_1 , b_2 \; ; \; x ,y )  \right)      \nonumber \\
&\quad= a_1 \left(  F_4 ( a_1 +1 , a_2  \; ; \; b_1 , b_2 \; ; \; x ,y ) - F_4 ( a_1 , a_2 \; ; \; b_1 , b_2 \; ; \; x ,y ) \right)
\label{eqn:F4recurrence}
\end{align}
Less obvious is the crossing relation,
\begin{align}
&F_4 \left( a_1 , a_2 \; ; \; b_1 , b_2 \; ; \; \frac{k_1^2}{k_3^2} , \frac{k_2^2}{k_3^2} \right)     \\
&=  \frac{ \Gamma (b_2 ) \Gamma (a_2 - a_1 ) }{  \Gamma ( a_2 ) \Gamma (b_2 - a_1) } \left( e^{i \pi} \frac{k_2^2}{k_3^2}  \right)^{-a_1}
 F_4 ( a_1 , a_1 - b_2 +1 \; ; \;  b_1 , a_1 - a_2 + 1  \; ; \;  \frac{ k_1^2 }{ k_2^2 } \; ; \; \frac{ k_3^2 }{k_2^2} )  \nonumber \\
 &+  \frac{ \Gamma (b_2 ) \Gamma (a_1 - a_2 ) }{  \Gamma ( a_1 ) \Gamma (b_2 - a_2) } \left( e^{i \pi} \frac{k_2^2}{k_3^2}  \right)^{- a_2 }
 F_4 ( a_2 , a_2 - b_2 +1 \; ; \;  b_1 , a_2 - a_1 + 1  \; ; \;  \frac{ k_1^2 }{ k_2^2 } \; ; \; \frac{ k_3^2 }{k_2^2} ) \nonumber
\end{align}
which exchanges $k_2 \leftrightarrow k_3$, and can be used to show that the expression \eqref{eqn:tripleHans} is crossing symmetric under the exchange of $2$ and $3$.  
 
In practice the above representation \eqref{eqn:F4sum} is of little use because the series only converges for $ | x|^{1/2} + |y|^{1/2} < 1$, which in terms of $k_1$ and $k_2$ corresponds to $k_1 + k_2 < k_3$ (which is inconsistent with the triangle inequality required by momentum conservation), and little is known about the analytic structure of $F_4$ beyond this domain. 
However, for certain values of $\left\{ \lambda, \nu_1, \nu_2, \nu_3 \right\}$, the Appell function may be expressed in terms of more elementary functions (products of two hypergeometric functions of a single variable), which can be analytically continued in a straightforward way to the physical region $k_1 + k_2 > k_3$.  
 
In particular, the fourth Appel function $F_4$ reduces to the first Appel function $F_1$ whenever $a_2 = b_2$,
\begin{align}
 &F_4 ( a_1 , b_2 \; ; \;  b_1 , b_2 \;  ; \;  \frac{-x}{(1-x)(1-y)} ,  \frac{-y}{(1-x)(1-y)} )   \nonumber \\
 &= ( 1- x)^{a_1} (1-y)^{a_1} \; F_1 \left( a_1  \; , \; b_1 - b_2 , 1 + a_1 - b_1  \;  ; \; b_1 \; ; \; x , x y   \right) \, . 
\end{align}
When $b_1$ also coincides with $b_2$, 
\begin{align}
 F_1 \left( a_1  \; , \; 0 , 1 + a_1 - b_2  \;  ; \; b_2 \; ; \; x , x y   \right) &= 
 {}_2 F_1 \left( a_1  \; , 1 + a_1 - b_2  \; ; \;  b_2 \; ; \;  x y   \right)
\end{align}
and when $b_1$ coincides with $1 + a_1 - b_2$, 
\begin{align}
 &F_1 \left( a_1  \; , \; 1 + a_1 - 2 b_2  \; , \;  b_2  \;  ; \; 1 + a_1 - b_2 \; ; \; x , x y   \right)   \\
 &=
(1- x)^{-a_1}  {}_2 F_1 \left( a_1  \; , b_2 \; ; \; 1 + a_1 - b_2  \; ; \;  - \frac{x (1-y)}{1-x}   \right)  \, .
\nonumber 
\end{align}

For example, when computing the correlators of conformally coupled scalar fields (i.e. $\nu_J = 1/2$ and $\lambda = d/2$) one needs the Appell functions, 
\begin{align}
F_4 \left(  1 , \frac{1}{2}     \; ; \;  \frac{3}{2} , \frac{1}{2} \; ; \; \frac{k_1^2}{k_3^2} , \frac{k_2^2}{k_3^2}    \right) &=  \frac{k_3}{4 k_1}    \log \left(  \frac{ k_{123} ( k_{13} - k_2   ) }{ ( k_{12} - k_3 )( k_{23} - k_1 )  }  \right)      \\
F_4 \left(  1 , \frac{3}{2}     \; ; \;  \frac{3}{2} , \frac{3}{2} \; ; \; \frac{k_1^2}{k_3^2} , \frac{k_2^2}{k_3^2}    \right) &=  \frac{k_3^2}{4 k_1 k_2}    \log \left(  \frac{ ( k_{13} - k_2 ) ( k_{23} - k_1 ) }{ k_{123} ( k_{12} - k_{3}  ) }  \right)      
\end{align}
and the expansion of,
\begin{align}
& F_4 \left(  \frac{ d/2  - 3 \nu }{2}  ,  \frac{d/2 - \nu}{2}     \; ; \;  1 - \nu  , 1 - \nu \; ; \; \frac{k_1^2}{k_3^2} , \frac{k_2^2}{k_3^2}    \right)  \nonumber  \\
 &=  1 -  \frac{1}{4} \left( \frac{d-3}{2}  -   3 (\nu - \tfrac{1}{2} )  \right)  \log \left(  \frac{ (k_1 - k_{23} ) ( k_3 - k_{12} ) ( k_2 - k_{13} )  k_{123}   }{ k_3^4 }   \right) + ...     
\end{align}
near $d=3$ and $\nu=1/2$. This gives a triple-$H$ integral,
\begin{align}
 \frac{ 8 i }{ \sqrt{2 \pi^3  k_1 k_2 k_3} } \left[ \frac{1}{ d-3 - 6 \left( \nu-\tfrac{1}{2} \right) } \left( 1 - \frac{d-3}{2} \gamma_E - \left(  \nu - \tfrac{1}{2} \right) \log ( 8 k_1 k_2 k_3 )     \right)   - \frac{1}{2} \log \left( - i k_T  \right)      \right]  \, . 
\end{align}
Once multiplied by $N_1 N_2 N_3$, it gives a three-point wavefunction coefficient,
\begin{align}
\alpha_3 =   2 i  \left[ \frac{1}{ d-3 - 6 \left( \nu-\tfrac{1}{2} \right) }  - \frac{1}{2} \gamma_E   - \frac{1}{2} \log \left( - i k_T  \right)      \right]  \, .  
\end{align}
The pole depends on the scheme, but there is always an anomalous violation of scale invariance due to the $\log (k_T)$ which is scheme-independent. See e.g. \cite{Bzowski:2015pba} for a general demonstration of the scheme-independence of the anomaly in 3d scalar CFTs.

  
When $\nu_2$ and $\nu_3$ are both conformally coupled scalars (but $\nu_1$ is arbitrary), the relevant Appell $F_4$ is given by the identity,
\begin{align}
2\,  F_4 \left( a_1 , a_1 + \frac{1}{2} \; ; \; b_1 , \frac{1}{2} \; ; \; x , y  \right)  &=
 \left( 1 + \sqrt{y} \right)^{-2a_1} \;  {}_2 F_1 \left( a_1 , a_1 + \frac{1}{2} \; ; \; b_1 \; ; \;  \frac{x}{(1 + \sqrt{y} )^2 }  \right)  \nonumber  \\
&+  \left( 1 - \sqrt{y} \right)^{-2a_1} \;  {}_2 F_1 \left( a_1 , a_1 + \frac{1}{2} \; ; \; b_1 \; ; \;  \frac{x}{(1 - \sqrt{y} )^2 }  \right) \; . 
\label{eqn:F4identity3}
\end{align}
Finally, we also note that when $\lambda = 1$ (i.e. an integral over three $H_{\nu_J} ( k_J t)$ with no overall factor of $t$), the Appell function factorises, since
\begin{align}
 F_4 ( a_1 , a_2 \; ; \; b_1 , b_2 \; ; \;  X(1-Y) ,  Y (1-X) )= {}_2 F_{1} \left(  a_1 , a_2 \; ; \; b_1  \; ; \;  X \right) \; {}_2 F_{1} \left(  a_1 , a_2  \; ; \;  b_2  \; ; \;  Y \right) \label{eqn:F4identity4}
\end{align}
when $a_1 + a_2 +1 = b_1 + b_2$. Both \eqref{eqn:F4identity3} and \eqref{eqn:F4identity4} may then be used to analytically continue into the physical region $k_1 + k_2 < k_3$.

\bibliographystyle{elsarticle-num}
\bibliography{references.bib}
\end{document}